\documentclass{aa}  

\usepackage{graphicx}
\usepackage{txfonts}
\usepackage[colorlinks = true,
            linkcolor = blue,
            urlcolor  = blue,
            citecolor = blue,
            anchorcolor = blue]{hyperref}
\begin{document}

   \title{Stellar populations with MEGARA: the inner regions of NGC 7025}

   %\subtitle{I. Overviewing the $\kappa$-mechanism}

   \author{M. Chamorro-Cazorla\inst{1,2},
          A. Gil de Paz\inst{1,2},
          A. Castillo-Morales\inst{1,2},
          B. T. Dullo\inst{1,2},
          J. Gallego\inst{1,2},
          E. Carrasco\inst{3},
          J. Iglesias-P\'aramo\inst{4},
          R. Cedazo\inst{5},
          M. L. Garc\'ia-Vargas\inst{6},
          S. Pascual\inst{1,2},
          N. Cardiel\inst{1,2},
          A. P\'erez-Calpena\inst{6},
          P. G\'omez-\'Alvarez\inst{6},
          I. Mart\'inez-Delgado\inst{6}
          \and
          C. Catal\'an-Torrecilla\inst{7}
          }

   \institute{\inst{1}Departamento de F\'isica de la Tierra y
  Astrof\'isica, Universidad Complutense de Madrid, E-28040 Madrid,
  Spain\\
  \email{mchamorro@ucm.es}\\
  \inst{2}Instituto de F\'isica de Part\'iculas y del Cosmos IPARCOS, Facultad de Ciencias F\'isicas, Universidad Complutense de Madrid, E-28040 Madrid,
  Spain\\
  \inst{3}Instituto Nacional de Astrof\'isica, \'Optica
  y Electr\'onica, Luis Enrique Erro No.1, C.P. 72840, Tonantzintla,
  Puebla, Mexico\\
  \inst{4}Instituto de Astrof\'isica de Andaluc\'ia-CSIC,  Glorieta de
la Astronom\'ia s/n, 18008, Granada, Spain\\ 
  \inst{5}Universidad Polit\'ecnica de Madrid, Madrid, Spain\\
  \inst{6}FRACTAL S.L.N.E. C/ Tulip\'an 2, p13, 1A. E-28231 Las Rozas de Madrid (Spain)\\
  \inst{7}Centro de Astrobiolog\'ia (CAB, CSIC-INTA), Carretera de Ajalvir km 4, E-28850 Torrej\'on de Ardoz, Madrid, Spain
             }

   \date{\today}

% \abstract{}{}{}{}{} 
% 5 {} token are mandatory
 
  \abstract
   {This paper aims to determine the capabilities of the MEGARA@GTC instrument integral-field unit to study stellar populations and exploit its combination of high spectral (R $\sim$ 6,000, 12,000 and 20,000) and spatial (0.62\arcsec) resolutions within its 12\farcs5 $\times$ 11\farcs3 field of view. We do so by analysing the commissioning data of the nearby S0a galaxy NGC~7025.}
   {We pursue to establish a systematic method through which we can determine the properties of the stellar populations in the observations made with MEGARA, more specifically within the MEGADES legacy project and, for this paper in particular, those of the stellar populations of NGC~7025.}
   {We use MEGARA observations of galaxy NGC~7025 taken during the commissioning phase of the instrument. We apply different approaches to estimate the properties of the stellar populations with the highest possible certainty. Besides the specific study on NGC~7025 and in the context of the MEGADES survey, we have carried out a number of tests to determine the expected errors (including potential biases) in such star formation history (SFH) derivations as a function of these parameters, namely spectral setup, signal-to-noise ratio, $\sigma$ and the SFH itself.}
   {All the studies we conduct (both full spectral fitting and absorption line indices) on the stellar populations of NGC~7025 indicate that the stars that form its bulge have supersolar metallicity and considerably old ages ($\sim$10\,Gyr), in general. We determined that the bulge of NGC~7025 has a mild negative mass-weighted age gradient using three different combinations of MEGARA spectral setups. Regarding its more detailed star formation history, our results indicate that, besides a rather constant star formation at early epochs, a peak in formation history of the stars in the bulge is also found 3.5-4.5 Gyr ago, partly explaining the mass-weighted age gradients measured.}
   {The scenario presented in NGC~7025 is that of an isolated galaxy under secular evolution that about 3.5-4.5\,Gyr ago likely experimented a minor merger (mass ratio 1/10) that induced an increase in star formation and also perturbed the morphology of its outer disc. Besides these specific results on NGC~7025, we report on different lessons learned for the ongoing exploitation of the MEGADES survey with GTC such as the need to obtain combined observations in the LR-B + LR-V setups and a signal-to-noise ratio of at least 20 per $\AA$.}

   \keywords{Galaxies: bulges --
            Galaxies: evolution --
            Galaxies: stellar content --
            Techniques: imaging spectroscopy --
            Instrumentation: spectrographs
               }
\authorrunning{M. Chamorro-Cazorla et al.}
   \maketitle

%
%________________________________________________________________

\section{Introduction}

The study of stellar populations provides vital information about the processes that have shaped the galaxies from their formation to the present day. Availability of spatially-resolved information on the age and metallicity of stars, both radially and in two dimensions, enables us to investigate how the different parts of galaxies have assembled. The different mechanisms that contribute to the formation and evolution of galaxies (monolithic collapse, major and minor galaxy mergers, secular processes, etc.) leave a distinct imprint on the 2D photometric and chemodynamical properties of galaxies of galaxies and, thanks to the latest developments in instrumentation, we are ready to further explore their roles.

One of these latest instrumental developments is MEGARA ({\it Multi-Espectrógrafo  en  GTC de  Alta  Resolución  para  Astronomía}), a new optical Integral-Field Unit (IFU) and Multi-Object Spectrograph (MOS) of the 10.4\,m GTC (\citealt{armando_2018SPIE};  \citealt{carrasco_2018SPIE}). In this paper we have used commissioning observations with the MEGARA IFU, which covers a region of 12.5\,$\times$\,11.3\,arcsec$^{2}$, to explore the 2D stellar content of the bulge component of the nearby S0a galaxy NGC~7025. The 567 hexagonal fibres encompassed the central 4.375\,$\times$\,3.955\,kpc$^{2}$, assuming a sampling scale of 0.350\,$\pm$\,0.001\,kpc\,arcsec$^{-1}$ \citep{Rizzo_2018}.

Aside from providing clues on the star formation history of the central regions of early-type spiral galaxies, the analysis of the stellar populations in NGC~7025 constitutes a pilot study that will help us to determine what we can learn from the ongoing MEGARA Guaranteed Time observations that are being obtained as a part of the MEGARA Galaxy-Disks Evolution Survey (MEGADES; Chamorro-Cazorla et al$.$ 2021, in prep.). The first stage of this project encompasses the analysis of the central regions of a total of 50 nearby galaxies. Upon completion the MEGADES legacy project fully exploits 400 hours of  MEGARA observations. Currently, the inner regions of a total of 41 galaxies have already been observed for $\sim$130 hours, including both Guaranteed Time observations and three related open-time programs (GTC153-18B, GTC147-19A, GTC117-19B). The initial goal with MEGADES is to provide a detailed study of the inner regions of nearby disk galaxies, both in terms of their spectro-photometric and chemical evolution and their dynamical characterisation, by disentangling the contribution of in-situ and ex-situ processes to the history of star formation and effective chemical enrichment of these regions. At a later stage, we will extend this study further out in galactocentric distance.

In addition to the study of their stellar populations and chemical enrichment, the dynamical characterisation of these inner regions will also allow us to identify and study potential galactic winds (GWs) to be found in these regions. Galactic winds constitute an important mechanism for redistributing dust and metals both in galaxies and towards the intergalactic medium \citep{Veilleux_2005}. It is also a mechanism that has been invoked to reproduce the scale relationships observed in galaxies \citep{Dutton_2009}, as well as to understand the apparent discrepancies between the theoretical and observed luminosity functions \citep{Springel_2003} and to understand the evolution of galaxies (especially of high-redshift objects) through the green valley \citep{Hopkins_2008}.

In order to carry out this pilot study for MEGADES, we use data taken during the commissioning of the MEGARA instrument on the galaxy NGC~7025. This galaxy is classified as an unbarred S0a galaxy and it was among the first objects observed with the MEGARA IFU. We selected this galaxy from the MEGADES sample because of the presence of residual interstellar \ion{Na}{i}, previously detected with the CALIFA survey \citep{sanchez2012}. The main incentive in studying galaxies with this feature is our interest in studying galaxies that are possible candidates for hosting galactic winds. NGC~7025 shows bar-like non-circular flows despite the fact that, photometrically at least, it is not considered to be a barred galaxy \citep{Holmes_2015}. Besides,  NGC~7025 is considered as a merger remnant with evident tidal features, yet relatively isolated \citep{Barrera_Ballesteros_2015}. The photometric and kinematic analyses by \citet{Dullo_2019} have shown that the bulge is a fast rotator with a low Sérsic index n $\sim$ 1.80. However, detailed stellar population analysis of the bulge is crucial if we are to properly understand this galaxy's formation.

\begin{table}
\caption{Global properties of NGC~7025.}              % title of Table
\label{table:NGC7025_properties}      % is used to refer this table in the text
\centering                                      % used for centering table
\begin{tabular}{c c c}          % centered columns (4 columns)
\hline\hline                     % inserts two horizontal lines 
\noalign{\smallskip}
Property & NGC~7025  & References \\    % table heading
\hline                                   % inserts single horizontal line
\noalign{\smallskip}
    Morphology & S0a & (1) \\      % inserting body of the table
    Redshift & 0.0172 $\pm$ 0.0001 & (2) \\
    R.A. (J2000) & 21$^\textup{h}$07$^\textup{m}$47$^\textup{s}$\hspace{-1mm}.34 & (3) \\
    Dec. (J2000) & +16$^\textup{d}$20$^\textup{m}$09$^\textup{s}$\hspace{-1mm}.1 & (3) \\
    $\log(M_{\star}/M_{\odot})$ & 11.26 & (4) \\
    $\log(M_{\star,\text{bul}}/M_{\odot})$ & 10.65 & (5) \\
    $r_{e}$ (\arcsec) & 23.0 & (6) \\
    $r_{e,\text{bul}}$ (\arcsec) & 4.68 & (5) \\
    $m_{g}$ & 12.85 & (1)/(7) \\
    $M_{i,\text{bul}}$ & -22.59 & (5) \\
    $D_{L}$ (Mpc) & 74.7 $\pm$ 0.9 & (2) \\
    PA (deg) & 44.78 & (2) \\
    $L_{H_\alpha}$ (erg\,s$^{-1}$)& 36.1 $\times$ 10$^{39}$ & (8) \\
\hline       %inserts single line
\end{tabular}
\parbox{87mm}{\footnotesize References. (1): \cite{Garcia_Benito_2015}. (2) \cite{Rizzo_2018}. (3) NASA/IPAC Extragalactic Database. (4) \cite{Barrera_Ballesteros_2015}. (5) \cite{Dullo_2019}. (6) \cite{Sanchez_Blazquez_2014}. (7) Petrosian magnitudes as given by SDSS DR7 database corrected for Galactic extinction. (8) \cite{Gomes_2016}.}
\end{table}

\begin{figure}[h]
	\centering
    \includegraphics[width=1\linewidth]{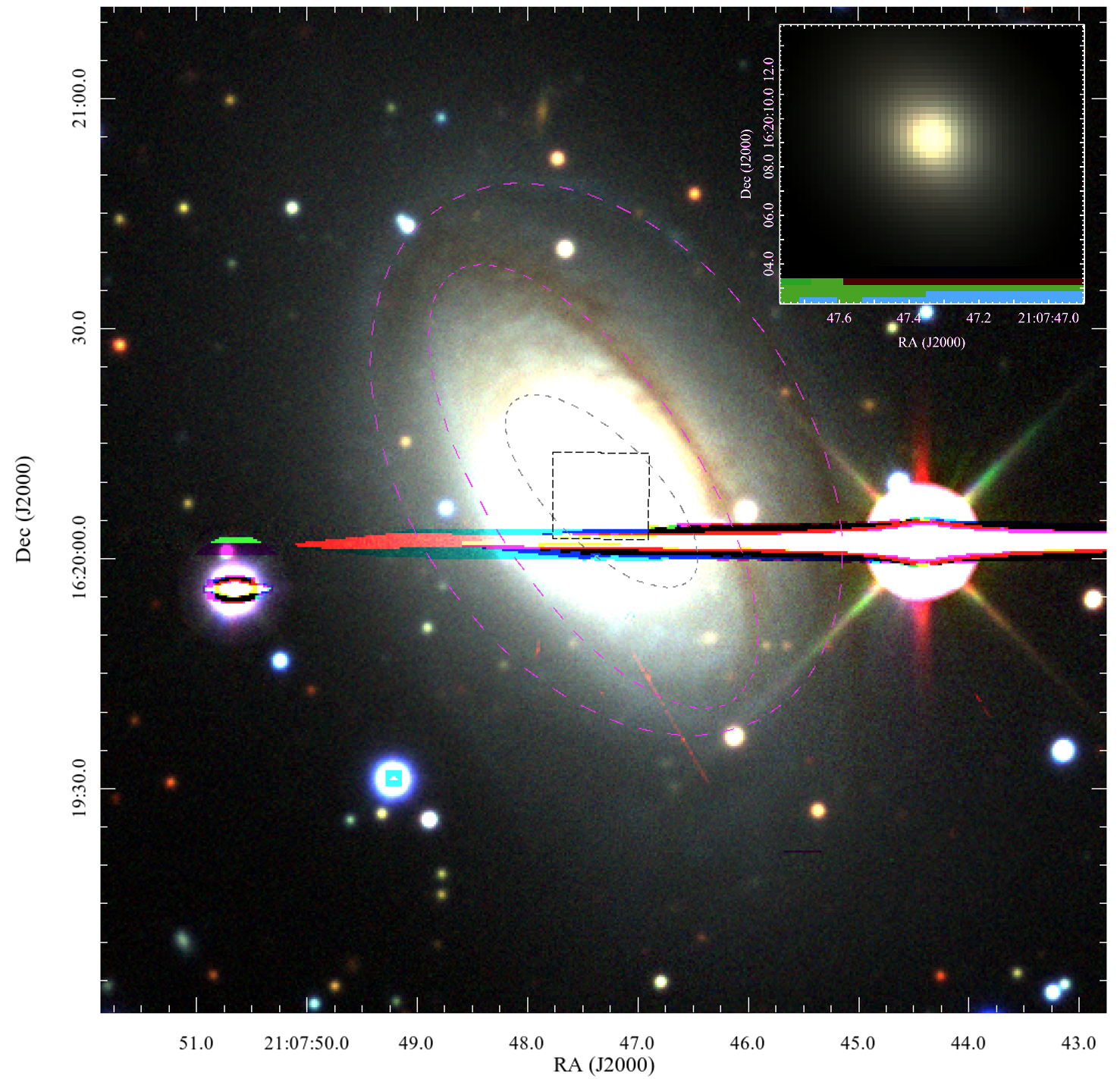}
	\caption{RGB false-colour DECam (g, r and z filters) coadded image of NGC~7025. The black dashed-line rectangular region represents the MEGARA field of view, shown in detail at the upper right-hand side of the figure. The grey dashed-line ellipse traces the inner star-forming ring identified by \citet{Barrera_Ballesteros_2015} and the magenta dashed-line ellipses trace two outer dust-lanes/tidal features clearly visible in this image. Note that while the innermost of these two features clearly resembles a dust lane with the NE half of the disk being in the foreground, the outermost one appears as a more diffuse tidal feature, which, along with its isophotal twist, led \citet{Barrera_Ballesteros_2015} to classify this object as a merger remnant.}
	\label{fig:NGC7025_ellipses}
\end{figure}

In Figure \ref{fig:NGC7025_ellipses} we present a false-colour DECam Legacy Imaging Surveys of NGC~7025 \citep{Dey_2019}. Overploted, we highlight the field of view (FoV) of MEGARA (central rectangular region with black dashed line), shown in detail at the upper right-hand side of the figure, some outer dust-lanes/tidal features (dashed-line grey and magenta ellipses) either visible in this image or previously reported \citep{Barrera_Ballesteros_2015}.

Throughout this paper, we assume a cosmological model with H$_{0}$ = 70\,km\,s$^{-1}$\,Mpc$^{-1}$, $\Omega_{\Lambda}$ = 0.7 and $\mathrm{\Omega_{m}}$ = 0.3. From the redshift of the galaxy (z=0.0172; \citealt{Rizzo_2018}), we obtain a luminosity distance of 74.7\,Mpc for NGC~7025.

\section{Observations}

\subsection{MEGARA spectroscopy}

MEGARA provides good enough spatial resolution at the distance of our object for the purpose of the objectives of the MEGADES survey. Each fibre captures the light of an hexagonal spaxel circumscribed in a circle of 0.62\,arcsec in diameter (217\,pc at the distance of NGC~7025). Additionally, apart from the 567 fibres covering the inner regions of our target, the MEGARA IFU pseudo-slit includes 56 additional fibres that are distributed in 8 different bundles placed between 1.75-2\,arcmin around the IFU and that are devoted to measuring the sky background simultaneously to the observations. The spectral range covered by the MEGARA instrument ranges from 3653 to 9700\,\r{A} with low (LR), medium (MR) and high spectral (HR) resolutions of R $\sim$ 6,000, 12,000 and 20,000; respectively, thanks to the use of 18 different VPHs (Volume Phase Holographic gratings) that are available to both the IFU and MOS modes.

In preparation for the optimisation of the MEGADES survey, we carried out commissioning observations on NGC~7025 taken with all 18 MEGARA VPHs. After the initial analysis conducted by \citet{Dullo_2019} we decided to focus, for both this work and for the observations of the inner regions of the MEGADES sample, on the low spectral resolution (LR; R$\sim$6,000) VPHs since they allow us to cover a wider spectral range with still good enough spectral resolution.  Besides, the high central velocity dispersion of this galaxy, $\sigma$\,$\sim$\,250\,km\,s$^{-1}$ \citep{Dullo_2019}, limits the advantage (in terms of information content) of using MR and HR VPHs. For the study of the disks of the MEGADES sample, especially in the case of low-inclination galaxies, we might have to also rely on medium (MR; R$\sim$12,000) and high-resolution (HR; R$\sim$20,000) data.

\begin{figure*}[h]
    \center
	\includegraphics[trim={0 1.1cm 0 1.2cm}, clip, width=0.4\linewidth]{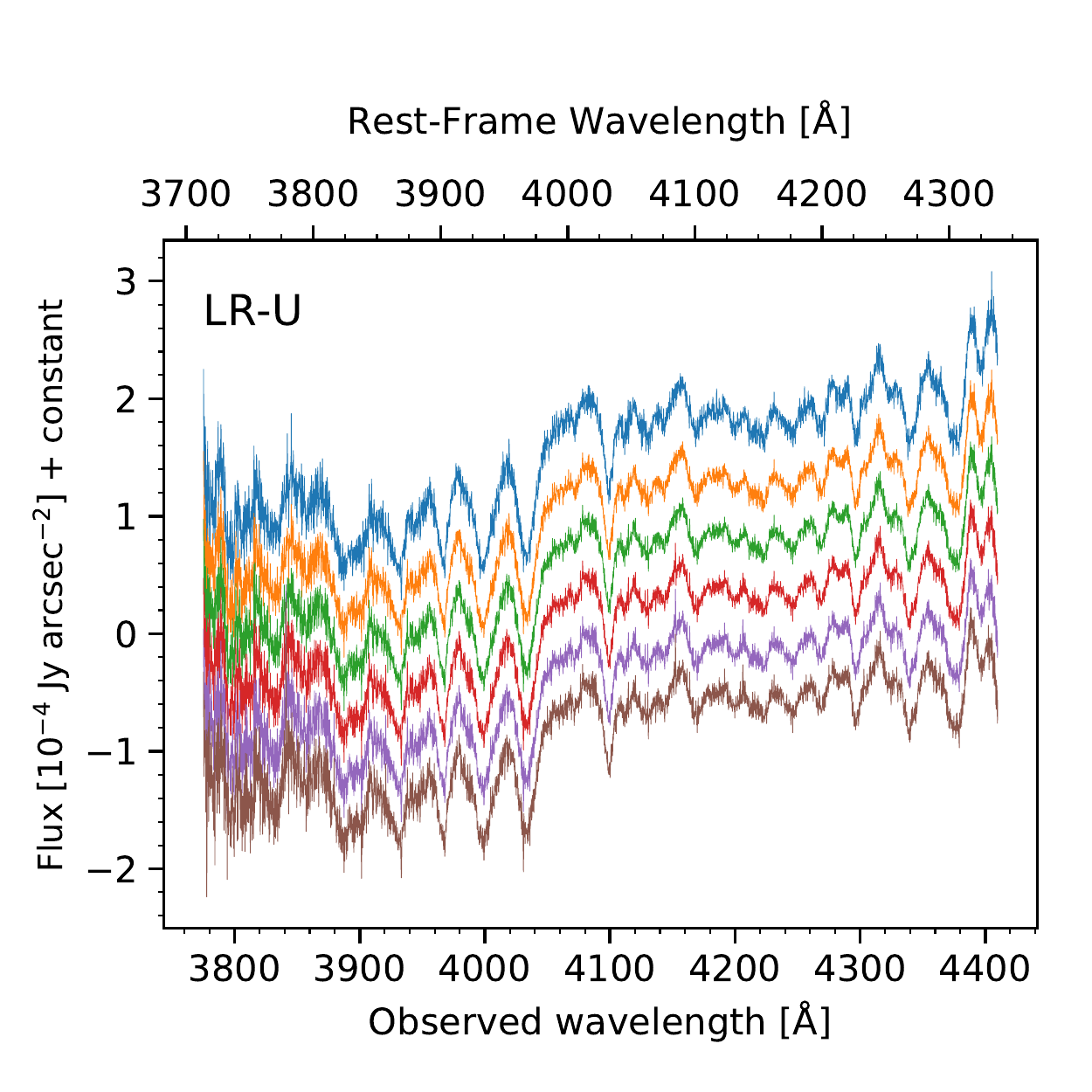}
	\includegraphics[trim={0 1.1cm 0 1.2cm}, clip, width=0.4\linewidth]{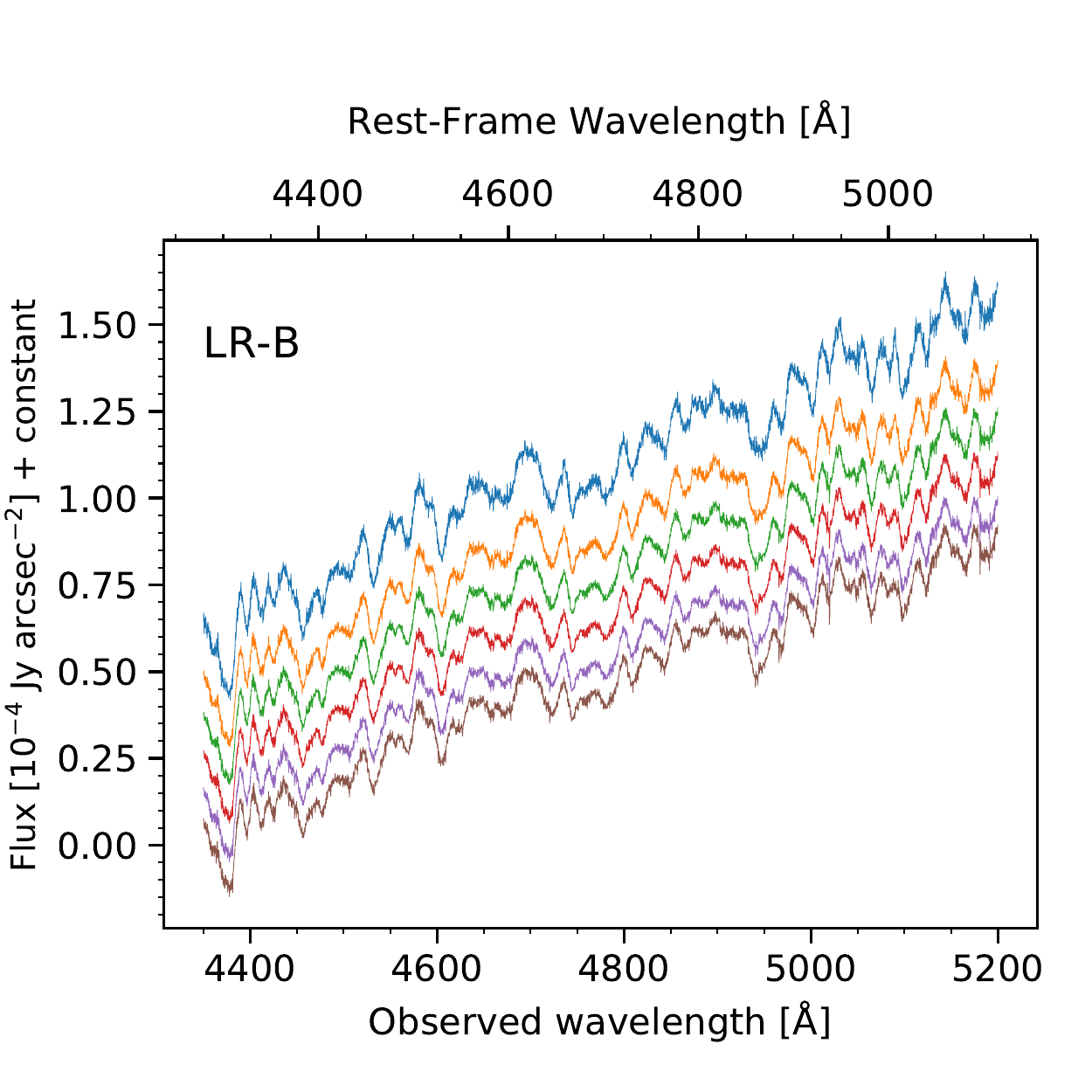}
    \includegraphics[trim={0 0 0 2cm}, clip, width=0.4\linewidth]{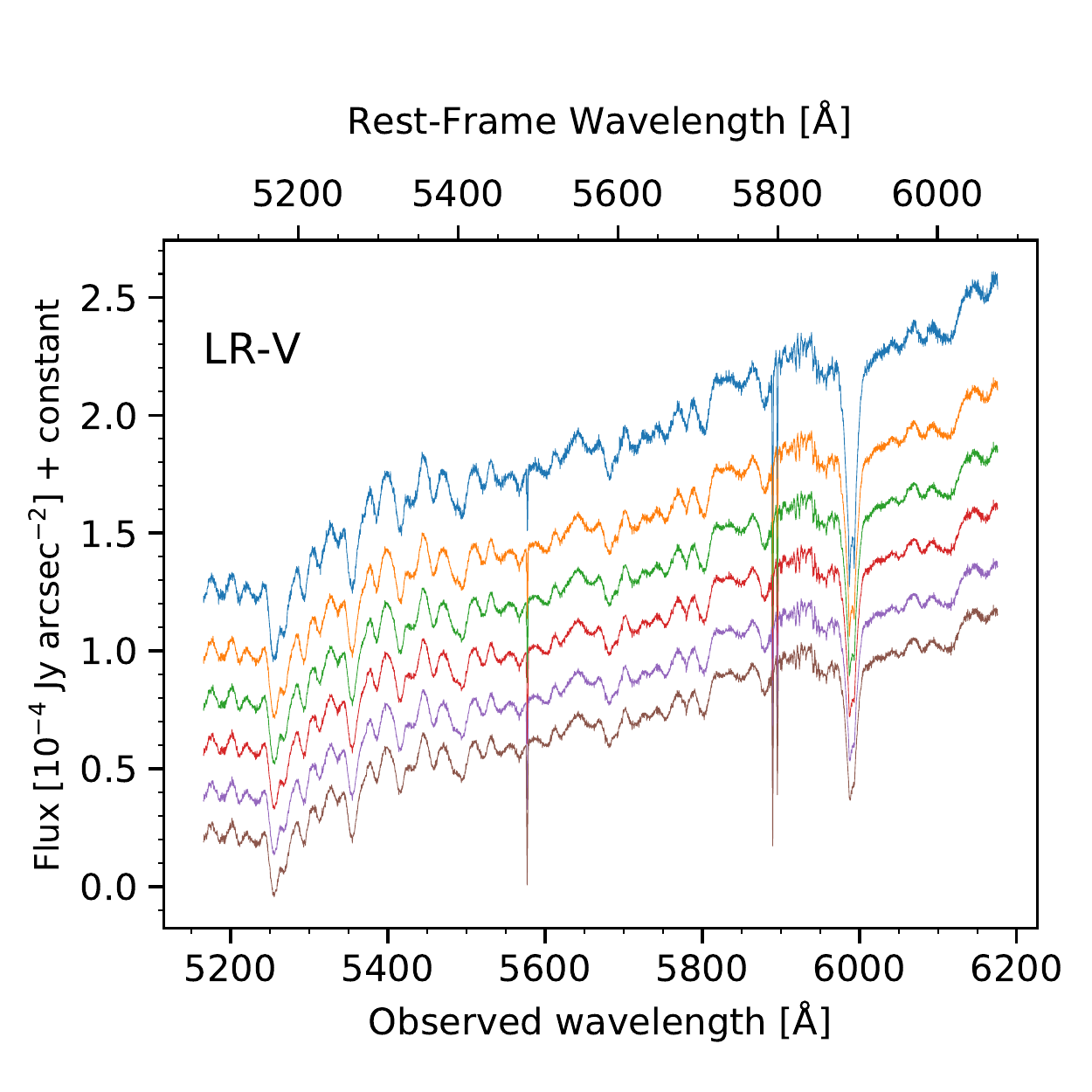} % Para recortar imágenes: [trim={<izquierda> <abajo> <derecha> <arriba>},clip]
	\includegraphics[trim={0 0 0 2cm}, clip, width=0.4\linewidth]{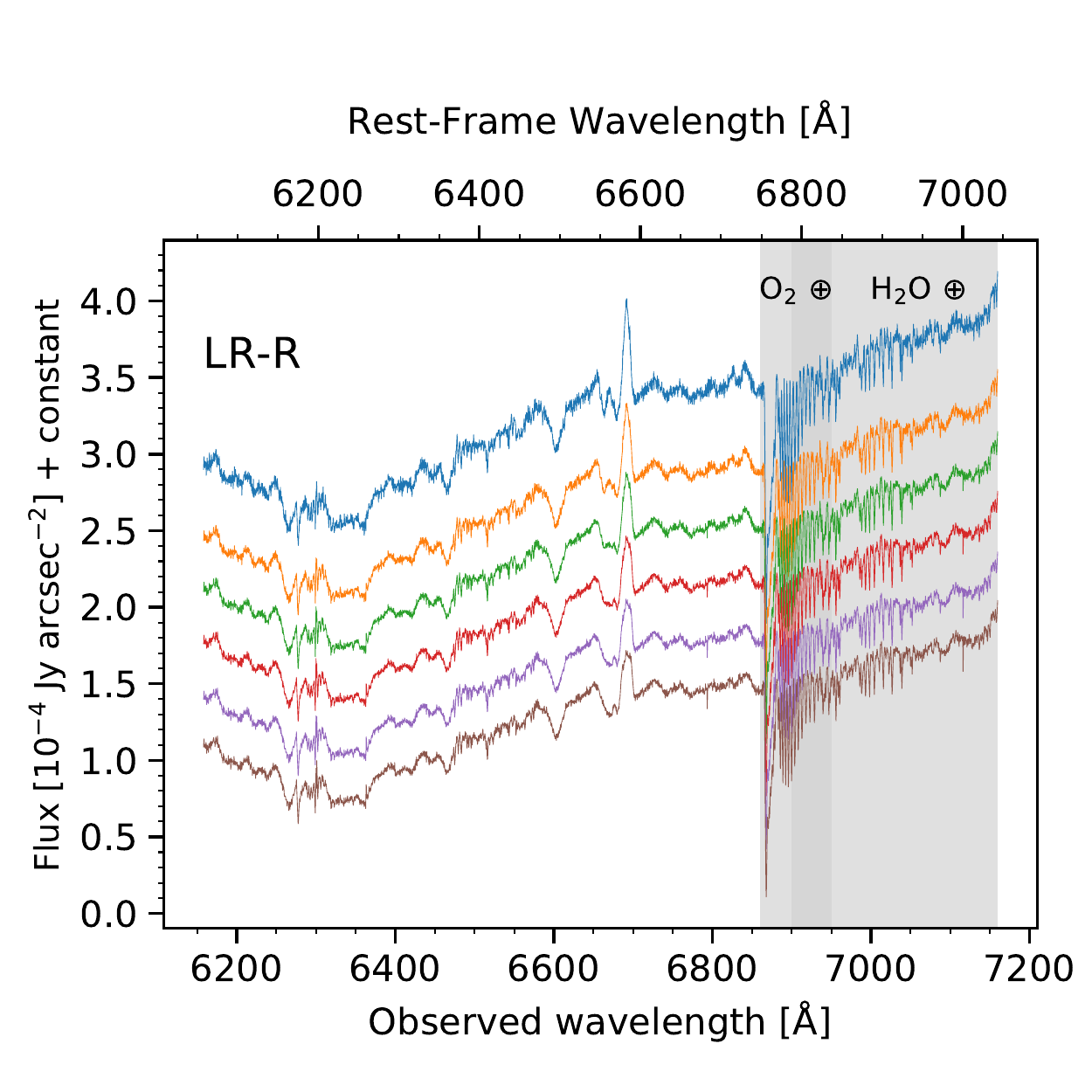}
	\includegraphics[trim={1cm 0 0 2cm}, clip, width=0.8\linewidth]{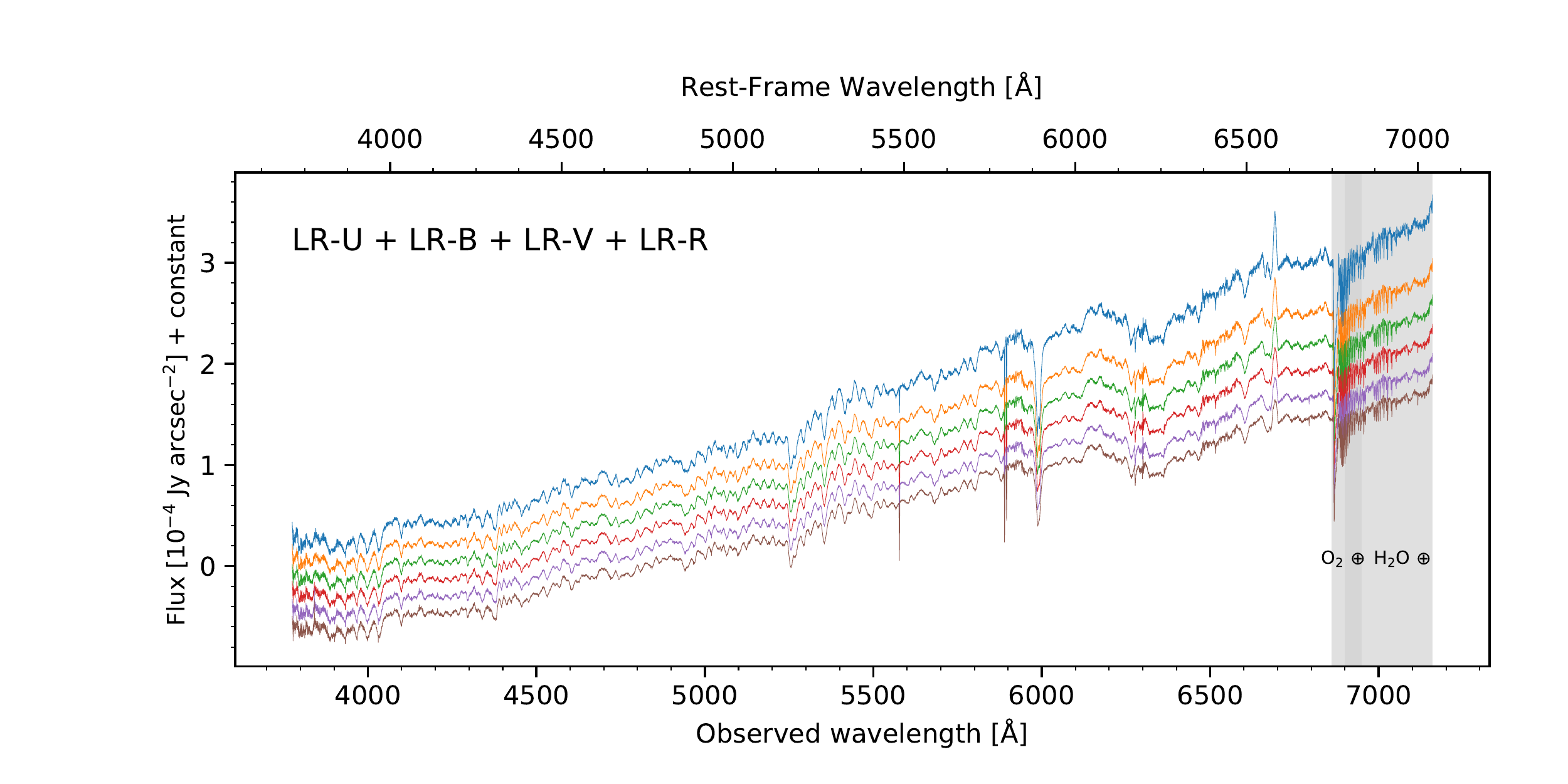}
	\caption{Integrated spectra, corresponding to the different elliptical rings depicted in Figure \ref{fig:elliptical_rings}. Each of the top four panels corresponds to one of the four different VPHs employed in this work. The bottom panel shows the spectra obtained by combining the spectra from the other four observations. Within each panel, the spectra have been vertically shifted downwards for display purposes, and are sorted from the innermost to the outermost radius (in colour order: blue, orange, green, red, purple and brown). For the spectra obtained with the LR-R VPH, we also marked the region where the spectra are severely affected by telluric absorption bands.}
	\label{fig:elliptical_spectra}
\end{figure*}

Among the low resolution VPHs, we have used the observations obtained with the four bluest LR VPH gratings of all those available in MEGARA. These VPHs are: VPH405-LR (henceforth LR-U), VPH480-LR (LR-B), VPH570-LR (LR-V) and VPH675-LR (LR-R) (see Figure \ref{fig:elliptical_spectra} for the different NGC~7025 spectra taken with these instrument setups and the concatenation of all of them). The data for each of these VPHs are shown in Table \ref{table:VPH_characteristics}. The selection of these VPHs is motivated by the fact that redwards of LR-R, telluric absorptions and the residuals from bright sky emission lines prevent improving the results of our stellar population analysis at the signal-to-noise ratios reached by our data. Although all four LR VPHs were analysed (see Appendix~\ref{section: appendix}), for the detailed study of the stellar populations in NGC~7025 we have mainly focused on the LR-B and LR-V VPHs since these setups yield better signal-to-noise ratios for a given exposure time (compared to LR-U) and include multiple spectral features sensitive to the star formation and chemical histories (compared to the more limited information content in this regard of the LR-R spectral range). Nevertheless, the addition of LR-R data will be key for the study of Galactic Winds in emission, one of the main objectives of MEGADES as a survey (Chamorro-Cazorla et al$.$ 2021, in prep.). 

\begin{table}
\caption{MEGARA VPHs characteristics.}              % title of Table
\label{table:VPH_characteristics}      % is used to refer this table in the text
\centering                                      % used for centering table
\begin{tabular}{c c c c}          % centered columns (4 columns)
\hline\hline                     % inserts two horizontal lines 
\noalign{\smallskip}
VPH & Spectral Coverage  & Rec.Disp. & $\mathrm{\Delta\lambda_{FWHM,c}}$ \\    % table heading
 & [$\AA$] & [$\AA$ pix$^{-1}$] & [$\AA$] \\
(1) & (2) & (3) & (4) \\
\hline                                   % inserts single horizontal line
\noalign{\smallskip}
    LR-U & 3728 - 4342 & 0.186 & 0.67 \\      % inserting body of the table
    LR-B & 4351 - 5251 & 0.230 & 0.79 \\
    LR-V & 5166 - 6176 & 0.270 & 0.94 \\
    LR-R & 6158 - 7288 & 0.310 & 1.11 \\
\hline                                             %inserts single line
\end{tabular}
\parbox{87mm}{\footnotesize (1) VPH name; (2) Wavelength coverage common to all fibres (in \AA); (3) Reciprocal dispersion (in \AA\,pix$^{-1}$); (4) Spectral resolution at the corresponding central wavelength of each setup.}
\end{table}

The spectra were obtained on the night of 2017 August 1st. For LR-U, LR-B and LR-V we took three exposures of 3$\times$900\,s and for LR-R three exposures of 3$\times$600\,s. The observing log for these observations is given in Table~\ref{table:log}. We have used different standard stars for the flux calibration. In particular, we have employed spectra from the standard star BD+33~26~42 for the flux calibration of LR-B (3$\times$60 s exposures), LR-V (3$\times$30 s exposures) and LR-R (3$\times$15 s exposures). For the flux calibration of LR-U we used data from the standard star BD+174708 (3$\times$15 s exposures). These spectra were obtained on the night of 2017 July 30. We also made use of the corresponding calibration data, including Th-Ne and Th-Ar arc-lamps, halogen lamps and twilight spectra.

\begin{table}
\caption{NGC~7025 MEGARA Observing log.}              % title of Table
\label{table:log}      % is used to refer this table in the text
\centering                                      % used for centering table
\begin{tabular}{c c c c}          % centered columns (4 columns)
\hline\hline                     % inserts two horizontal lines 
\noalign{\smallskip}
VPH & Date & Exp. Time & Seeing ('') \\    % table heading
(1) & (2) & (3) & (4) \\
\hline                                   % inserts single horizontal line
\noalign{\smallskip}
    LR-U & 01 Aug. 2017 & 3x900 s & 0.6 $\pm$ 0.2 \\      % inserting body of the table
    LR-B & 01 Aug. 2017 & 3x900 s & 0.6 $\pm$ 0.2 \\
    LR-V & 01 Aug. 2017 & 3x900 s & 0.6 $\pm$ 0.2 \\
    LR-R & 01 Aug. 2017 & 3x600 s & 0.6 $\pm$ 0.2 \\
\hline                                             %inserts single line
\end{tabular}
\parbox{87mm}{\footnotesize (1) VPH name; (2) Observing date; (3) Total number of exposures and time per exposure (in seconds); (4) Seeing, as provided in the GTC log files (in arcsec).}
\end{table}
  
\subsection{Data reduction}
 
Data reduction has been performed using the
MEGARA Data Reduction Pipeline version 0.6.1
\citep[DRP hereafter]{sergio_pascual_2018_1240478}, described in \cite{africa_castillo_morales_2020_3932063}\footnote{\href{https://doi.org/10.5281/zenodo.1974953}{DOI: 10.5281/zenodo.1974953}} for the MEGARA 2D spectroscopic observations. 

The first step, before executing any of the MEGARA DRP reduction recipes, is to cleanse the images of those cosmic rays. Although three different raw images were initially obtained for each setup with the idea of getting rid of cosmic rays through the median combination of these images, some affected pixels still survived to this procedure. In a second step, we used the CLEANEST software ~\citep{Cardiel_2020} to interactively interpolate the remaining bad pixels. After this step, we continued with the data reduction using the DRP. Firstly, the bad pixels are masked and the bias subtraction is performed. This bias subtraction is done by using a Master Bias obtained with the \textit{MegaraBiasImage} task. After this step is completed, the path of the light coming from each fibre is traced through the CCD using the DRP tasks \textit{MegaraTraceMap} and \textit{MegaraModelMap} on a series of consecutive flat halogen lamp frames. Once we trace the spectra, the wavelength calibration is carried out with the routine \textit{MegaraArcCalibration}, using ThAr or ThNe arc-lamp frames depending on the wavelength range, to an accuracy of 0.03\,$\AA$ and 0.01\,$\AA$ (rms), respectively. We then correct for variations in sensitivity, from blue-to-red and global (i.e$.$ from fibre-to-fibre), with the \textit{MegaraFiberFlatImage} task and we implement an illumination correction with the \textit{MegaraTwilightFlatImage} task, using halogen and twilight flat frames. The absolute flux calibration is performed using the standard star frames mentioned before, whose reference data was extracted from from the CALSPEC calibration database\footnote{\href{https://www.stsci.edu/hst/instrumentation/reference-data-for-calibration-and-tools/astronomical-catalogs/calspec}{https://www.stsci.edu/hst}} \citep{Bohlin_2020}. To determine in which area of the detector the standard star is located, we use the \textit{MegaraLcbAcquisition} routine. After that position is determined, we use the task \textit{MegaraLcbStdStar} along with the La Palma extinction curve published by \citet{King_1985} to produce the sensitivity curve. We then run the \textit{MegaraLcbImage} recipe, whose final result is a set of two Row-Stacked Spectra (henceforth RSS) FITS frames. The two RSS FITS files produced by \textit{MegaraLcbImage} have 623 spectra with a common flux calibration and wavelength solution plus a constant reciprocal dispersion for all fibres. One of these two frames includes the sky background spectrum while the other is obtained after subtracting the median spectrum of all 56 sky fibres that are located at the edges of the MEGARA IFU+MOS field of view. 

\section{Analysis}
\label{Analysis}
Traditionally, the approach taken to study stellar populations in nearby galaxies has been to analyse different spectral features along their semi-major axis, using long-slit spectroscopy. Bearing this in mind, but taking advantage of the 2D spectroscopic information afforded by the MEGARA IFU, we have drawn elliptical rings around the central point of NGC~7025 and extracted their spectra to derive radial spectral information of the galaxy. The centre of these rings is the brightest fibre for each observation. We have considered a constant ellipticity of 0.21 for all of them and a position angle of 46.78\,$^{\circ}$ \citep{Dullo_2019}. Each of the rings has a width of 1\,arcsec. The analysis of the spectra of NGC~7025 using elliptical annuli not only preserves most of the information content on the spatial variation of the stellar population properties but it also provides improvement in signal-to-noise ratios over both the long-slit and pixel-by-pixel 2D studies and, therefore, reduces uncertainties. 

In Figure \ref{fig:elliptical_spectra} we plot the spectra for every elliptical rings extracted from our observations with the MEGARA IFU for the different low spectral resolution VPHs analysed in this work. The flux of the spectra is shifted vertically so all spectra can be properly displayed. The signal-to-noise ratios of the spectra has been estimated using the residuals of the regions fitted best in these spectra by Penalized Pixel-Fitting software (pPXF) by \citealt{Cappellari_2004} (see also \citealt{Cappellari_2017}) and vary across VPHs. For LR-U we reach a signal-to-noise ratio per angstrom of $\sim$36, the lowest of all. This is because MEGARA is least sensitive in this spectral range. For the rest of the VPHs the signal-to-noise values derived are significantly larger, averaging 180 for LR-B, 250 for LR-V and 140 for LR-R. Within each VPH setting, the spectra do not reveal any obvious difference possibly because the region covered by the MEGARA IFU does not reach beyond the NGC~7025 bulge, so the stellar populations are rather homogeneous. 

\begin{figure}[h]
	\centering
    \includegraphics[trim={0 0 0 1cm}, clip, width=1\linewidth]{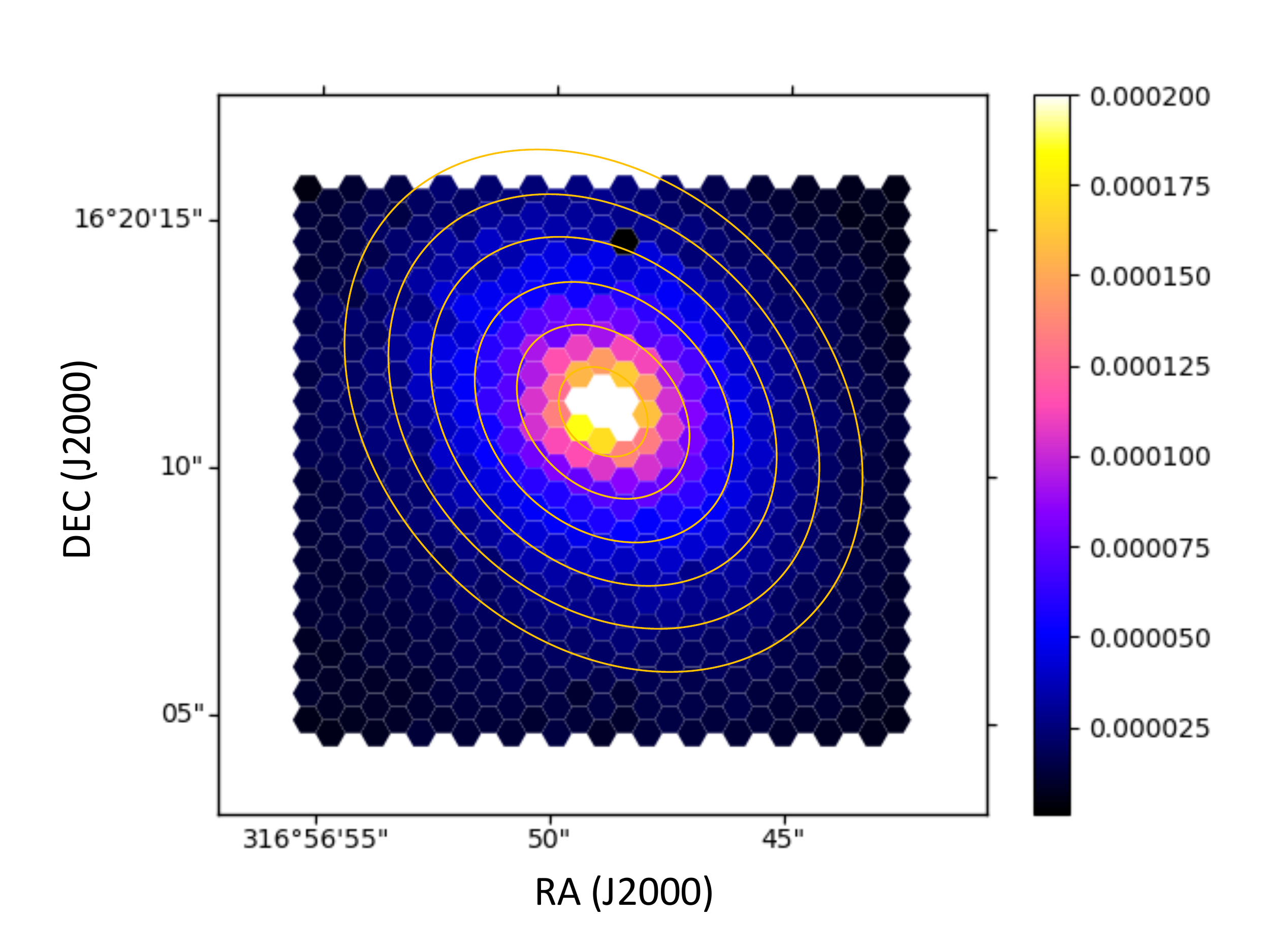}
    \vspace{-5mm}
	\caption{LR-V continuum intensity distribution in the bulge of NGC~7025 after averaging the spectral range between 5600 and 5870\,\AA. The elliptical rings from which the galaxy spectra are extracted (see the text for further detail) are overplotted. Flux units are Jy.}
	\label{fig:elliptical_rings}
\end{figure}

In Figure \ref{fig:elliptical_rings} we show a continuum image of NGC 7025 taken with the MEGARA IFU. This image is obtained by collapsing the LR-V spectra within the range between 5600 and 5870\,\AA\ (see Table~\ref{table:VPH_characteristics} for the spectral coverage of each VPH). We also show the elliptical regions from which we have extracted the spectra that we will use to perform the radial study of the stellar populations in NGC~7025.

\subsection{Stellar population synthesis models}

The stellar population synthesis models used for this study are the ones included in the MILES simple stellar population (SSP) grid \citep{Vazdekis_2010} which are based on the Padova+00 stellar libraries (\citealt{SanchezBlazquez_2006} and \citealt{FalconBarroso_2011}). These models cover the range 3525$-$7500\,$\AA$ at 2.5\,$\AA$ spectral resolution (Full Width at Half Maximum; FWHM). Although the spectral resolution of MEGARA is better than the spectral resolution of these models, in the case of NGC 7025, the mismatch does not have much of an impact since the relatively large velocity dispersion of inner regions of this galaxy broadens the lines to values greater than the spectral resolution of the models anywhere within the MEGARA IFU field of view. These MILES SSPs cover a range in age from 0.063\,Gyr to 17.78\,Gyr and in metallicity from $-$2.32 to $+$0.22 ([M/H]). Note that the ages of the SSP models extend beyond the age of the universe in order to mitigate possible edge effects at very old ages \citep{gildepaz_2002}. We have used an unimodal initial mass function (IMF hereafter) with logarithmic slope of 1.3 \citep{Salpeter_1955}, as suggested by \citet{Dutton_2013} for massive spiral galaxies. Note that these SSP models will be broadened to match the velocity distribution of the stars in the observed spectra. For the spectral indices (Section~\ref{lineindices}) this is achieved by making use of a Gaussian kernel whereas for the spectral analysis to be performed with pPXF \citep{Cappellari_2017} a Gauss-Hermite method is employed (see more details in Section~\ref{section: fitting}). 

\subsection{Spectral line indices}
\label{lineindices}

As an initial approach to the problem of stellar population modelling, we carry out an analysis based on spectral indices, as this is the simplest (and also the most classic) and serves as a reference for what we can expect in later more complex analyses. We examine different indices covering a wide spectral range in order to evaluate the impact of different wavelengths and specific features on the age and metallicity derived for the stellar populations of the galaxy.

\begin{figure*}[h]
    \center
	\includegraphics[trim={0 0 0 1cm}, clip, width=0.49\linewidth]{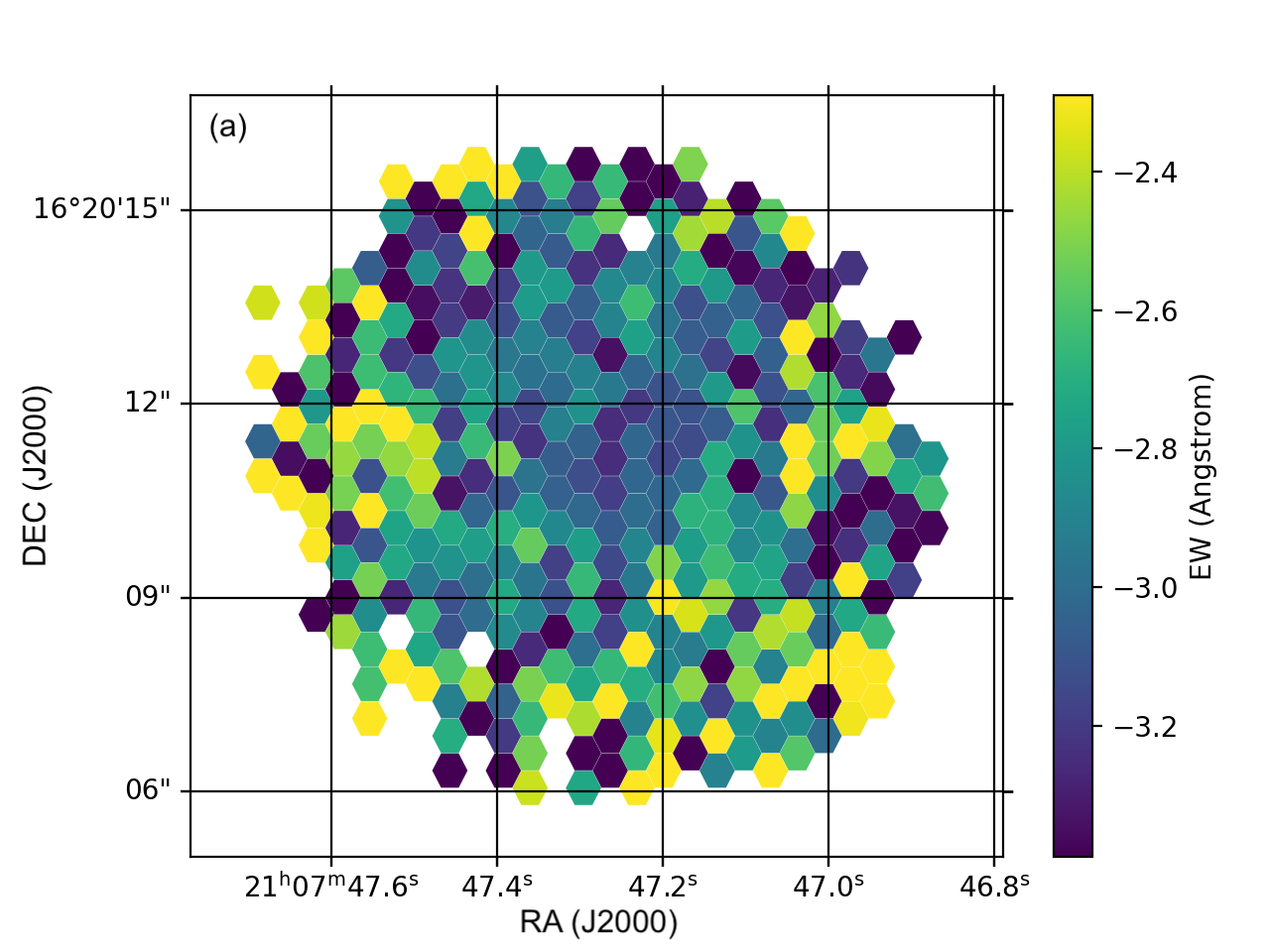}
	\includegraphics[trim={0 0 0 1cm}, clip, width=0.49\linewidth]{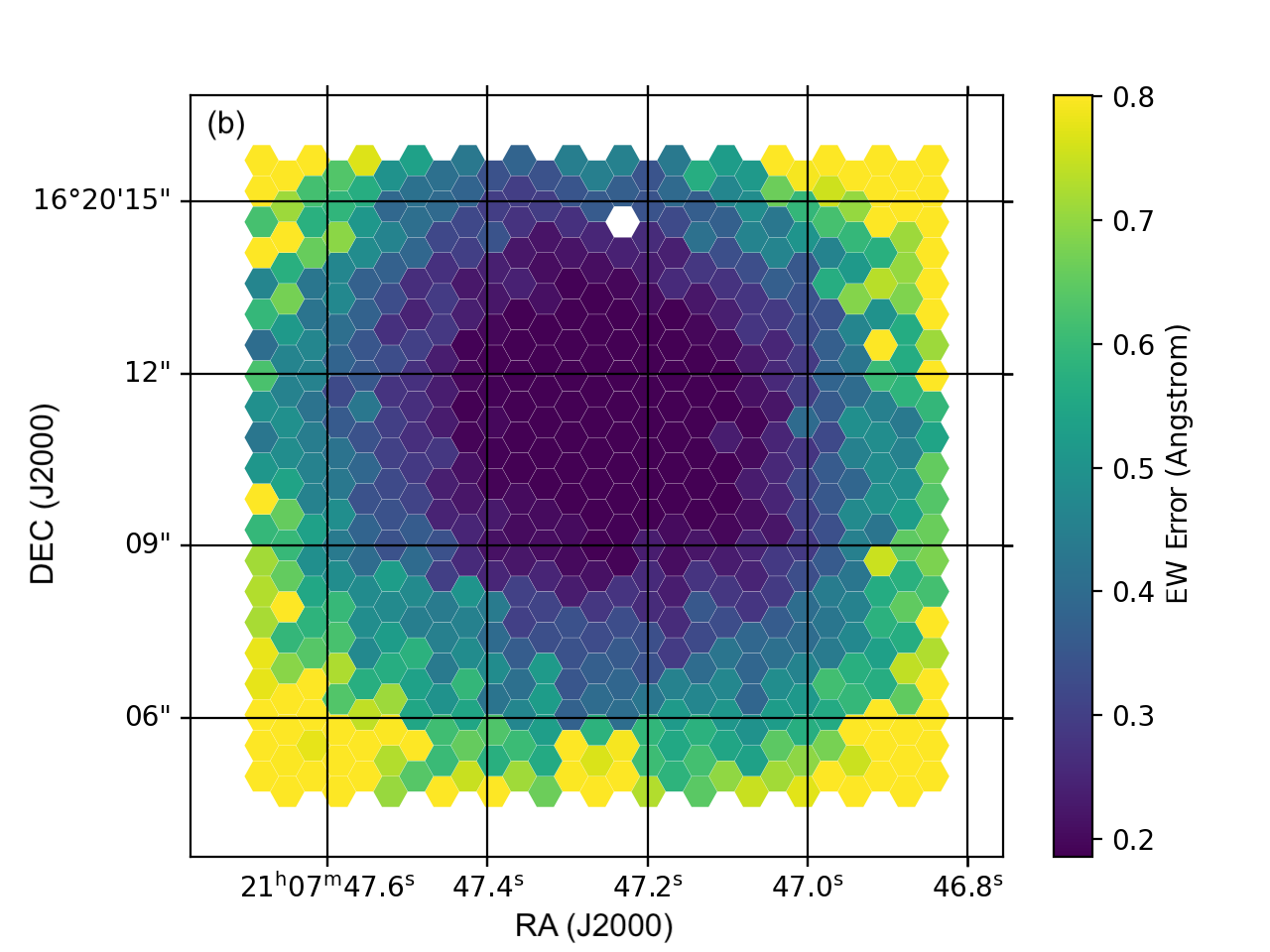}
	\caption{Left panel (a): the two-dimensional map of the $\langle$Fe$\rangle$-index for NGC~7025. Right panel (b): error map associated with the $\langle$Fe$\rangle$-index measurement.}
	\label{fig:indices_2Dmap}
\end{figure*}

Fitting Gaussian kernel to our data, we measure eleven different spectral indices in total. Seven of them were taken from the work of \cite{Worthey2014}, namely H$\beta$, H$\delta_{\mathrm{F}}$, $\mathrm{Fe}4383$, $\mathrm{C_{\mathrm{2}}4668}$, $\mathrm{CN1}$, $\mathrm{Mgb}$, $\mathrm{Ti}4296$. The $\langle$Fe$\rangle$ = (Fe5270+Fe5335)/2 index was taken from \cite{Gonzalez_1993} while the $\mathrm{Ca}3934$, $\mathrm{Fe}4045$, and $\mathrm{Mg}4480$ indices were given in the paper by \cite{Lino2020}. Spectral indices errors are computed using the line flux uncertainty obtained as in \citealt{Tresse_1999}. In Figure~\ref{fig:indices_2Dmap} we show the two-dimensional maps for the $\langle$Fe$\rangle$ index (left panel) and its error (right panel). Only  measurements whose error is less than 0.5\,\AA\ are shown in the left panel. The comparison between the indices measured and those predicted by the SSP models (at the same level of broadening) will be presented in Section \ref{section: Results: Spectral line indices}.

\subsection{Stellar population and kinematical fitting}
\label{section: fitting}

\begin{figure*}[h]
    \center
	\includegraphics[trim={0 15mm 5mm 20mm},clip,width=0.49\linewidth]{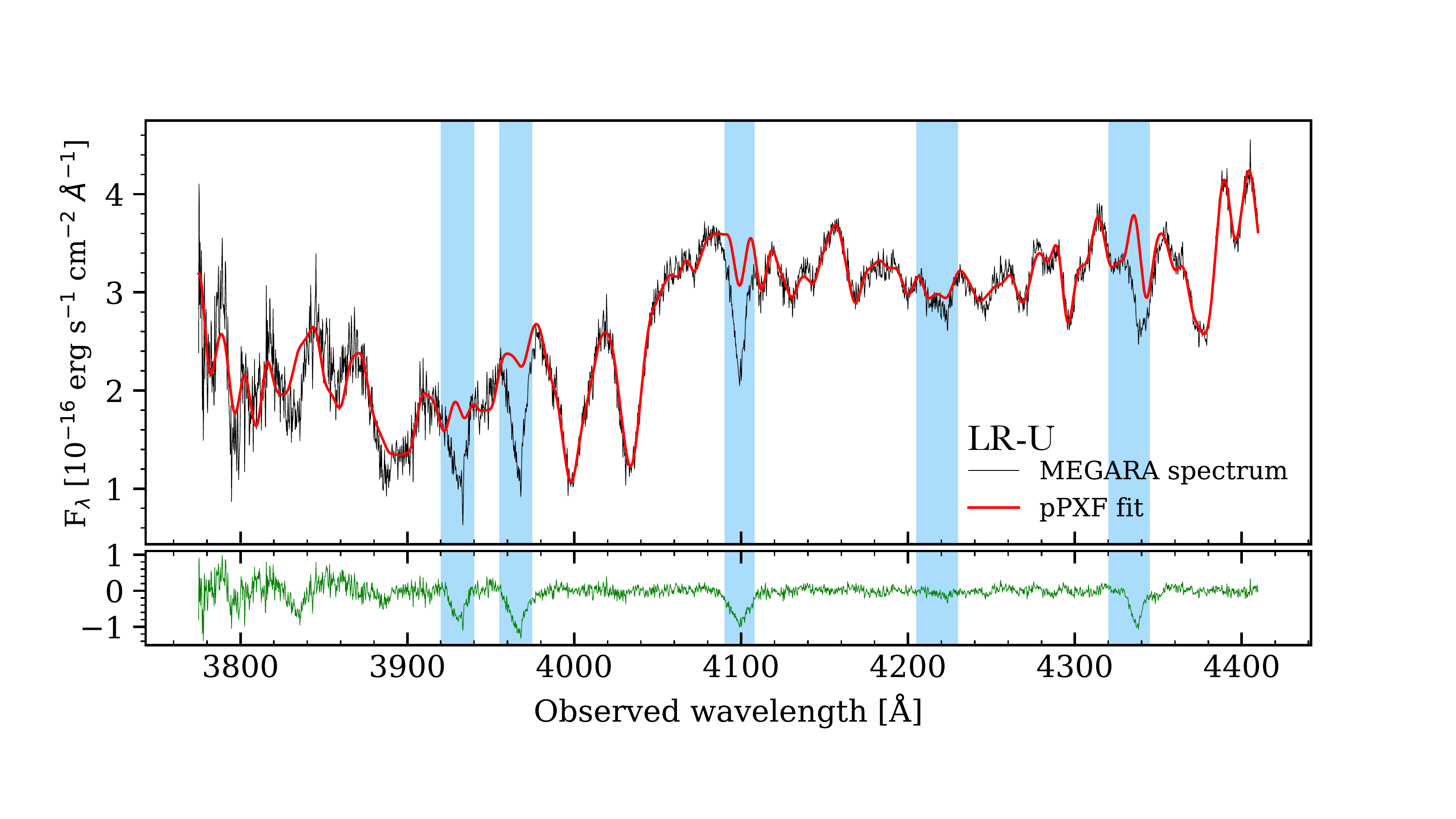}
	\includegraphics[trim={0 15mm 5mm 20mm},clip,width=0.49\linewidth]{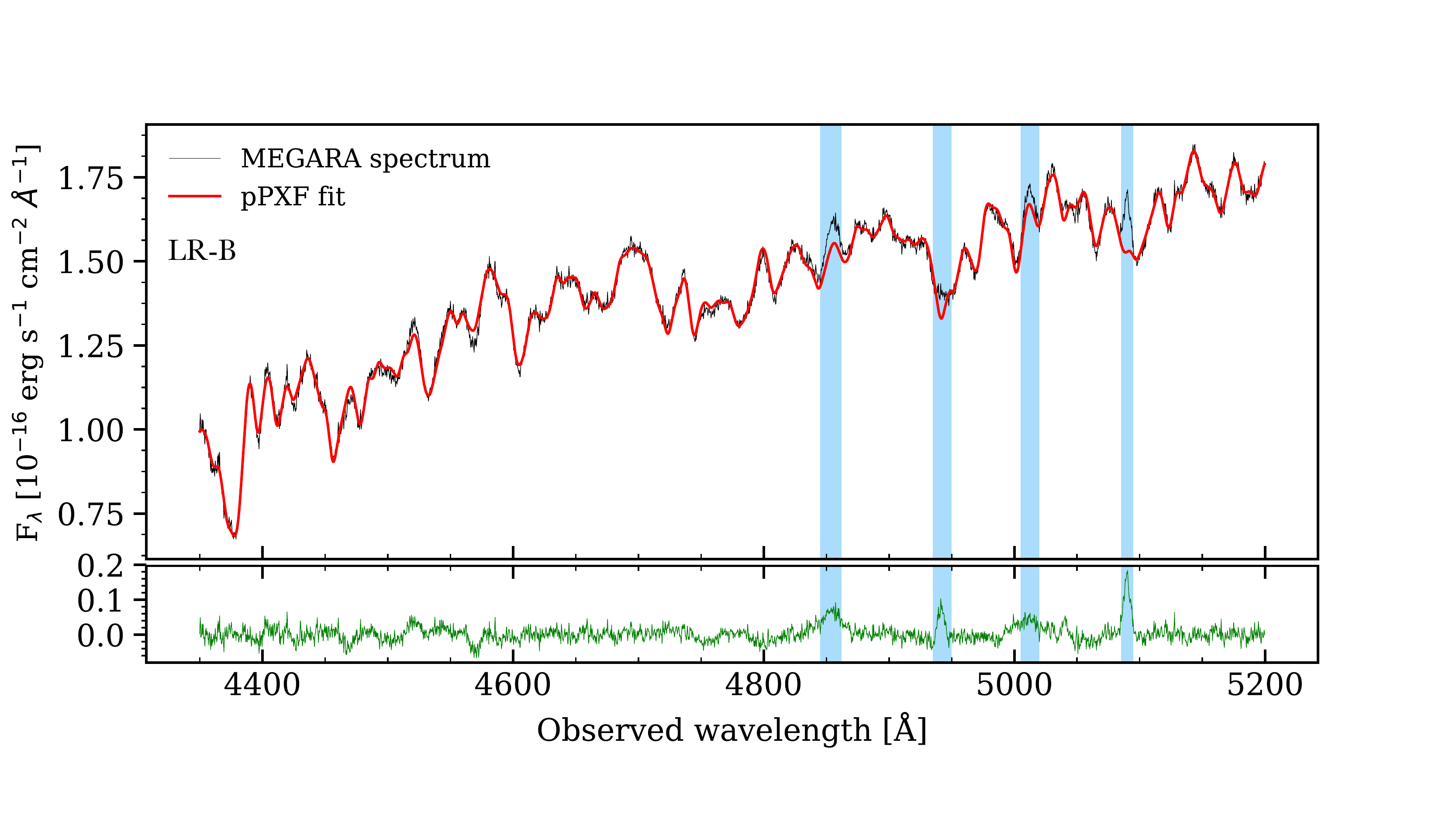}
    \includegraphics[trim={0 0 5mm 20mm},clip,width=0.49\linewidth]{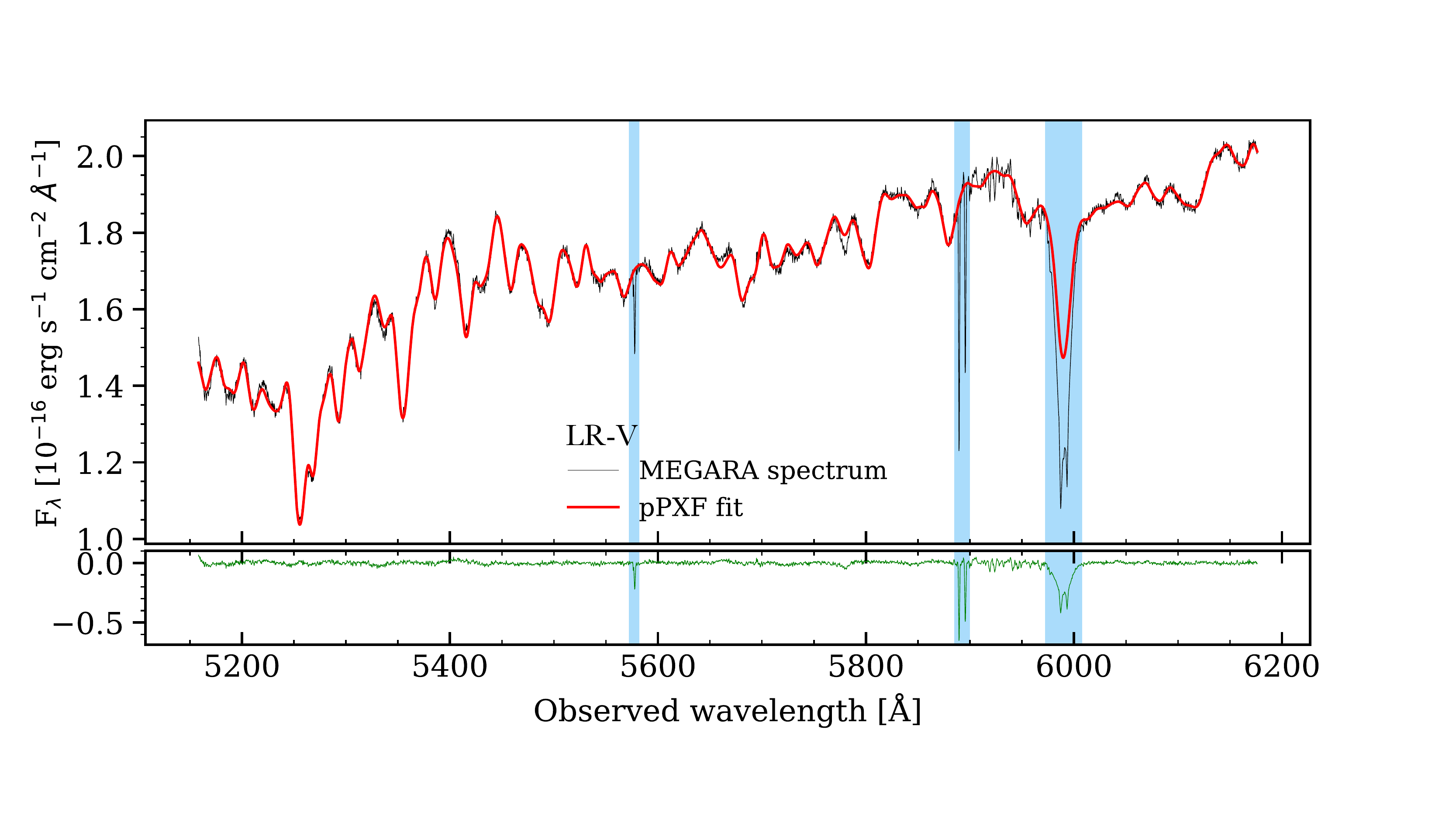} % Para recortar imágenes: [trim={<izquierda> <abajo> <derecha> <arriba>},clip]
	\includegraphics[trim={0 0 5mm 20mm},clip,width=0.49\linewidth]{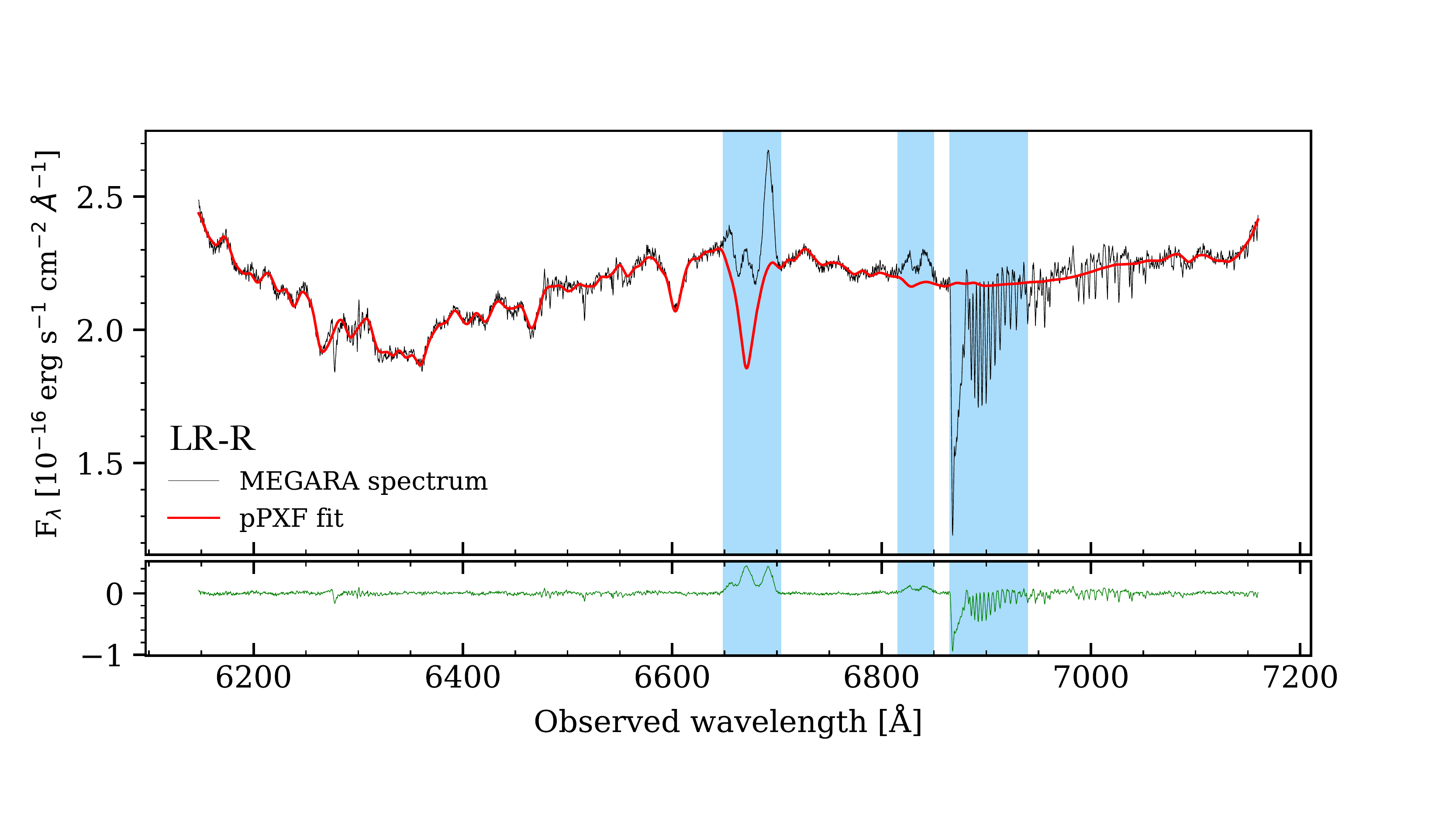}
	\caption{Spectrum of NGC~7025 within the first elliptical aperture (having a semi-major axis of 1 arcsec) in different VPHs (see Figure \ref{fig:elliptical_rings}). The red and green line curves show pPXF best fits to the spectra and the corresponding residuals, respectively. The blue vertical stripes indicate the masked regions omitted from the pPXF fitting.}
	\label{fig:ppxf_spectra_fitting}
\end{figure*}

We perform  full spectral fitting analysis of our spectra using pPXF by \citealt{Cappellari_2017}. This software enables the extraction of the stellar kinematics and stellar population properties by means of analysing the absorption spectrum of galaxies or specific regions within galaxies, including individual spaxels. In order to ensure that the emission lines do not affect the fitting of the absorption features, we masked all the emission lines present in the spectra as well as absorption lines of interstellar origin, such as NaD (see Figure \ref{fig:ppxf_spectra_fitting}). We also masked regions affected by telluric absorptions or residuals from bright sky lines (e.g$.$ [\ion{O}{i}]5577\,\AA).

For our fittings, we follow \cite{Kacharov_2018} and set pPXF to use 10$^{\mathrm{th}}$-order both additive and multiplicative polynomials and first-order regularisation with a low regularisation factor R=5. The use of these polynomials ensures that our analysis is free of any low-frequency effects that may be present in the spectra (e.g. reddening, residual blue-to-red variations not corrected by the data reduction, ...). We have also followed the method proposed by these authors for estimating the uncertainties in the derived stellar population properties, which is based on the wild bootstrapping method \citep{Wu_1986}. This estimation uncertainties  only allows us to calculate the statistical error coming from the variance of the spectra. The effect of possible errors coming from the models themselves and the spectral libraries cannot be included as we have no prior knowledge of them. To this must be added the unknown capability to reproduce a complex star formation and chemical history of any stellar population composed from these models. Briefly, the method is applied as follows. First, we fit the spectrum with pPXF and get the best fit and the residuals from this initial fitting. Once this initial fitting is completed, we resample the residuals obtained previously with the wild bootstrapping method. The resampled residuals are added to the best fit to obtain a new spectrum with the errors distributed in a different way. We repeat this resampling procedure 100 times for each spectrum analysed and fit these new spectra with the resampled errors. The set of 100 best-fitting solutions yield the probability distributions of the derived properties. Note that in the case of the kinematical parameters derived, pPXF already provides direct robust error estimates. 

In an attempt to understand the limitations of our analysis, we made extensive tests with pPXF using mock spectra. We show the results of all these tests in Appendix \ref{section: appendix}. Following the results from our rigorous test, and together with the assessment of the amount of information that can be extracted from each VPH in terms of absorption features as well as the relative sensitivity of the VPHs, we decided to focus on the results of the analysis conducted on the combination of LR-B+LR-V data. When concatenating the spectra of both observations, the first thing to check is whether or not they have the same pointing. In the case of LR-B and LR-V, the centre of the galaxy is shifted one spaxel away from the other. Since we do not have the same pointing, we can only concatenate the spectra extracted from the elliptical regions explained above in section \ref{Analysis}. Once we have the spectra separately, what we do is to identify a region of the spectrum without features in the region that both VPHs overlap in wavelength. What we do now is to match the two spectra using the flux ratio at that point. This has no impact on our analysis since pPXF subtracts the stellar continuum from the spectra with polynomials. The last step is to resample the whole new spectrum, considering that the original observations have different resolutions, to the resolution of the spectrum with the lowest resolution. We will also show the results obtained on each setup individually to highlight the impact of the spectral range under study on the stellar population properties derived. 

\section{Results}
\subsection{Spectral line indices}
\label{section: Results: Spectral line indices}

\begin{figure*}[h]
    \center
	\includegraphics[trim={0 1cm 1cm 1.2cm}, clip, width=0.33\linewidth]{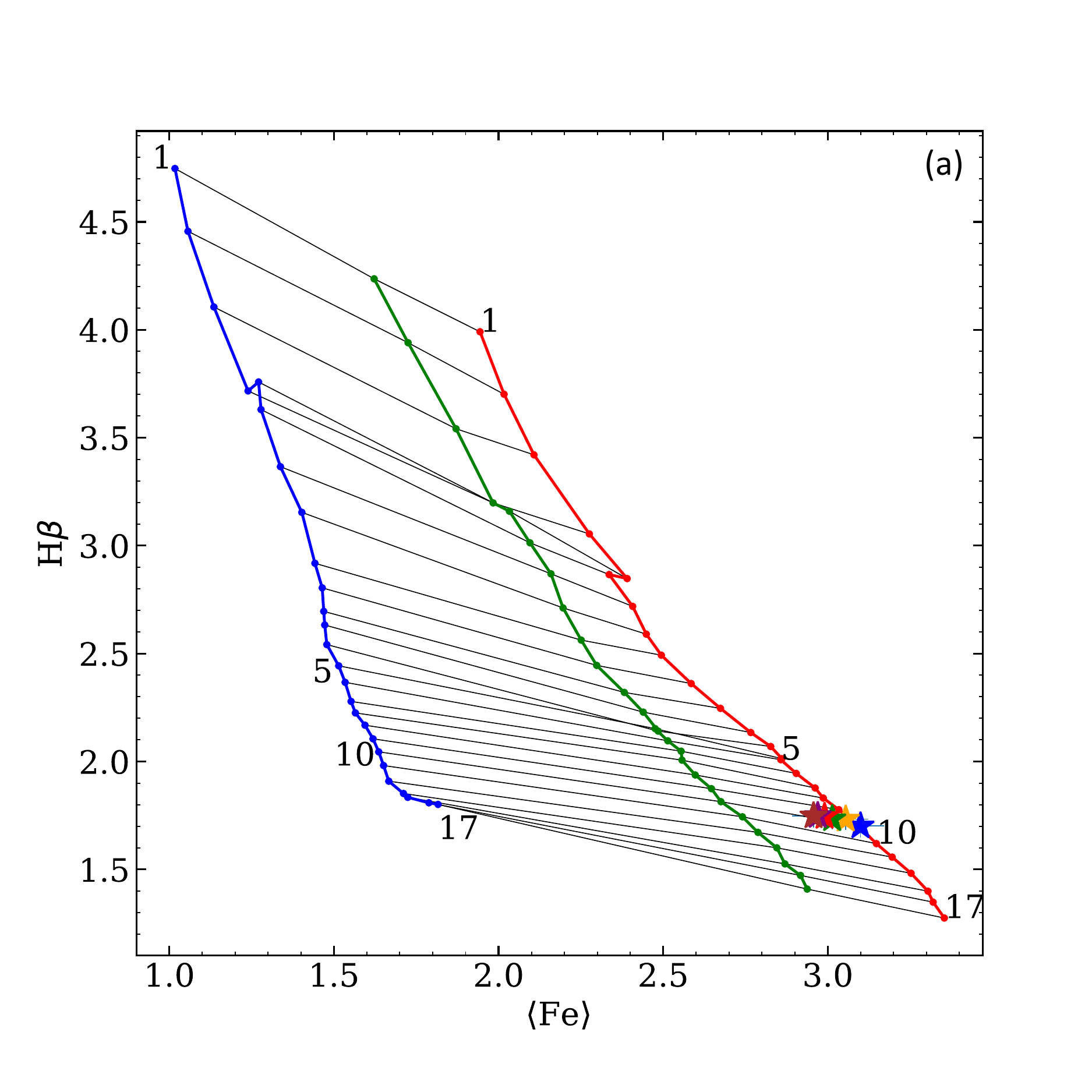}
	\includegraphics[trim={0 1cm 1cm 1.2cm}, clip, width=0.33\linewidth]{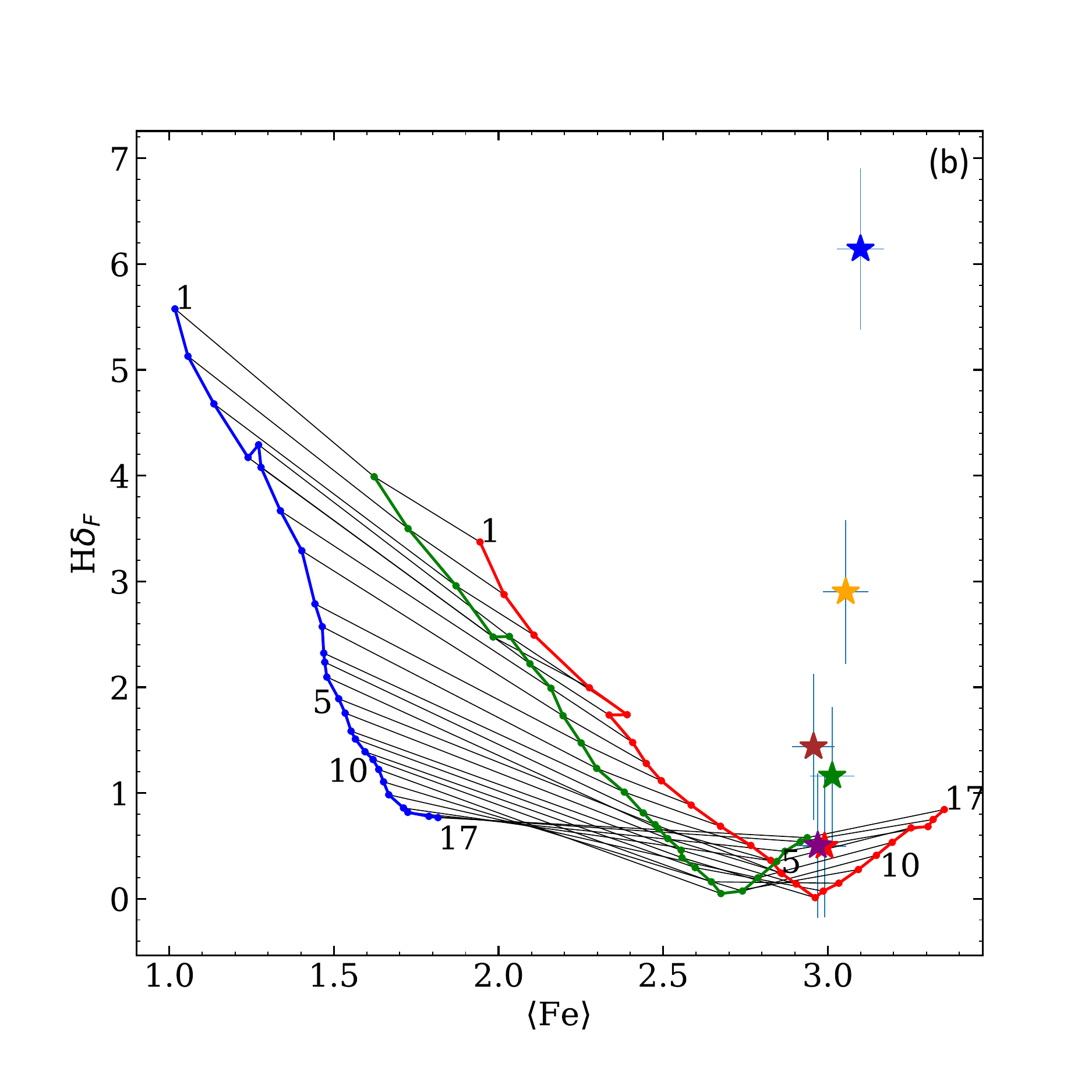}
    \includegraphics[trim={0 1cm 1cm 1.2cm}, clip, width=0.33\linewidth]{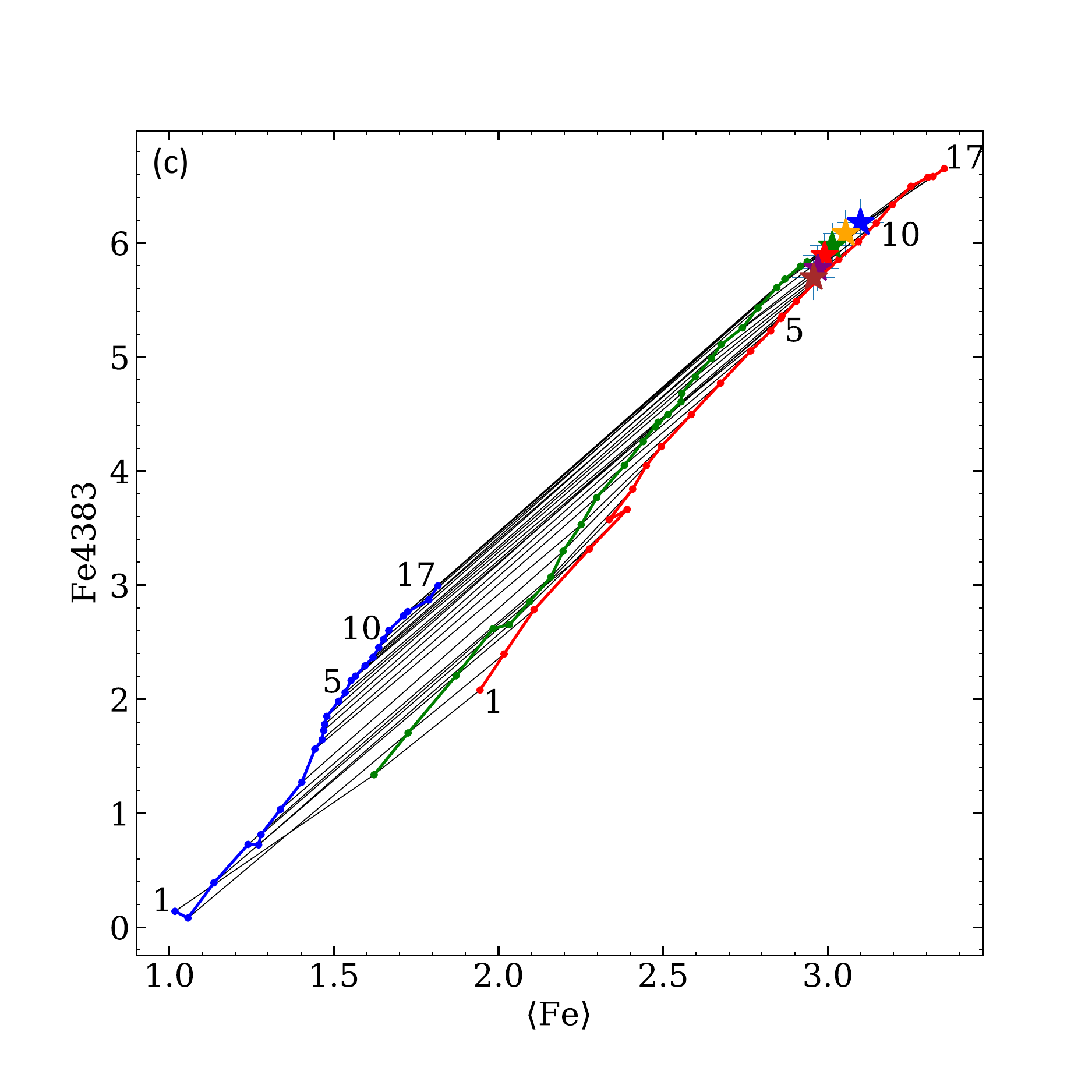} % Para recortar imágenes: [trim={<izquierda> <abajo> <derecha> <arriba>},clip]
	\includegraphics[trim={0 1cm 1cm 2cm}, clip, width=0.33\linewidth]{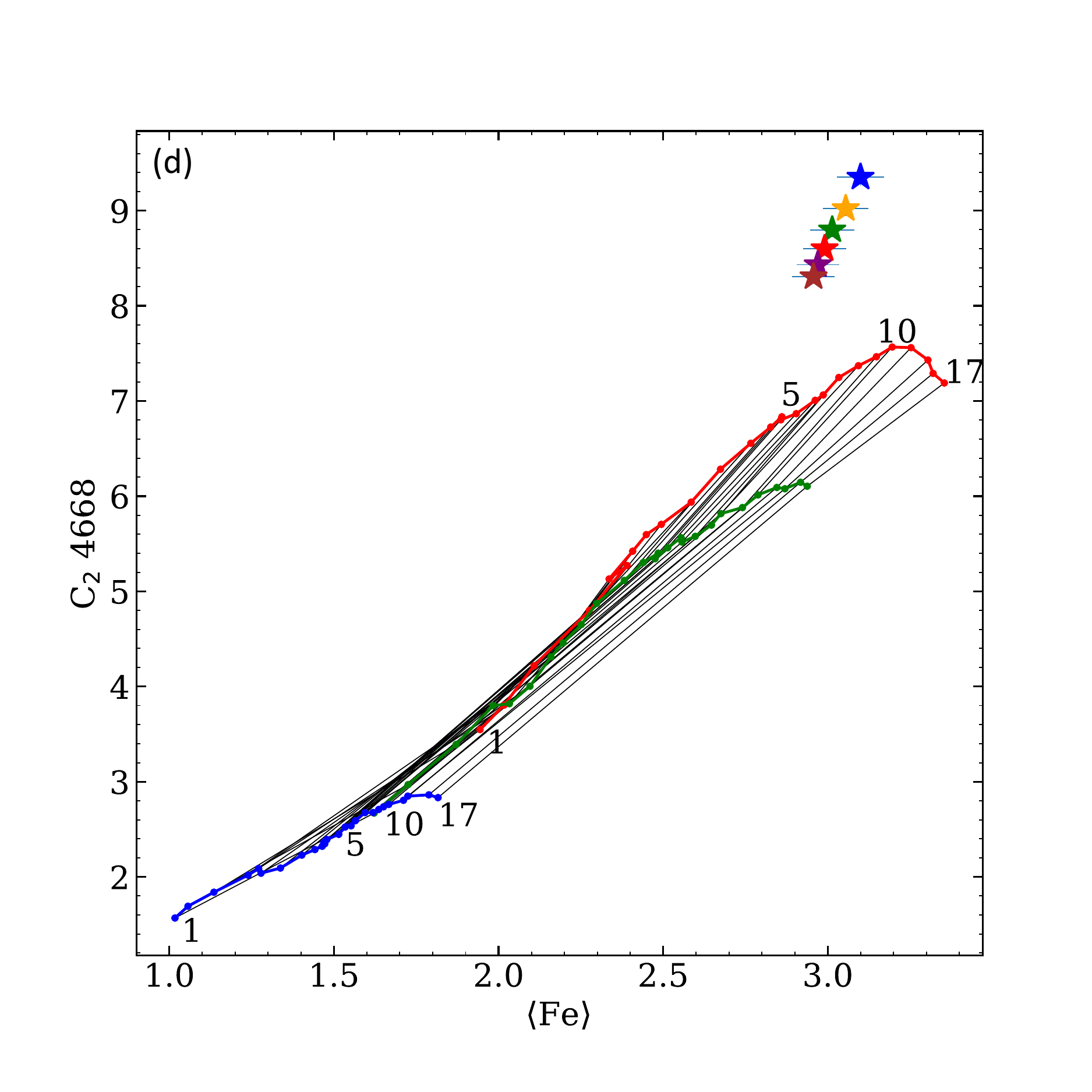}
	\includegraphics[trim={0 1cm 1cm 2cm}, clip, width=0.33\linewidth]{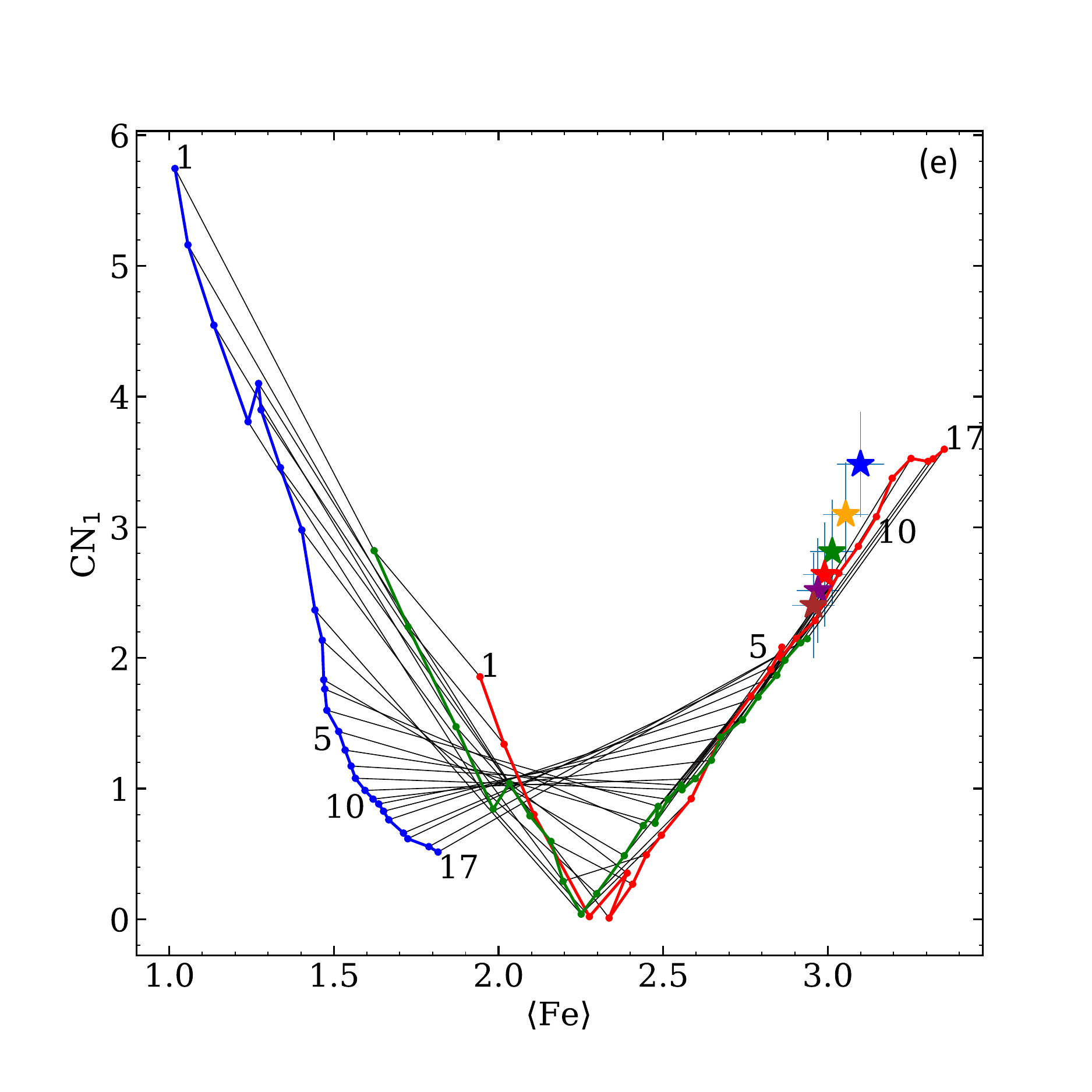}
	\includegraphics[trim={0 1cm 1cm 2cm}, clip, width=0.33\linewidth]{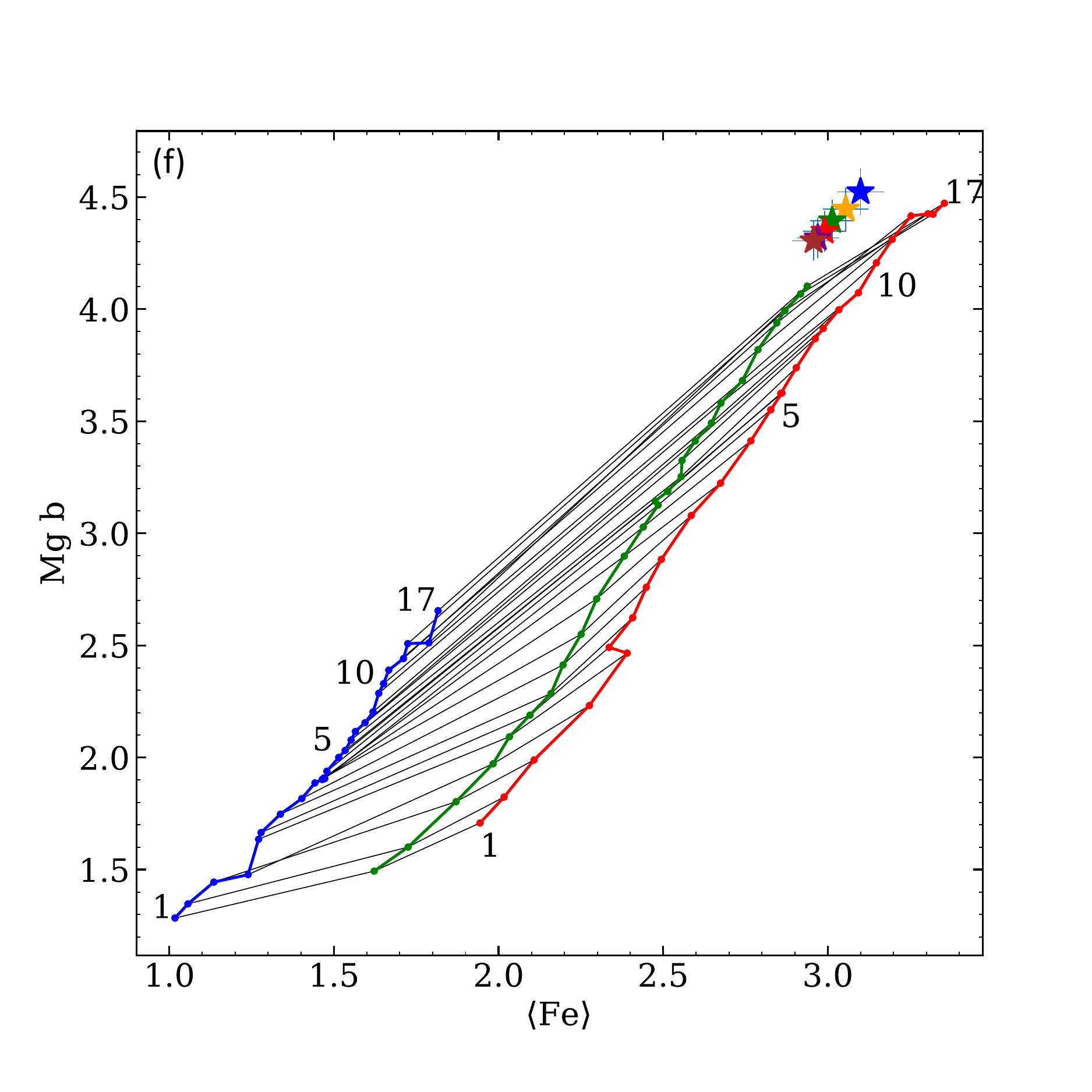}
	\includegraphics[trim={0 1cm 1cm 2cm}, clip, width=0.33\linewidth]{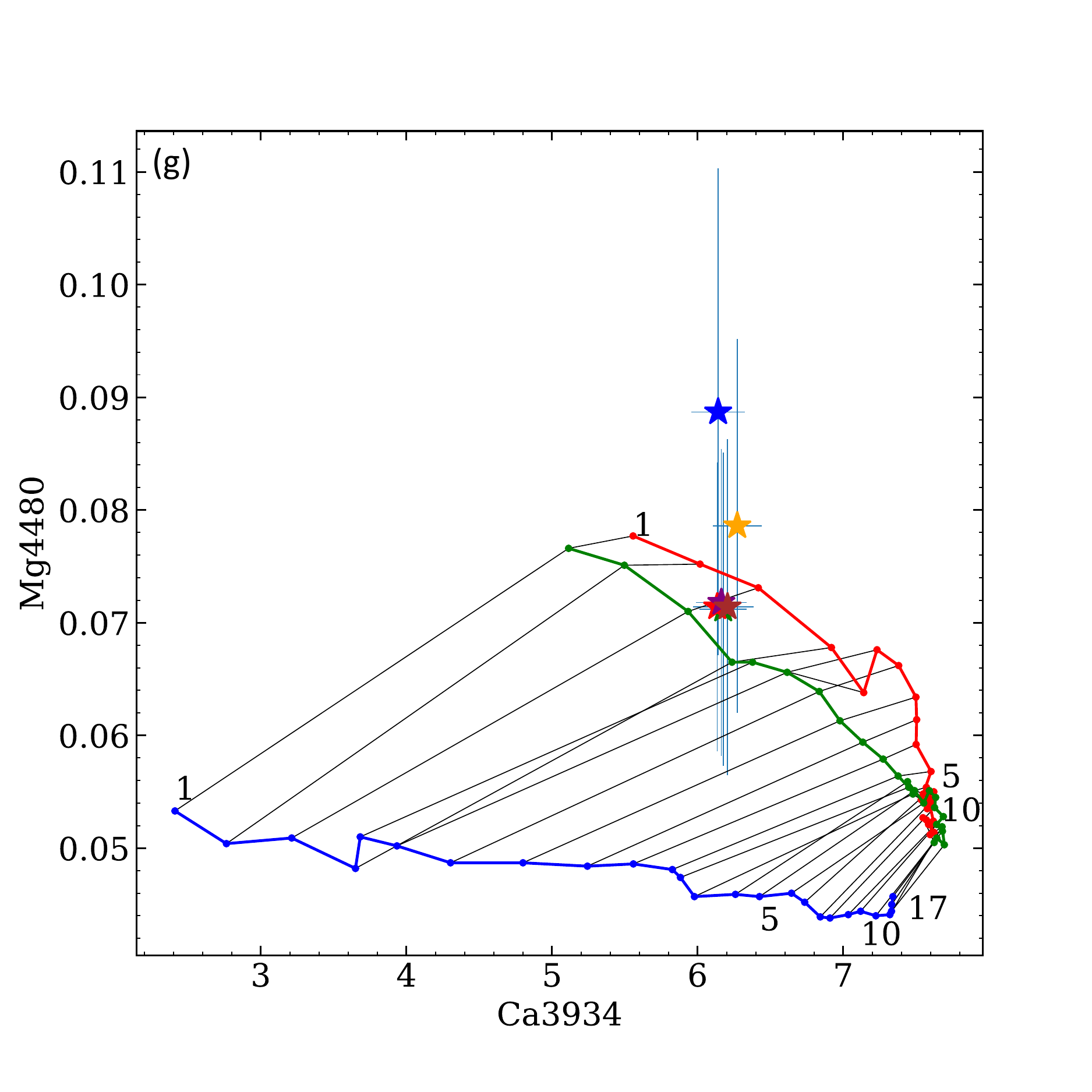}
	\includegraphics[trim={0 1cm 1cm 2cm}, clip, width=0.33\linewidth]{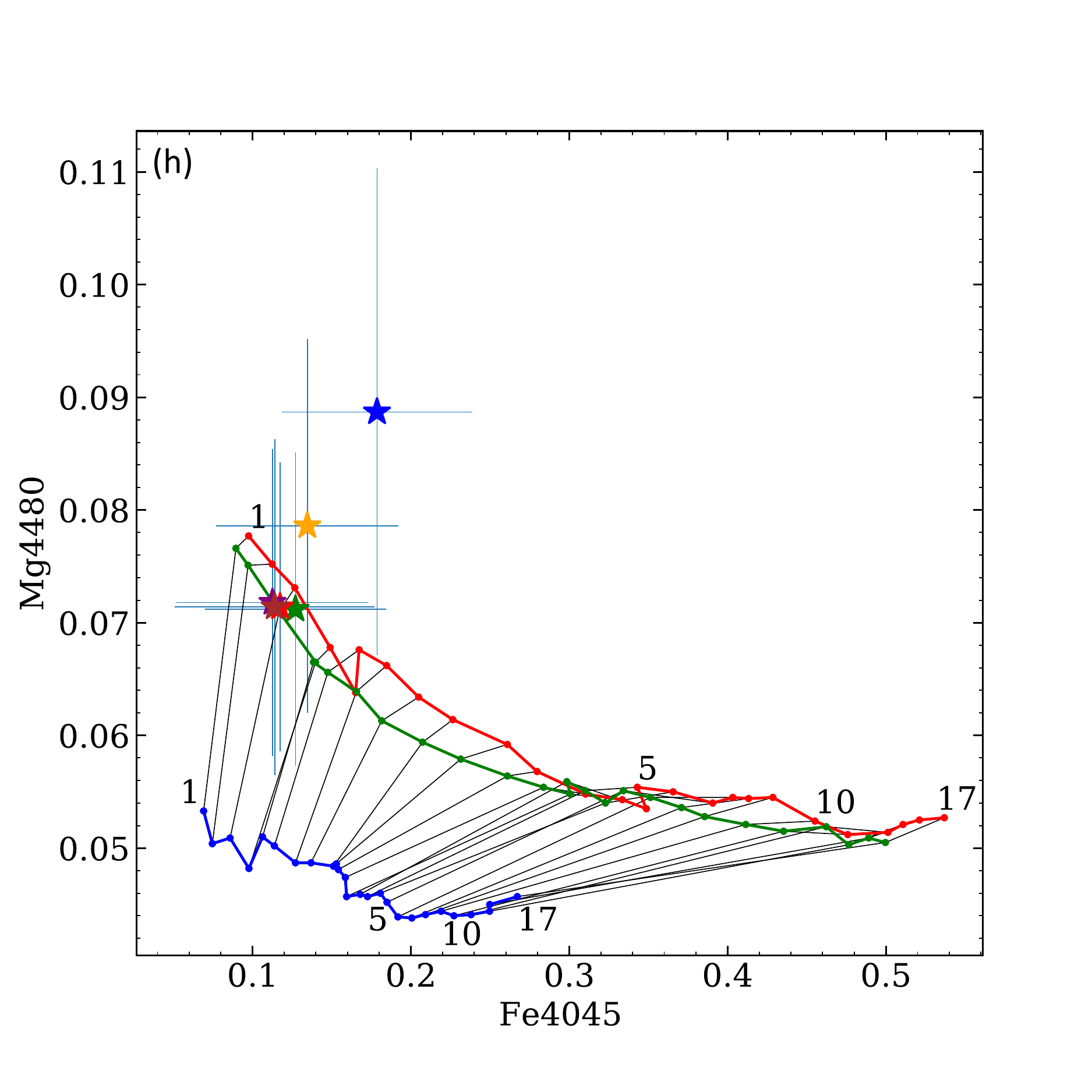}
	\includegraphics[trim={0 1cm 1cm 2cm}, clip, width=0.33\linewidth]{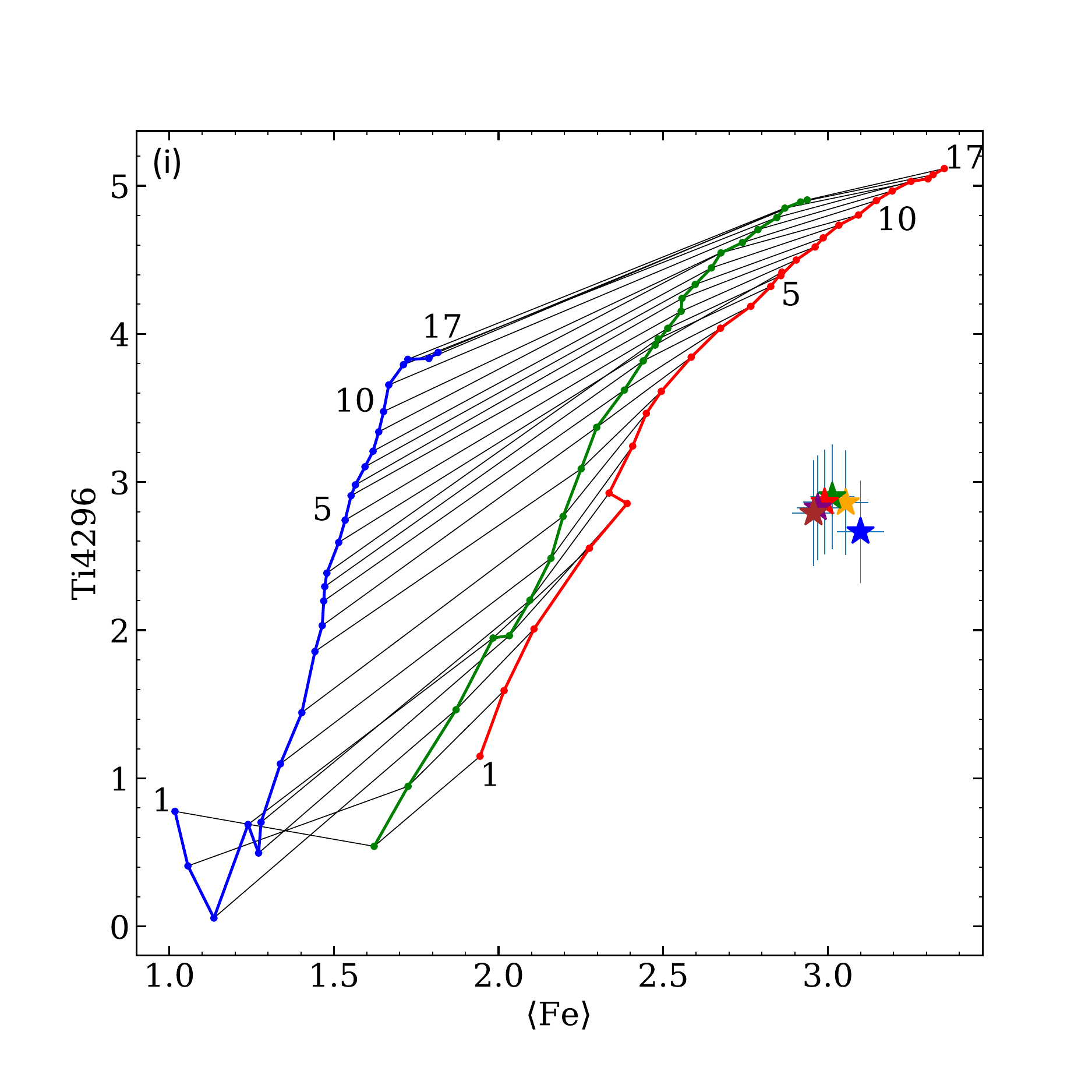}
	\caption{Spectral indices measured in the bulge of NGC~7025. The reference grid has been created using three different metallicities [M/H] = $-$0.71, 0.00 and 0.22 (blue, green and red solid lines, respectively). Ages range from 1 to 17\,Gyr with some of the models having their ages (in Gyr) marked in the plot. Star symbols, including their uncertainties, represent the indices measured in the spectra of the elliptical rings we extracted from NGC~7025. The rings from the innermost to the outermost radii are color coded as blue, orange, green, red, purple and brown.}
	\label{fig:indices}
\end{figure*}

All the indices, including the predictions from reference grid of SSP models (see Figure \ref{fig:indices}), were computed based on the definitions of these indices by \cite{Worthey2014}, \cite{Tang2014} and \cite{Lino2020}. To create a reference grid as close as possible to the NGC~7025 data, we broadened the MILES SSP models with a Gaussian kernel to match the velocity dispersion of 250\,km\,s$^{-1}$ measured by \citep{Dullo_2019} on NGC~7025.

In the two-dimensional $\langle$Fe$\rangle$-index map (Figure \ref{fig:indices_2Dmap}) we find a deeper index (larger in absolute value) within the innermost 2\,arcsec. This is true for other spectral indices as well since absorption lines are found to be deepest, in general, in the innermost regions of NGC~7025. As we move away from these central regions, the values we obtain become somewhat larger (shallower) but noisier. This is to be expected since, as we can see in the right panel of Figure \ref{fig:indices_2Dmap}, the errors become larger when the S/N of the galaxy declines by going further away from the centre of the MEGARA IFU and the signal of the galaxy becomes fainter. 

The 2D distribution shown in Figure \ref{fig:indices_2Dmap} needs to be put into context with the help of Figure \ref{fig:indices}, where we show the relation between the values of all indices measured and both the age and metallicity of the SSP MILES models. This figure represents the most classical approach to quantify the variation of the strength of spectral features with the properties of the stellar populations. The indices measured for each of the spectra of the elliptical rings are shown with a star symbol. The order of the elliptical annuli follows the same colour pattern as in Figure \ref{fig:elliptical_spectra}. From the innermost to the outermost, their colours are blue, orange, green, red, purple and brown, with an increment of 1\,arcsec in semi-major axis radius in between every consecutive annulus. Note that in the case of the central measurement (red star) this actually corresponds to an elliptical aperture, not an annulus, of 1\,arcsec in semi-major axis radius.

Figure~\ref{fig:indices} reveals several interesting insights. First, for most indices we favour old ages and high metallicities and an overall tendency for having deeper indices in the central regions (especially for indices reaching large EW values, such as Fe, C$_{\mathrm{2}}$, CN$_{\mathrm{1}}$, or Mg~b\footnote{Note that the Balmer indices have an opposite behaviour with age compared to the metal indices.}). However, we also find that the closest SSP model to each data point differs significantly depending on the index used. For example, we yield SSP ages in the range 1-2\,Gyr for Ca3934-Mg4480 or Fe4045-Mg4480 but ages close to 10\,Gyr for $\langle$Fe$\rangle$-H$_{\beta}$, $\langle$Fe$\rangle$-Mg~b, or $\langle$Fe$\rangle$-Fe4383. In this regard, it seems that the bluer the index, the more sensitive it is to the presence of a young stellar population, which leads to younger SSP age. This suggests that NGC~7025 has experienced an extended star formation history,  which explains the differences in the observed spectra compared to the predictions of Single (in terms of age and metallicity) Stellar Population models. In any case, all pairs of indices obtained are consistent with supersolar metallicities.

Another issue that we find is that, in some cases, the measurements on NGC~7025 fall outside the predictions of our SSP grid for any age or metallicity considered. This could be due to the fact that we did not include models with different $\alpha$-enhancement values when creating the reference grid since our default MILES SSP models (based on the Padova+00 isochrones; \citealt{Girardi_2000}) do not incorporate multiple $\alpha$-enhancements. For this reason we repeated the whole analysis procedure using models with different $\alpha$-enhancement values based on the BaSTI isochrones (\citealt{Pietrinferni_2004} and \citealt{Pietrinferni_2006}). Although these results are not shown in this paper, they do not solve the problem, but make it worse instead. The $\alpha$-enhancement models, that could potentially reduced those offsets, instead of increasing the abundances of the $\alpha$-elements, effectively decrease the contribution of iron. Besides, the sensitivity of the indices to this change in $\alpha$-enhancement is not homogeneous as it varies from $\alpha$-element to $\alpha$-element. Indeed, the enhancement in $\alpha$-elements appears in scenarios where there is a fast chemical evolution which does not seem to be the case here.

Another possible explanation for this behaviour is that we are using SSP models to generate the reference grid and this is not allowing us to estimate the impact of  multiple stellar populations on the indices measured. We can consider the effects of different episodes of star formation by using full spectral fitting tools such as pPXF (section \ref{section: fitting}). Additionally, in order to interpret the results of having different SSP age for different indices, we will perform the full spectral fitting analysis on VPHs covering different spectral ranges. We explore the results of these analyses in the following subsections. Unless explicitly stated, the ages quoted correspond to the median of the mass-weighted values (using the mass fractions and ages of all SSPs considered) of each of the 100 realisations obtained by wild bootstrapping as part of our pPXF fitting method (see Section~\ref{section: fitting}) while the error bars represent the size of the 1-$\sigma$ ellipses when the age is used as marginal variable. 

\subsection{Results from full spectral fitting}
\label{section: age gradients}

We now present the results from the stellar population analysis of NGC~7025 from the full spectral fitting method as detailed in section \ref{section: fitting}. It should be remembered that MEGARA FoV allows us to explore only the inner region (the central 4.4 x 4.0 \,kpc$^{2}$) of NGC~7025 so we are only probing the bulge of this galaxy \mbox{($r_{e,\text{bul}}$ = 4\farcs63; $r_{e,\text{bul}}$ = 1.48\,kpc)}.

\begin{figure*}[h]
    \center
	\includegraphics[width=0.66\linewidth]{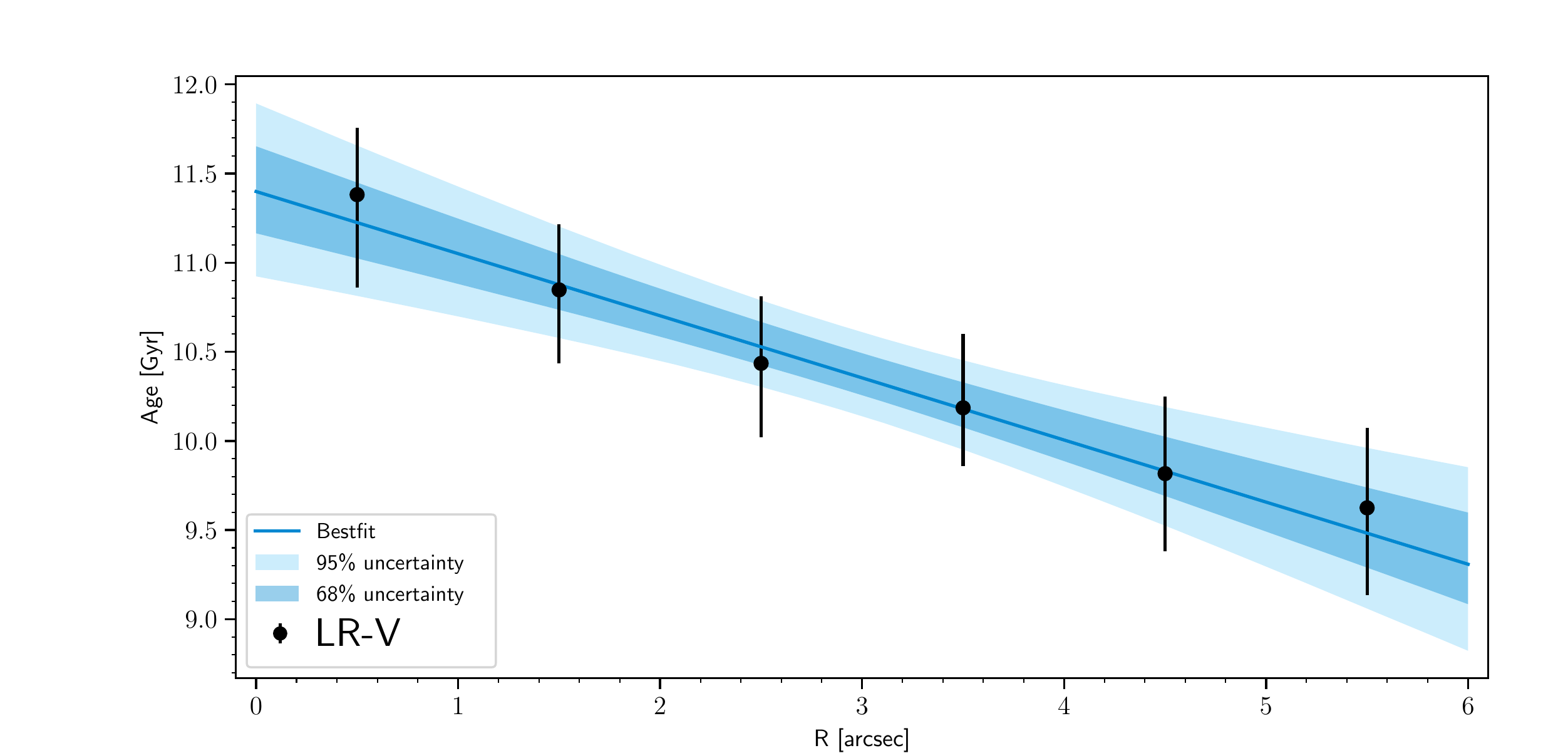}
	\includegraphics[width=0.3\linewidth]{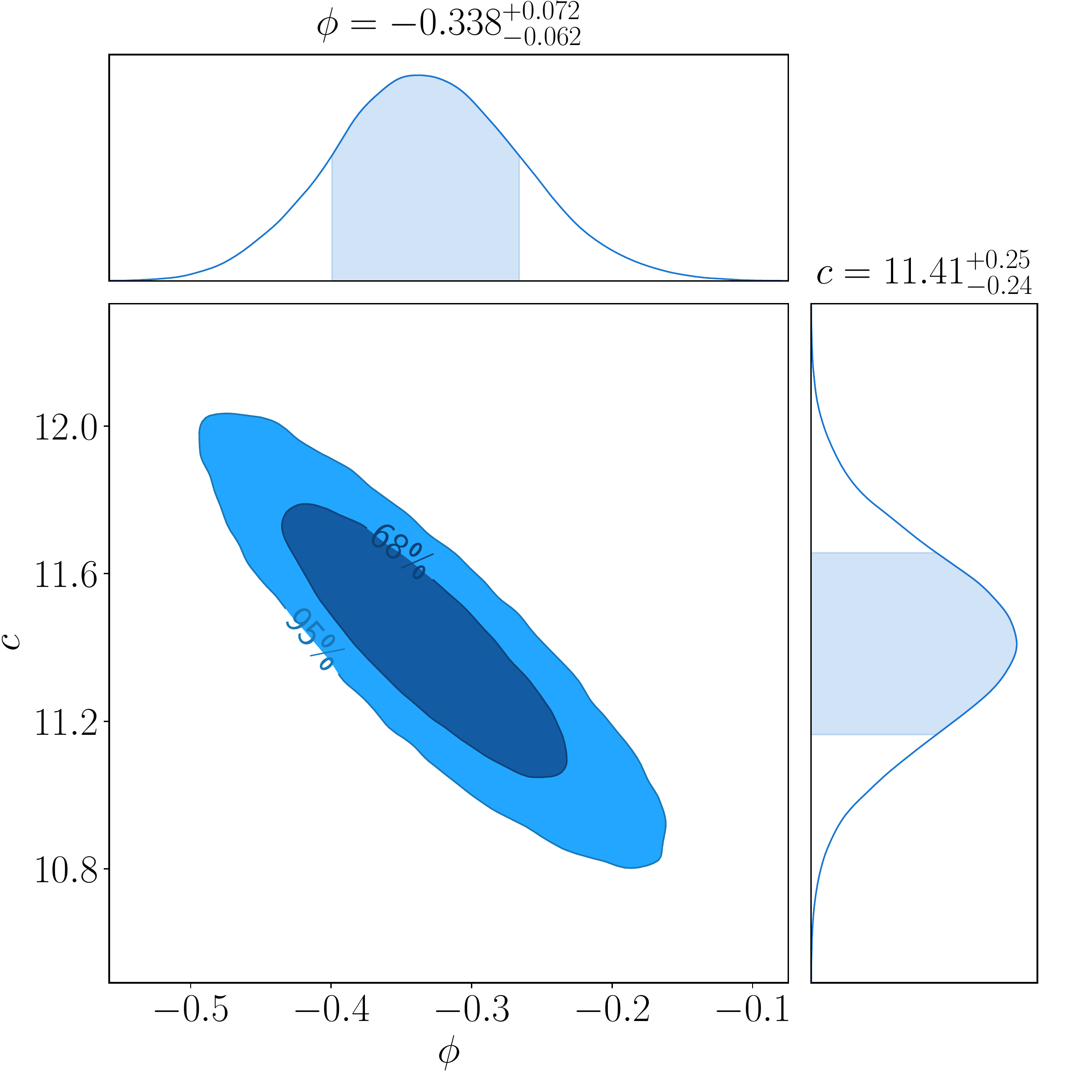}
% 	\caption{Age gradient for LR-V VPH fitted with MILES SSP.}
	\includegraphics[width=0.66\linewidth]{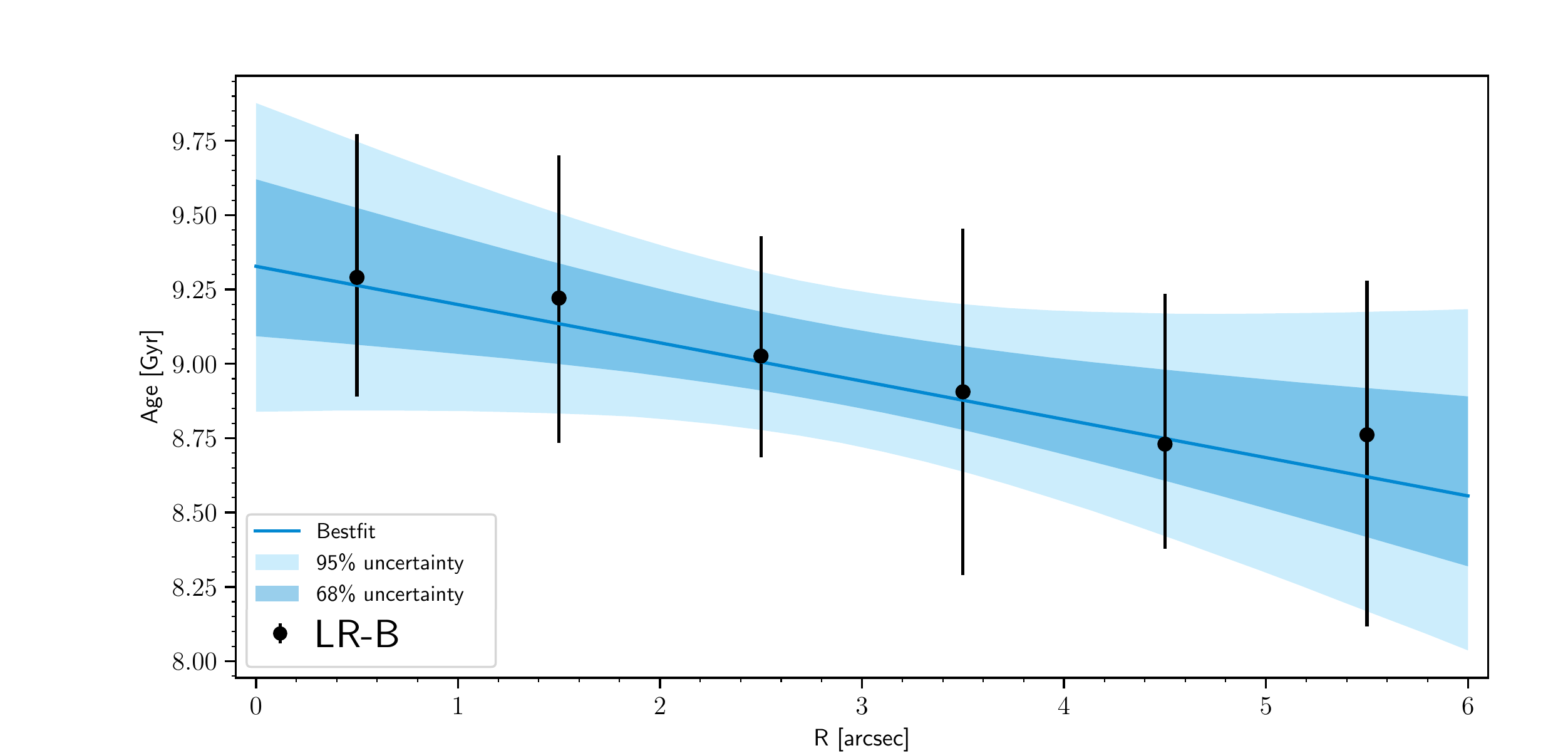}
	\includegraphics[width=0.3\linewidth]{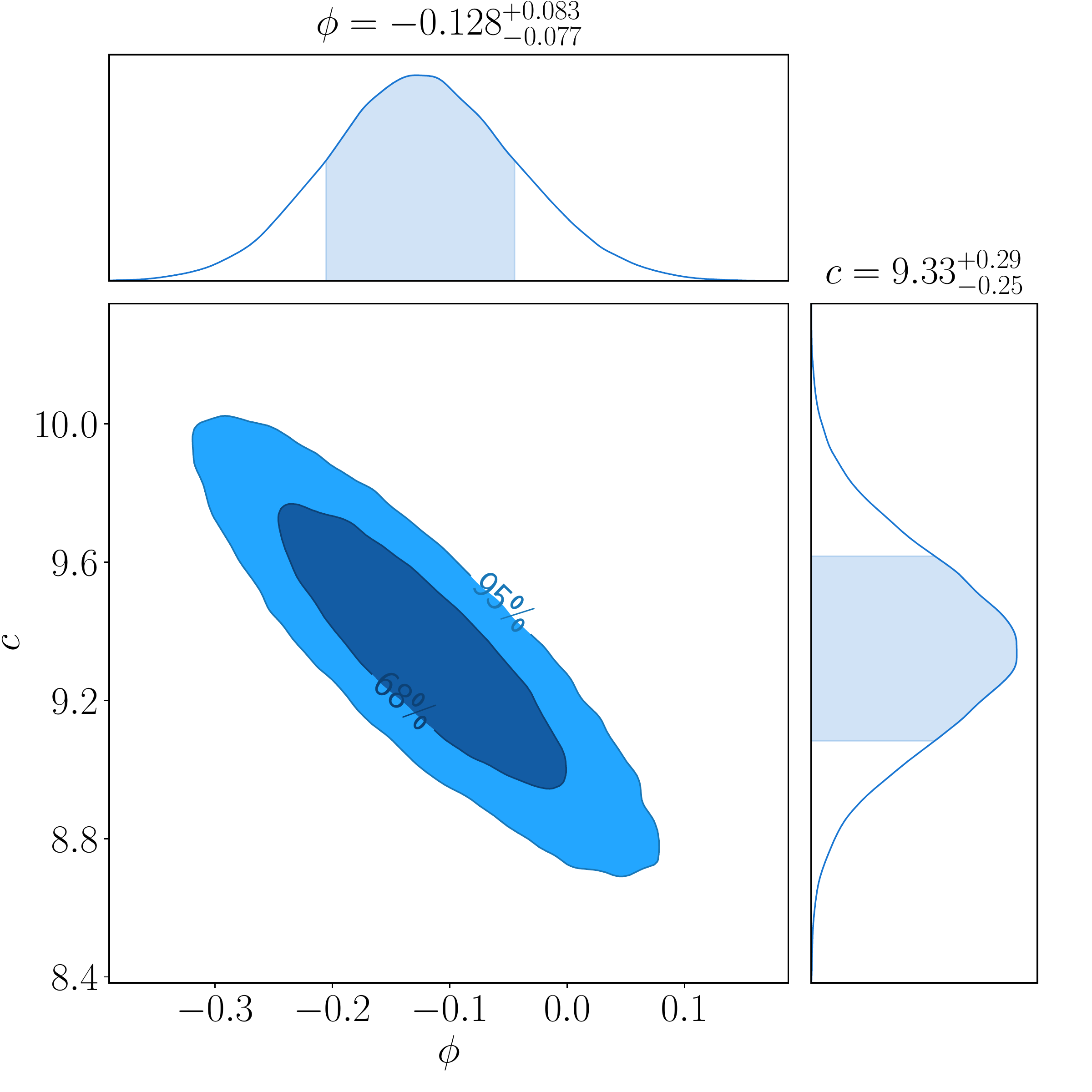}
% 	\caption{LR-B.}
	\includegraphics[width=0.66\linewidth]{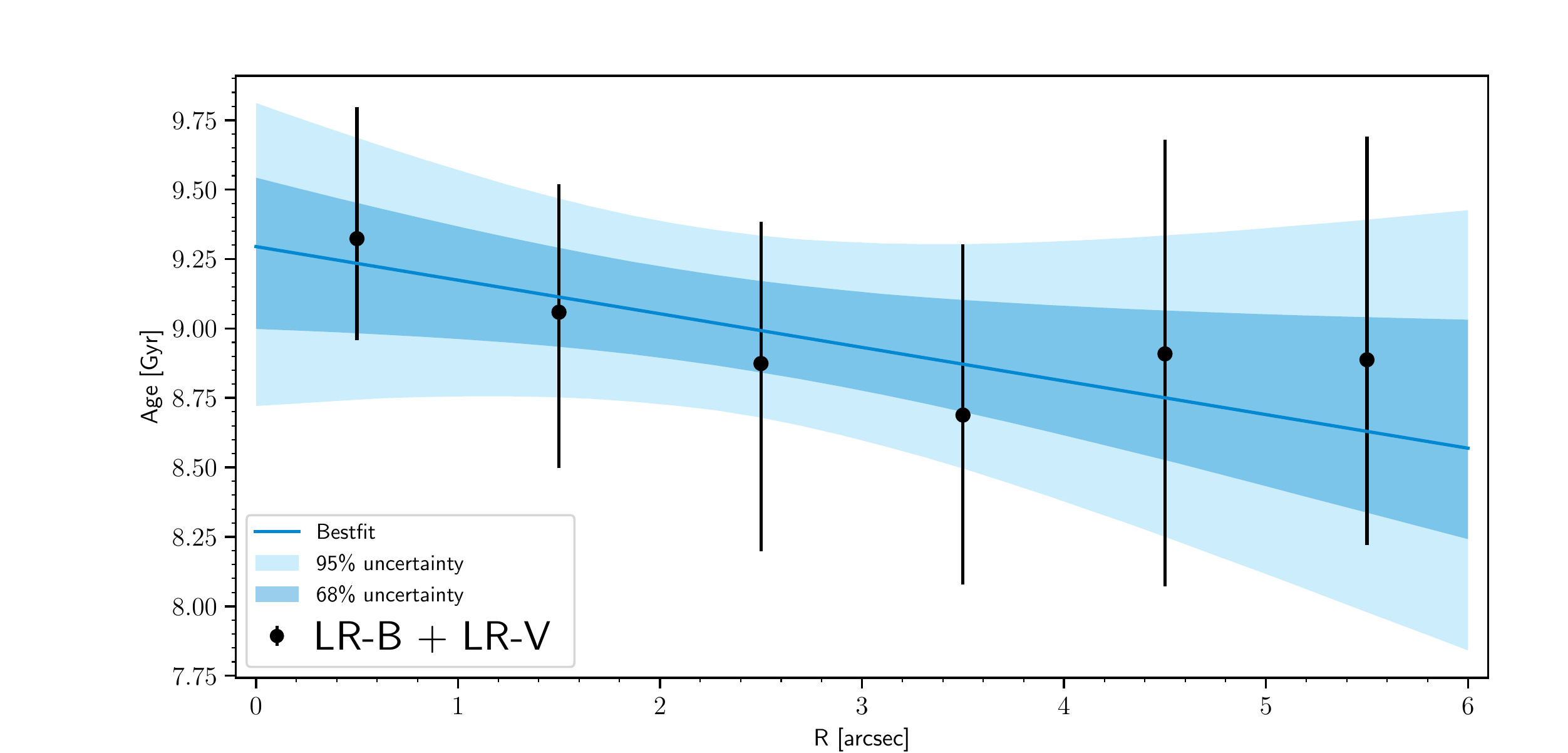}
	\includegraphics[width=0.3\linewidth]{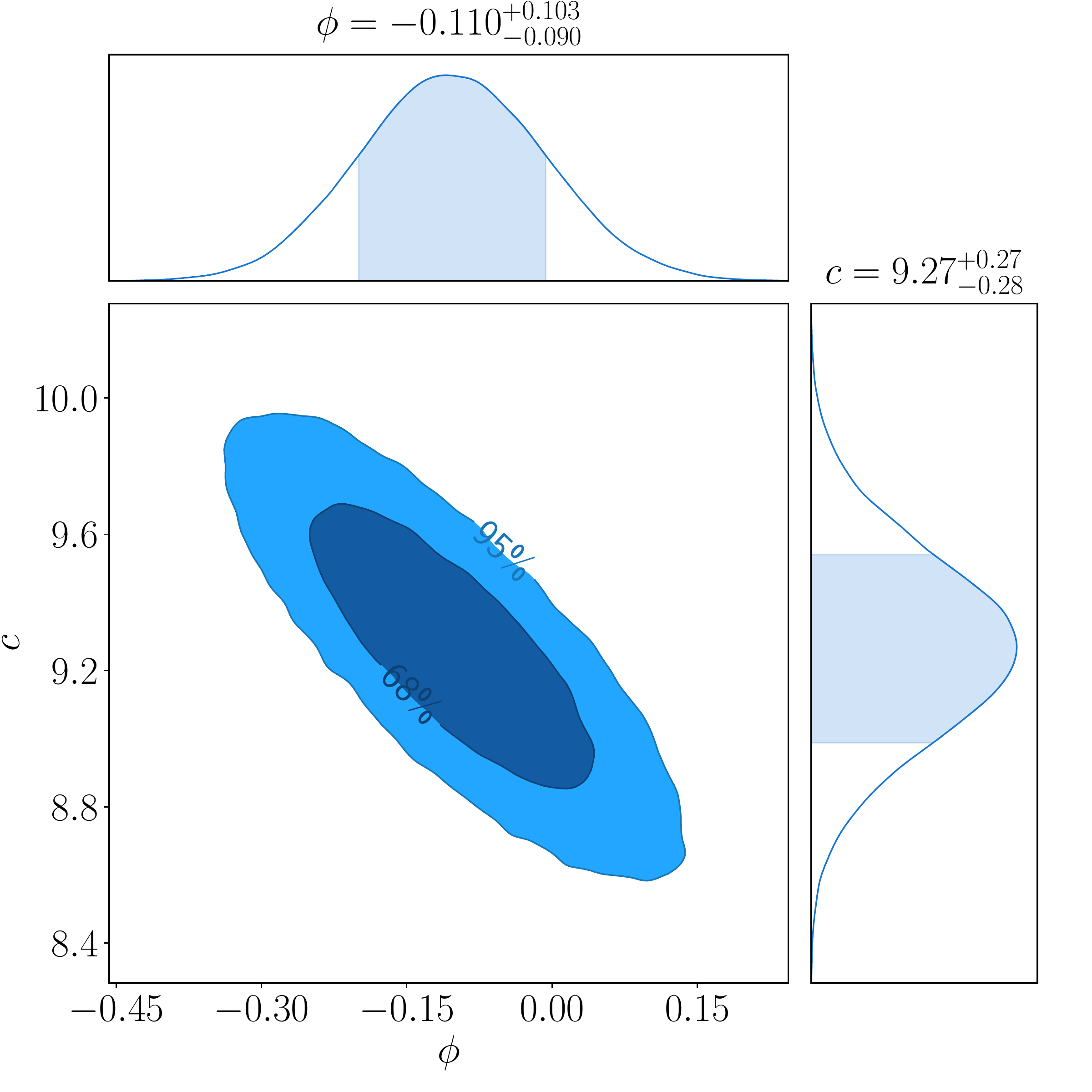}
	\caption{Left column from top to bottom: radial age profiles measured based on the data from LR-V, LR-B and the combination of both. Age data (and corresponding error bars) are the mass-weighted ages derived with pPXF (see text for details). Right column panels: probability distribution of the arc tangent of the slopes ($\phi$) and y-intercepts ({\it c}) obtained with the {\it ChainConsumer} MCMC Bayesian method.}
\label{fig:age_gradient_bayes}
\end{figure*}

In Figure \ref{fig:age_gradient_bayes} we show the age extracted frome the different elliptical rings that are applied to the LR-V, LR-B and LR-B+LR-V IFU data of NGC~7025. We employed the {\it ChainConsumer} Markov Chain Monte Carlo (MCMC) Bayesian method to estimate the age gradients using the Python code from \cite{Chainconsumer}. We have fitted our data linearly but with a small modification. Instead of fitting a linear relation described as $y= mx + c$, we have fitted a model such as $y = \tan(\phi)x + c$ following the recommendation by the code developer. By doing so the prior has a more uniform distribution. The median output values for the slope [$\tan(\phi)$] and the y-intercepts along with their corresponding marginalised 1-$\sigma$ errors are given in Table \ref{table:age_gradients}.

Figure \ref{fig:age_gradient_bayes} shows the radial age profile for NGC 7025 from the LR-V setup. Overall, it is evident that the bulge of the galaxy has an old age ($\sim$ 10.5\,Gyr) and a negative age gradient of \mbox{$-$0.93 $^{+0.21}_{-0.18}$}\,Gyr\,kpc$^{-1}$. In the right-hand panel, we show the corresponding probability distribution of the arc tangent of the slopes (see above) and y-intercepts that we obtained with the {\it ChainConsumer} MCMC Bayesian fitting method. Note that this plot excludes a flat age gradient ($\phi$=0) with high confidence.

Regarding the results on the analysis of the LR-B setup (middle panels of Figure \ref{fig:age_gradient_bayes}) two clear differences are noticeable. First, the absolute values for the mass-weighted ages of NGC~7025 are slightly different in this case ($\sim$ 9\,Gyr) from those derived for LR-V, as expected from the spectral line indices analysis. In addition to this, we find that the age gradient derived, \mbox{$-$0.36 $^{+0.24}_{-0.22}$}\,Gyr\,kpc$^{-1}$, is shallower than that obtained for LR-V. This difference is large enough for the flat age gradient solution to be now excluded only at a $\sim$1-2-$\sigma$
level. The comparison between these results shows how delicate the spectral range of use is when deriving the properties of composite stellar populations in galaxies. 

Finally, to try to shed some light on this issue, we examine what happens if we combine the LR-B and LR-V observations. Bottom panels of Figure \ref{fig:age_gradient_bayes} show the results obtained with this configuration. These results (see also Table \ref{table:age_gradients}) are much closer to those obtained by analysing only LR-B observations than by analysing LR-V observations. Here the trend with galactocentric distance seems to flatten out as we move away from the centre of NGC~7025, something we do not see in the LR-B data. This could be due to the increasing influence of the spectral features present in the spectral range covered by LR-V, perhaps due to the larger signal-to-noise ratio of LR-V (compared to LR-B) at these outer radii.

\begin{figure}[h]
	\centering
    \includegraphics[trim={0 0 0 1cm}, clip, width=1\linewidth]{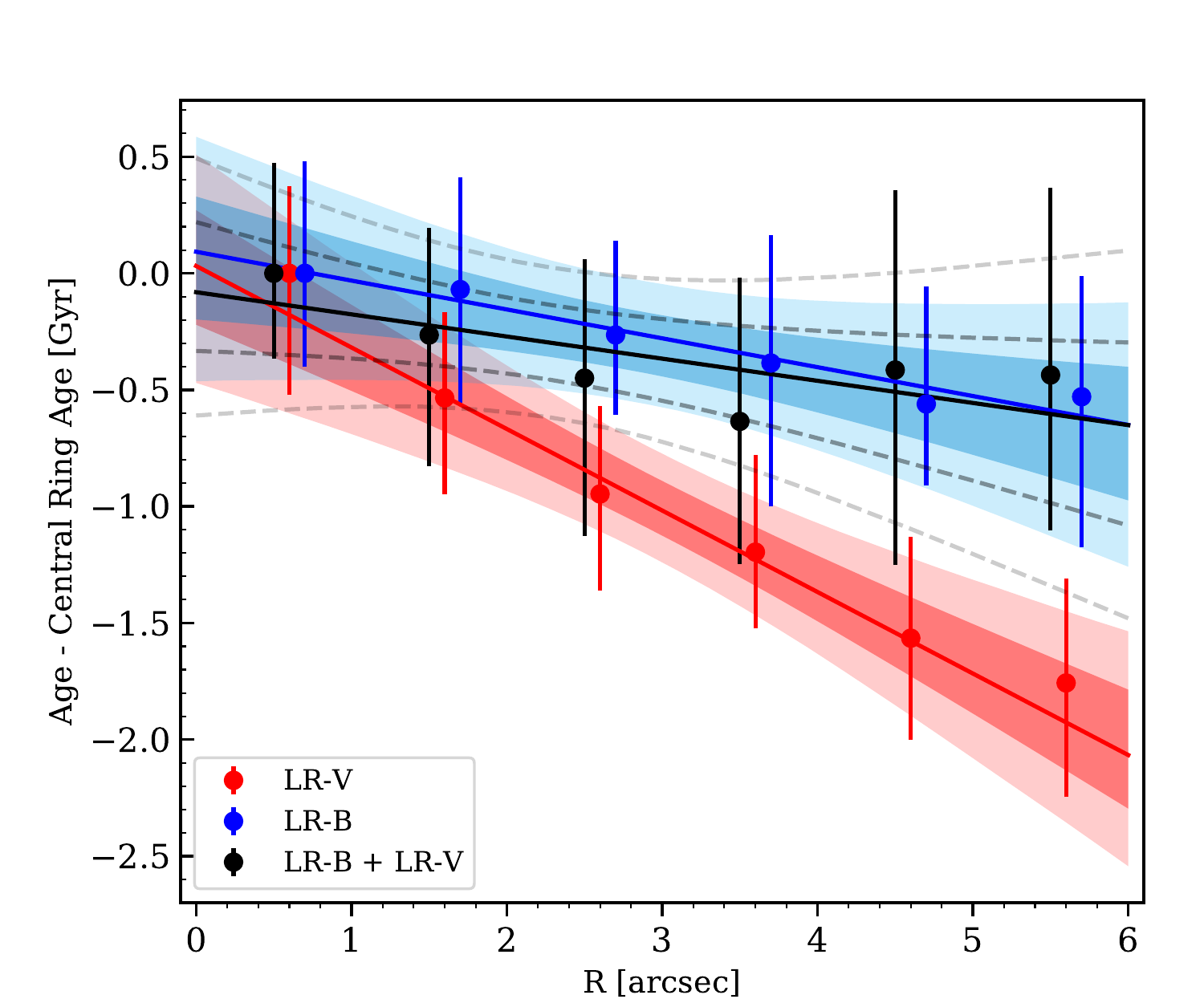}
	\caption{Age gradients for LR-V, LR-B and LR-B+LR-V spectral setups. In this plot we have subtracted the age of the central ring from the rest of the age measurements. The scatter points are shifted slightly on the x-axis for a better visualisation. Red and blue shaded areas show confidence levels for LR-V and LR-B setups. Dark and light dashed-lines also represent 1-$\sigma$ and 2-$\sigma$ confidence levels respectively for the LR-B+LR-V spectral setup.}
	\label{fig:age_gradients_3_redes}
\end{figure}

The aforementioned results strongly suggest that the absolute value of the age (even if it is a mass-weighted value) of a composite stellar population is markedly sensitive to the spectral range covered by the observations. However, we can rely more robustly on relative changes, such as age gradients (in our case a flat age gradient is excluded at $\geq$1-$\sigma$ confidence level in all cases). Figure~\ref{fig:age_gradients_3_redes} displays simultaneously the age gradients of the galaxy from the three VPH configurations. To remove the global offset in age between the different setups, we have subtracted from all measurements the value of the age measured in the central ring. For sake of a better visualisation, we have slightly shifted the data points along the x-axis. In this figure we appreciate that, within the errors, all the results obtained agree with a mild negative age gradient. However, the need for a careful analysis in order to draw firm conclusions should be emphasised, especially when dealing with composite stellar population. 

\begin{table}
\caption{Age gradients [$\tan(\phi)$] and y-intercepts derived from different spectral setups following the {\it ChainConsumer} MCMC Bayesian method. }              % title of Table
\label{table:age_gradients}      % is used to refer this table in the text
\centering                                      % used for centering table
\begin{tabular}{c c c}          % centered columns (4 columns)
\hline\hline                     % inserts two horizontal lines 
\noalign{\smallskip}
MEGARA VPH & Slope & Intercept  \\    % table heading
& [Gyr\,kpc$^{-1}$] & [Gyr]  \\
\hline                                   % inserts single horizontal line
\noalign{\smallskip}
    LR-V & $-$0.93 $^{+0.21}_{-0.18}$ & 11.41 $^{+0.25}_{-0.24}$  \\      % inserting body of the table
    \noalign{\smallskip}
    LR-B & $-$0.36 $^{+0.24}_{-0.22}$ & 9.33 $^{+0.29}_{-0.25}$  \\
    \noalign{\smallskip}
    LR-B + LR-V & $-$0.31 $^{+0.29}_{-0.26}$ & 9.27 $^{+0.27}_{-0.28}$ \\
\noalign{\smallskip}
\hline                                             %inserts single line
\end{tabular}
\end{table}  

\begin{figure}[h]
	\centering
    \includegraphics[trim={0 0 0 1cm}, clip, width=1\linewidth]{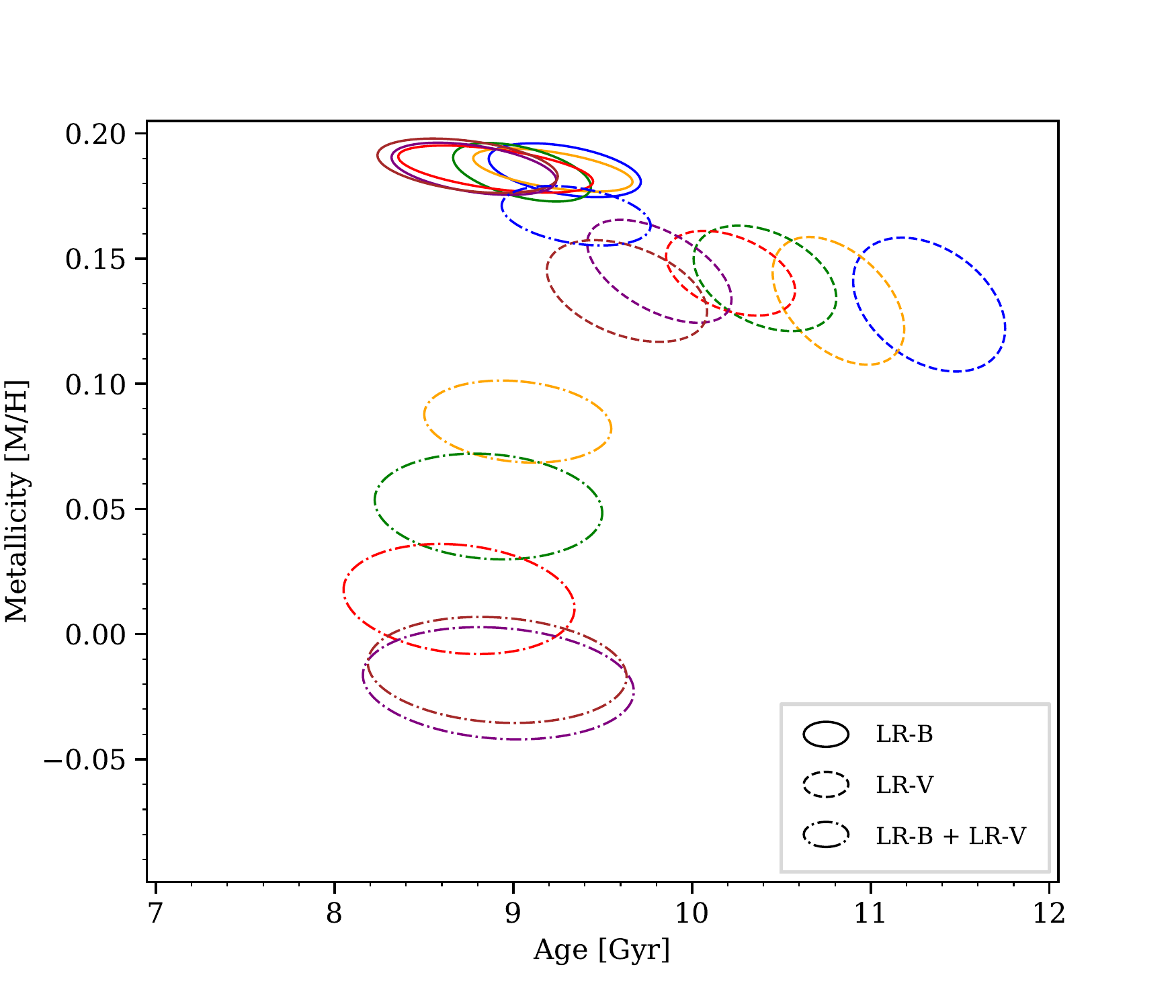}
	\caption{1-$\sigma$ ellipses of the age and metallicity probability distributions obtained for the 100 realisations for each VPH. The order of the rings follows the same colour pattern as in Figure \ref{fig:elliptical_spectra}, with the blue ellipse being the innermost and the brown one the outermost.}
	\label{fig:age_met_degeneracies}
\end{figure}

The evolution of the chemical abundances of elements constitutes one of the three components of galaxy evolution, along with spectro-photometric and dynamical evolution \citep{tinsley}. Therefore, the derivation of the overall metallicity of stellar populations, together with that of the gas in its different phases, is key for the study of stellar populations and galaxy evolution in general. Besides, given the well-known degeneracy between age and metallicity in stellar populations, especially when only photometry is available \citep{worthey}, we now address the analysis of the (mass-weighted) stellar metallicities derived from our pPXF analysis of NGC~7025. Thus, in Figure~\ref{fig:age_met_degeneracies} we show the results obtained when measuring the age and metallicity in rings at different galactocentric radii from the realisations explained in section \ref{section: fitting}. In this figure we witness that, as anticipated from the analysis of the spectral indices (Section~\ref{section: Results: Spectral line indices}), the galaxy exhibits a super-solar metallicity distribution for the most part. If we compare these results with those we obtained for the absorption line indices, we see that they are in best agreement with those inferred from the iron indices (see Figure \ref{fig:indices}). However, we should mention here that in the metallicity range between 0 (Solar) and 0.22\,dex no SSP model is available, which means that all intermediate points between these two values are the result of the different weights obtained in the fitting of only a few models with discrete metallicity values. This means that we have a rather poor sampling of this area of the space of parameters. This is a relevant issue since the impact of metallicity on the spectro-photometric output of the stellar populations is not linear. Nevertheless, this factor should not have a major impact on the age related results, at least for the derivation of relative properties, since we consistently derive super-solar metallicities. Furthermore, as expected we notice degeneracy associated with our results: a younger age can be partly compensated by a higher stellar metallicity.

It should be noted that our results for both age and metallicity are similar to those obtained by \citet{deAmorim_2017} using data from the CALIFA survey \citep{sanchez2012} analysed with the spectral synthesis code \textsc{starlight} \citep{Cid_Fernandes_2005}.

\subsection{Star formation history of NGC~7025}
\label{section:SFH}
The output from our pPXF fits also allows us to reconstruct the star formation history of NGC~7025 as a function of galactocentric distance using the same elliptical rings employed above. This provides us with complementary information to the one presented in section \ref{section: age gradients} where, in order to estimate the age gradients, we used the (mass-weighted) age of all the stellar populations present in each of the regions under study. In contrast, here we can analyse the contribution of each of them to the light we observe within the FoV of the MEGARA IFU. These data can help us to determine whether or not we are detecting stellar populations in our data that were born at different epochs throughout the history of the galaxy.

\begin{figure*}[h]
	\centering
	\includegraphics[width=0.45\textwidth]{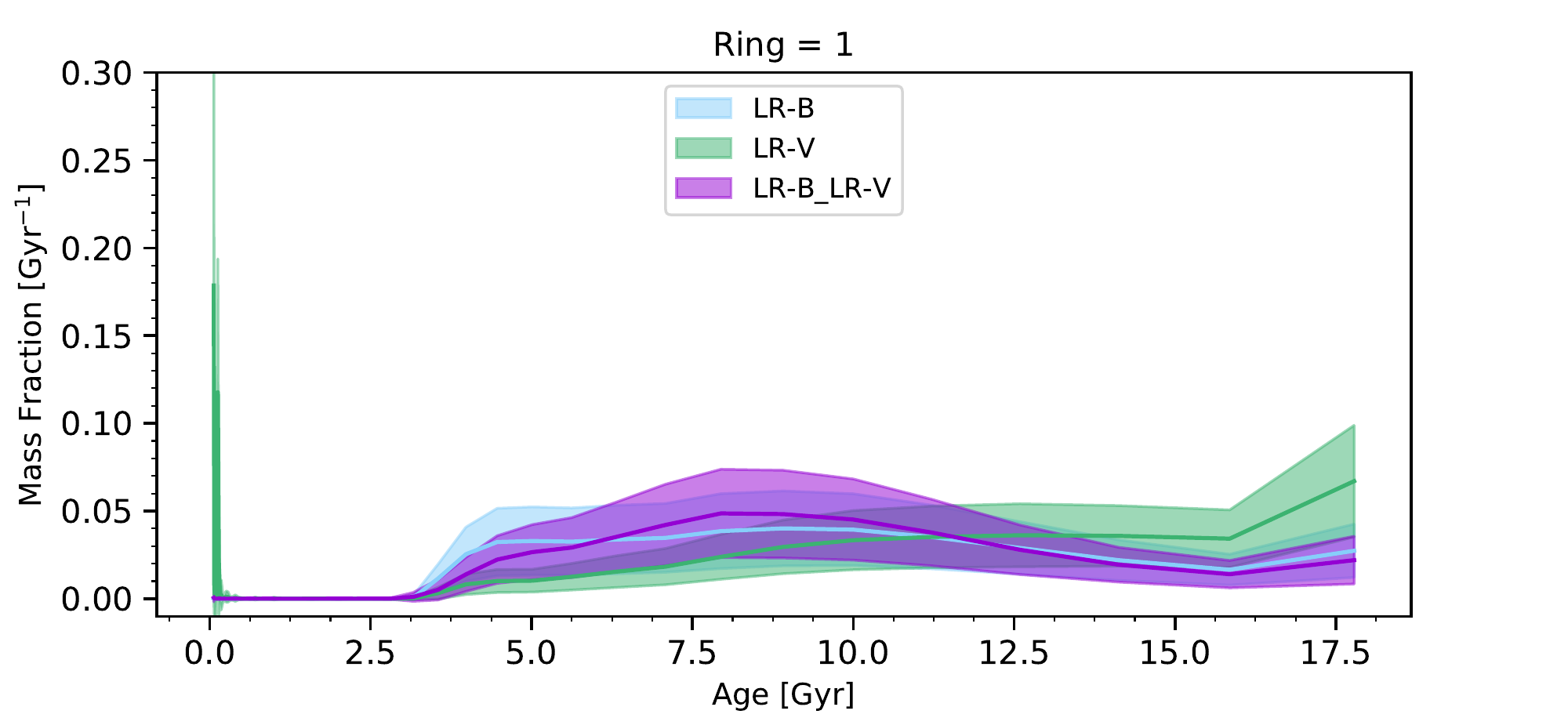}
	\includegraphics[width=0.45\textwidth]{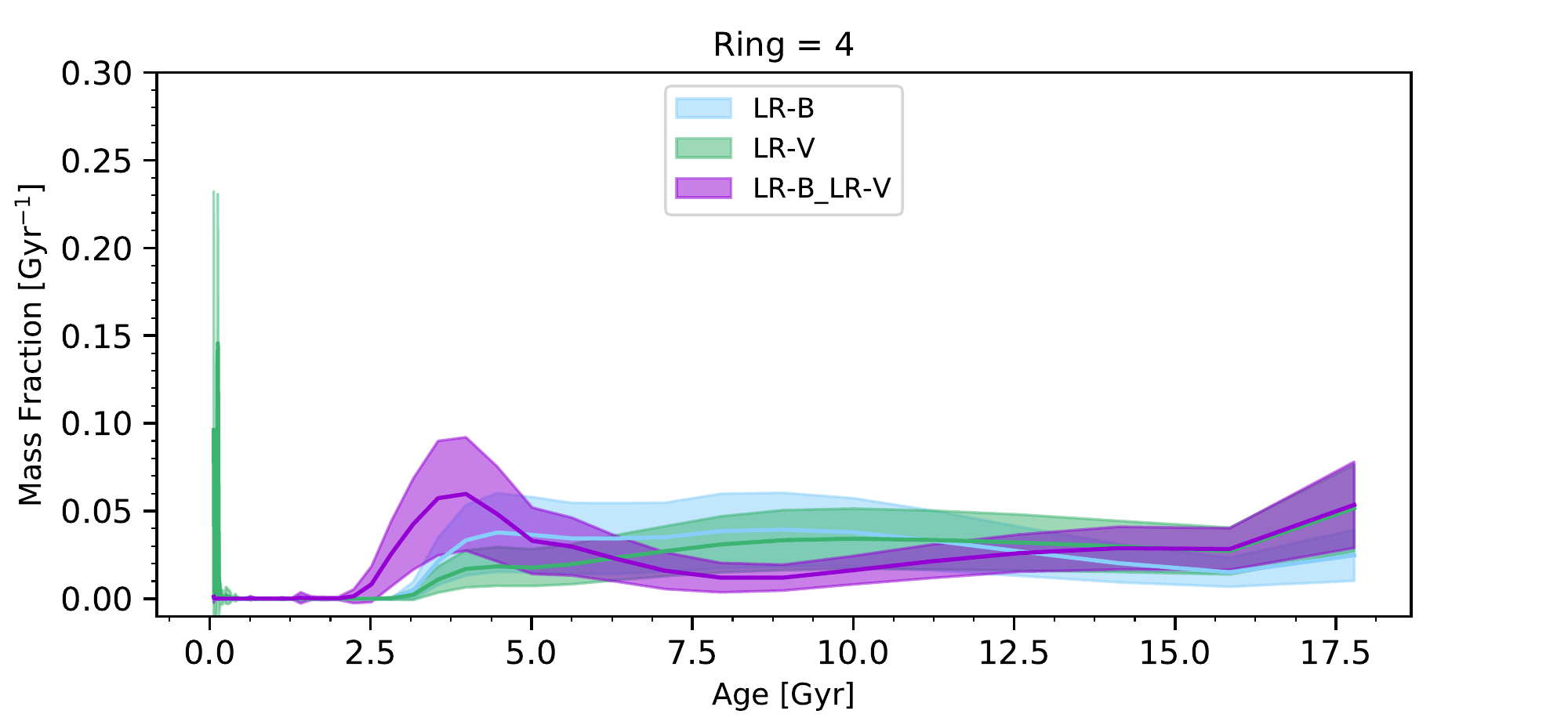}
	\includegraphics[width=0.45\textwidth]{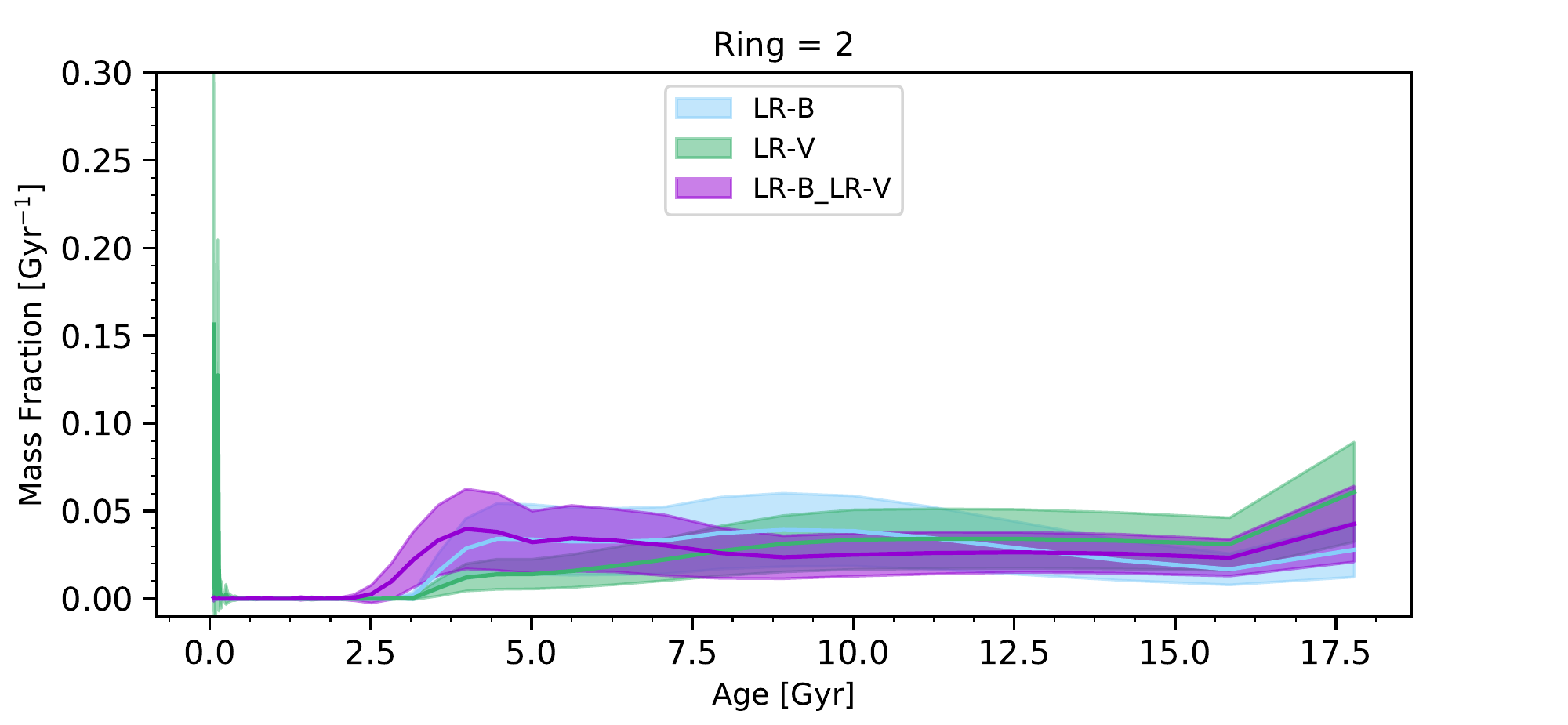}
	\includegraphics[width=0.45\textwidth]{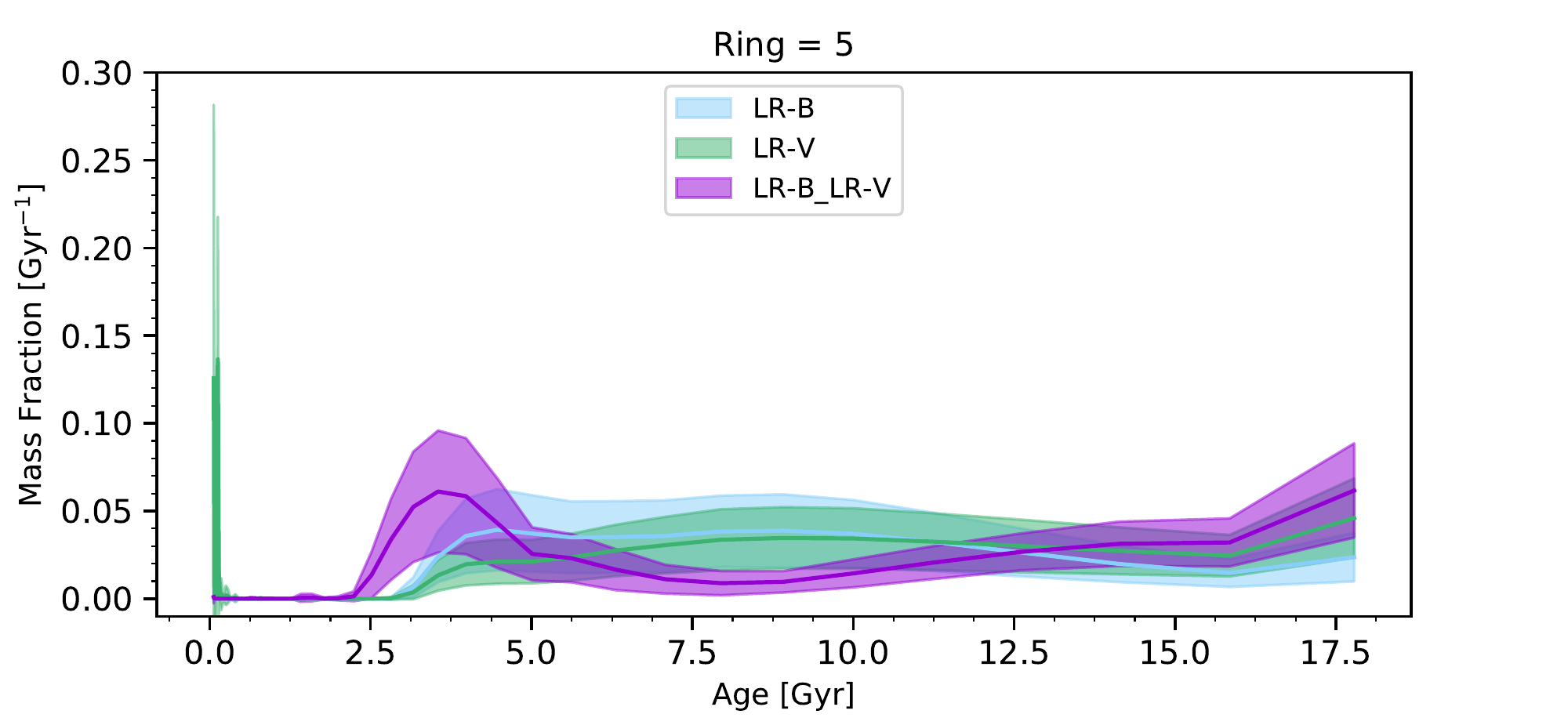}
	\includegraphics[width=0.45\textwidth]{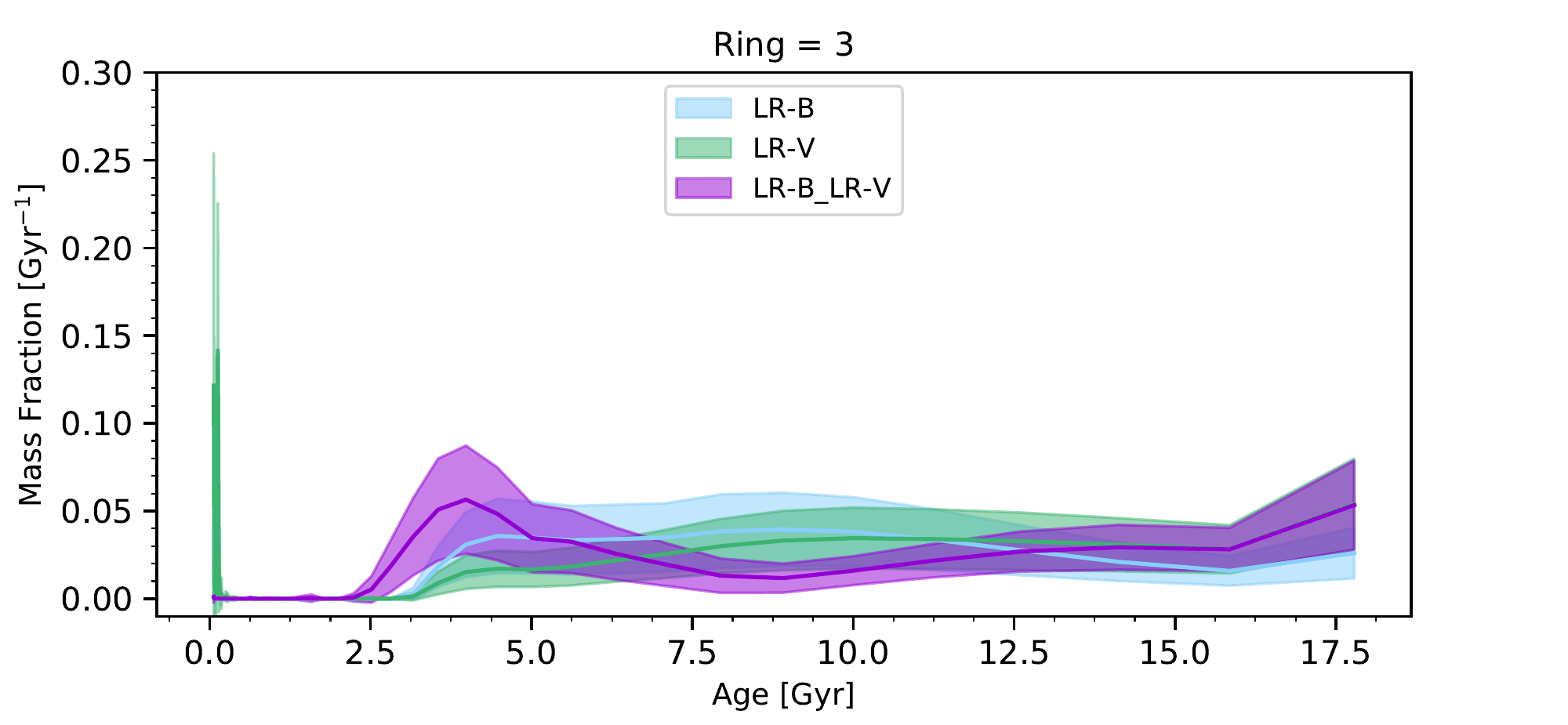}
	\includegraphics[width=0.45\textwidth]{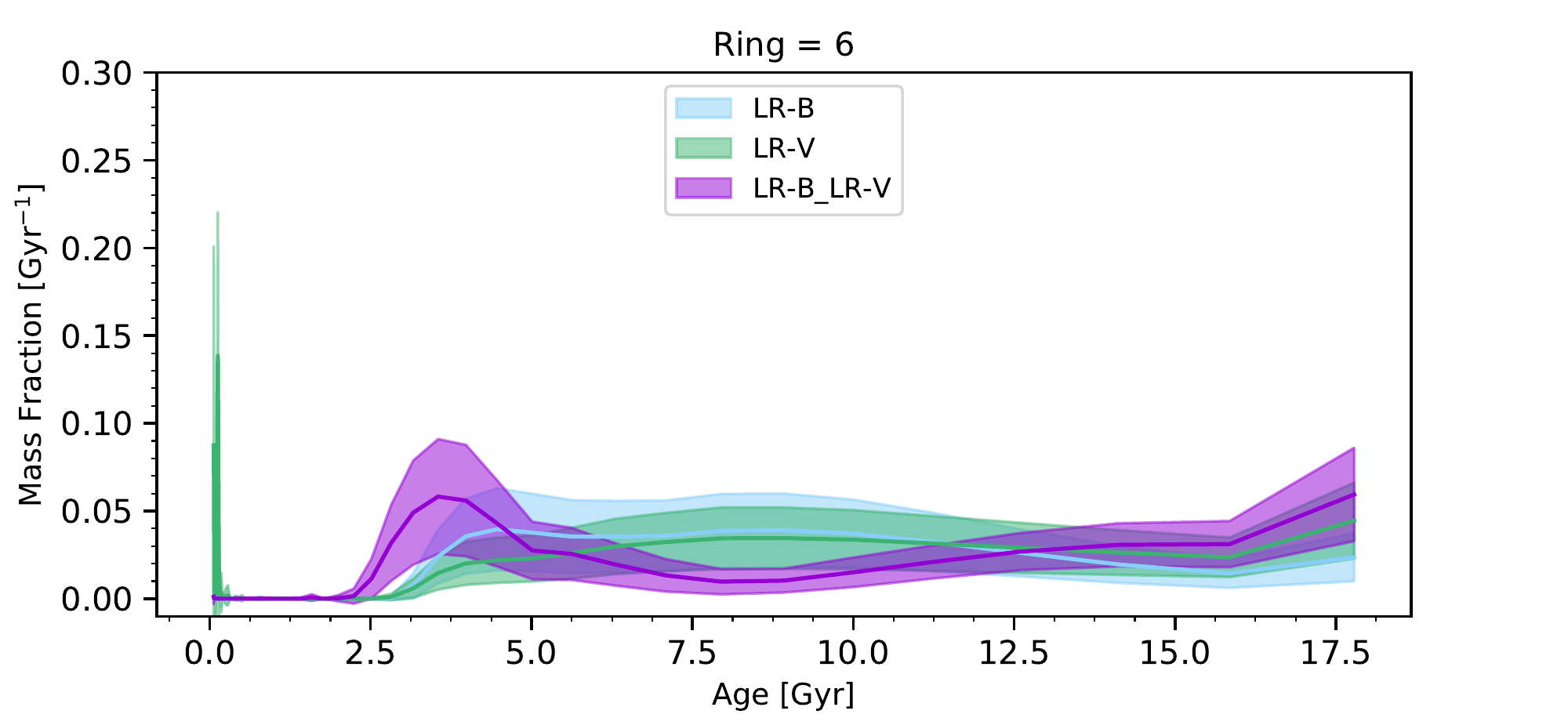}
	\caption{Mass fraction (equivalent to the fractional SFR) as a function of time for the different elliptical rings analysed in NGC~7025 (being ring 1 the central region and ring 6 the outermost region). Each panel shows the mass fraction from the LR-B, LR-V and LR-B+LR-V data together with the corresponding 1-$\sigma$ uncertainties.}
	\label{fig:SFH}
\end{figure*}

We did not limit ourselves to the best-fitting weights obtained for each spectrum (from which best-fitting SFHs are derived) but also included the realisations described in section \ref{section: fitting} to ensure that our results reflect the impact of observational uncertainties and (age-metallicity) degeneracies. In Figure \ref{fig:SFH} we show the average SFHs for all of these realisations, together with their uncertainties, for the observations carried out with the three different VPH setups explored in this paper. A total of six different SFHs are shown, which correspond to the six elliptical rings. Within each panel we show, in different colours, the SFHs for the different MEGARA configurations: LR-B, LR-V and LR-B+LR-V.

As for the SFH obtained from the LR-V data (green curves in Figure~\ref{fig:SFH}), we find that the fractional star formation rate (SFR) progressively increases from $\sim$4\,Gyr to $\sim$10\,Gyr (in look-back time), before it plateaus at older ages and then increasing again at the very old age end. As we move outwards, this rise in SFR at old ages gets progressively less pronounced while, simultaneously, a mound of SFR at ages $\sim$9\,Gyr develops. We also find a peak in the SFH at very young ages that weakens as we move away from the centre of the galaxy. Besides the little residual star formation that could be present within our MEGARA IFU pointing (see emission-line map in \citealt{Barrera_Ballesteros_2015}) this effect could be also due to the lack of sensitivity of the LR-V setup to intermediate ages. Thus, the LR-V spectrum of an  intermediate-aged population could be also reproduced by the combination of a relatively young burst and an older population.  

The age gradient reported in section \ref{section: age gradients} can be clearly deduced from these plots as a consequence of this behaviour.  In the case of the LR-B setup (blue curves in Figure \ref{fig:SFH}) the SFH is quite extended, being rather flat between ages 3 to 13\,Gyr, and decreasing in the last $\sim$3\,Gyr. This behaviour leads to the radial changes in the SFH being also more subtle. The only noticeable difference in between the SFH of the different rings, which could be behind the mild radial variation in mass-weighted age previously reported, is the change in relative strength and age between the rise in SFR at $\sim$4\,Gyr and the bump at $\sim$9\,Gyr.

The combination of the LR-B and LR-V data provides a more complete view of the SFH of the inner regions of NGC~7025 (purple curves in Figure \ref{fig:SFH}). In this case, we find that from the second elliptical region outwards an epoch of high SFR develops at ages around 3.5-4.5\,Gyr on top of a low-SFR plateau (that comes with relatively low uncertainties in fractional SFR) that starts to rise again beyond $\sim$11\,Gyr and ends with a peak at ages older than the age of the Universe. This result highlights the importance of obtaining SFHs derived from deep spectroscopic data when analysing composite stellar populations. Thus, should we have considered the age gradient estimates alone, which use the mass-weighted average of the SSP ages of all the models considered, the two periods of star formation in NGC~7025 would have remained unnoticed. In this regard, the age gradients derived for each setup (and the differences reported between them in Table~\ref{table:age_gradients}) can be interpreted now as the results of a complex history of star formation with multiple episodes of star formation and a relatively quiescent epoch in between them.

We might think that the differences between the results of the three different setups are large and that they do not behave as expected a priori. However, in Figure 10 we see that, if we consider the uncertainties present in each curve, there is not such a large discrepancy between the different setups and that the results are compatible over most of the star formation history of the galaxy.

\section{Discussion}

Once we have all the results of the study of the stellar populations in the central region of NGC~7025 conducted in this article, we now aim to reconstruct its evolutionary history. In \citet{Dullo_2019} they showed that the bulk of the stellar population in the bulge was formed through secular processes instead of by a major merger (this includes evidence for its fast rotation, low Sersic index, large-scale negative colour gradient...). Our results now suggest that, besides this early secular formation, about 3.5-4.5\,Gyr ago, something happened that caused the stellar populations in the bulge to rejuvenate, either through in-situ star formation or through the accretion of stars (from the inner disk or from a companion). 

NGC~7025 has traditionally been classified as an isolated galaxy \citep{Karachentseva_1973}. However, \cite{Barrera_Ballesteros_2015} classified this galaxy as a merger remnant with evident tidal features. Such a post-merger scenario for NGC~7025 nicely fits with the results we have obtained in this work, with NGC~7025 forming stars through secular processes until 3.5-4.5\,Gyr ago when a merger took place leading to a temporary increase in the star formation rate. 

\begin{figure}[h]
	\centering
    \includegraphics[width=1\linewidth]{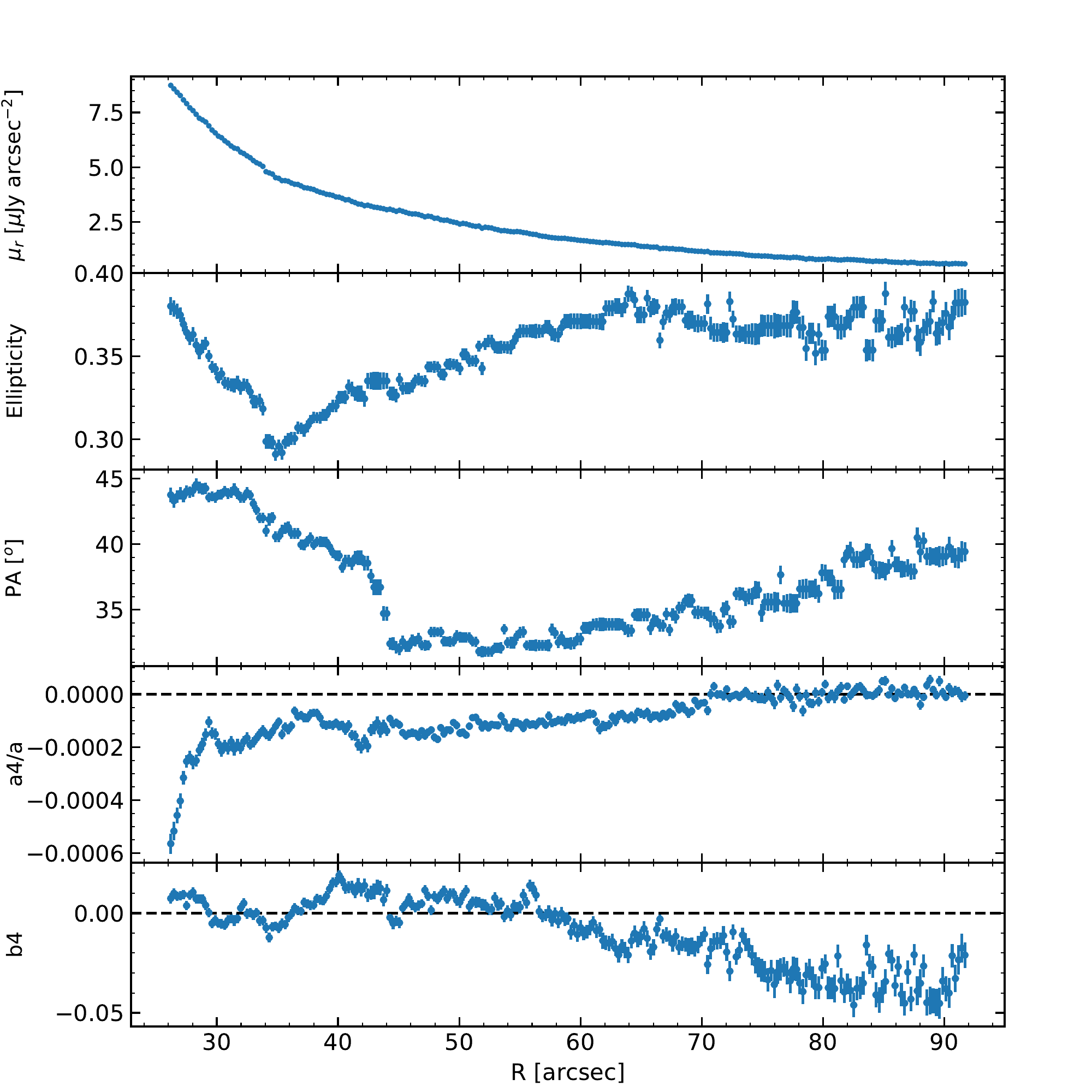}
	\caption{Radial profiles of NGC~7025 derived by running IRAF task {\sc ellipse} on the DECam $r$-band image of NGC~7025. From top to bottom: surface brightness, ellipticity, position angle, a4/a, and b4, including their corresponding errors.}
	\label{fig:perfiles}
\end{figure}

\begin{figure}[h]
	\centering
    \includegraphics[width=1\linewidth]{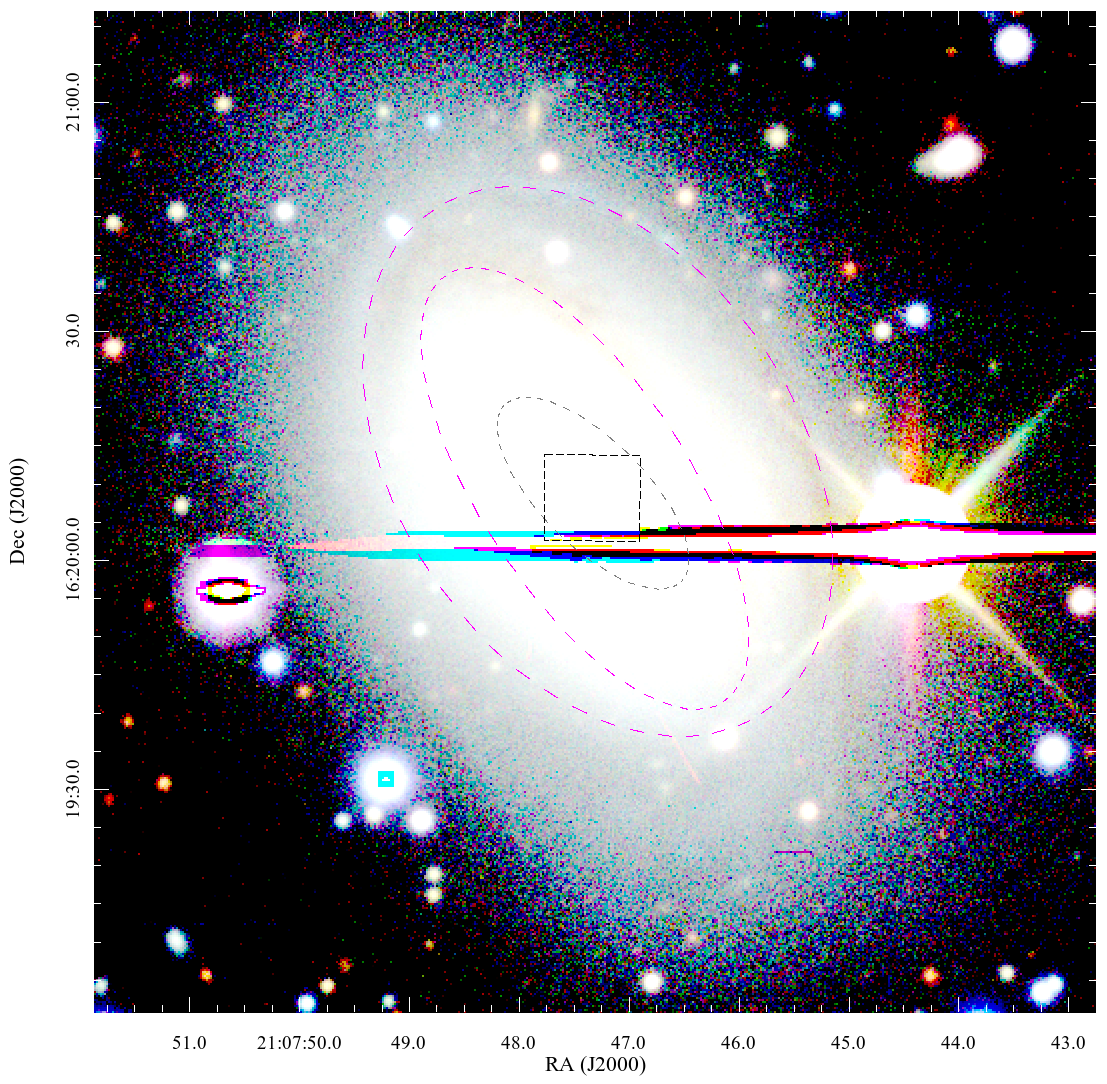}
	\caption{RGB false-colour DECam image of NGC~7025 scaling to better visualise low surface brightness features. Black dashed line represents the MEGARA field of view. Grey and magenta dashed-line ellipses follow the dust bands noticeable.}
	\label{fig:NGC7025_ellipses_b}
\end{figure}

Regarding the morphological evidence for a past merger in the case of NGC~7025, we show in Figure~\ref{fig:NGC7025_ellipses} a set of ellipses tracing some non-axisymmetric features. Some of them follow what appear to be obscuring dust lanes while others resemble extended stellar tidal features, each with a different axial ratio and position angle that can be interpreted as consequence of a warp in its outer disk. Warping can occur for several reasons, one of them being a merger. The presence of a well-formed disc together with the estimated time of the merger leads us to think that if the warp is due to a merger, it should be a minor one with a mass ratio between galaxies of 1 to 10 \citep{Lotz_2010}. Interestingly, such a minor merger could trigger an initial starburst that would cause the further heating of the gas causing the galaxy to stop forming new stars. This is compatible with the sharp drop in star formation history seen 2.5\,Gyr ago in the bulge of NGC~7025. Another argument supporting this scenario is the fact that this galaxy has boxy outer isophotes \citep{naab_2006} which can be seen in the high-contrast DECam false-colour image shown in Figure \ref{fig:NGC7025_ellipses_b}. Indeed, by running the IRAF task {\sc ellipse} on the DECam $r$-band image of NGC~7025 we derive $a_4/a$ and $b_4$ radial profiles (see Figure~\ref{fig:perfiles}). In the region where the outer isophotes can be measured with low uncertainties (between 60-70\,arcsec; see Figure \ref{fig:NGC7025_ellipses_b}) we find values of $-$1$\times$10$^{-4}$ and $-$0.2, respectively for $a_4/a$ and $b_4$, indicative of the presence of boxy isophotes. 

\begin{figure}[h]
	\centering
    \includegraphics[width=1\linewidth]{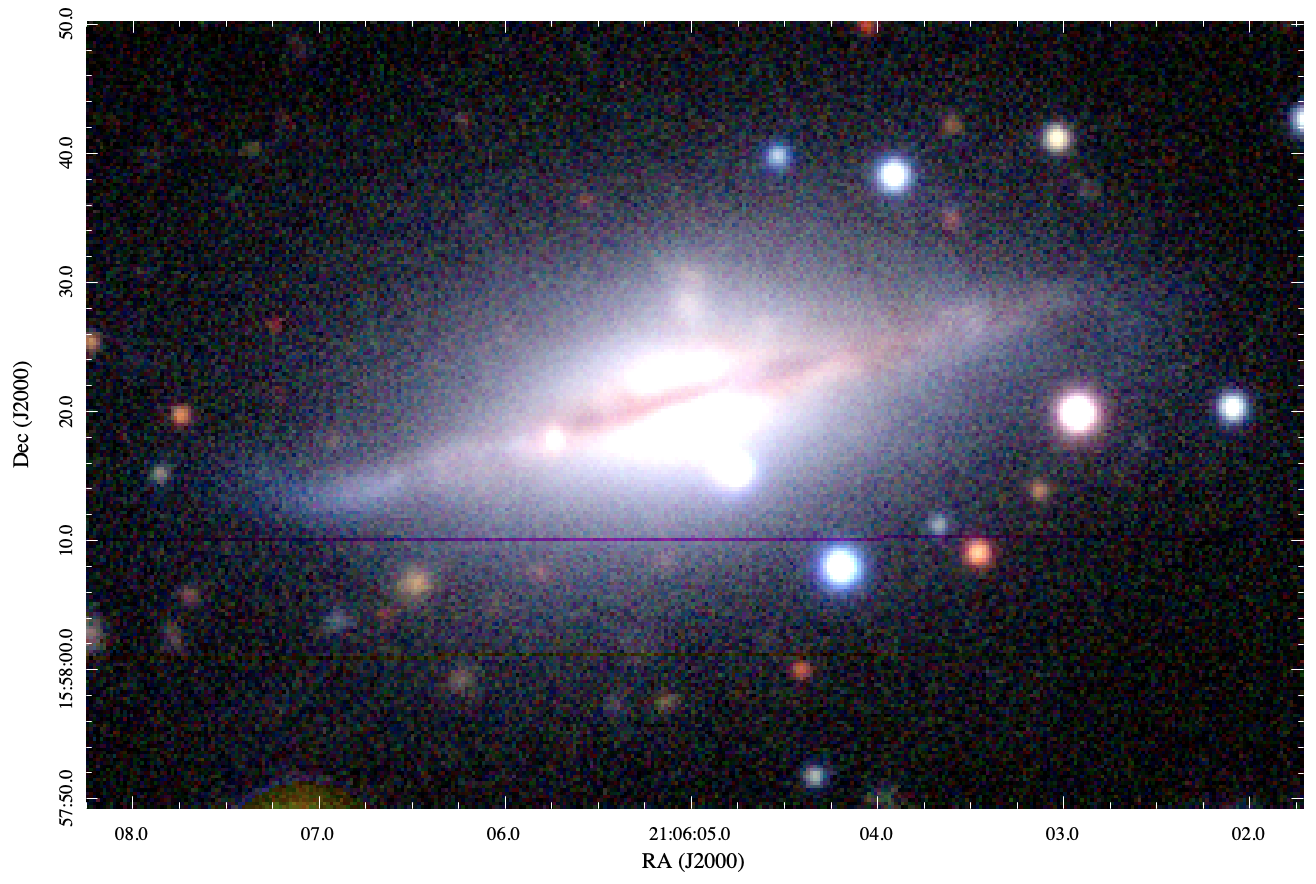}
	\caption{RGB false-colour DECam image of UGC~11677 in asinh scale using z, r and g filters (respectively).}
	\label{fig:UGC11677}
\end{figure}

An alternative scenario to this merger would be a fly-by encounter with another galaxy. An event of this kind could also lead to a gravitational perturbation strong enough to (1) cause a warp that could actually survive up to several billion years, depending mainly on the angle of incidence of the interaction \citep{Kim_2014}, and (2) trigger star formation at the time of the fly-by passage. The main problem with this scenario is the fact that we do not find any certain candidate galaxy for such an encounter. The only object that could have had a close encounter with NGC~7025 about 3.5-4.5\,Gyr ago is UGC~11677 (at a projected distance of 667\,kpc). UGC~11677, with $m_{B}$=16\,mag, is two magnitudes fainter than NGC~7025, with $m_{B}$=14.1\,mag \citep{Wenger_2000}, so if it ever had a noticeable gravitational interaction with NGC~7025, it should show stronger signs of perturbation than NGC~7025 itself. However, Figure \ref{fig:UGC11677} shows that UGC~11677 presents a rather regular morphology characteristic of Sa-type galaxy with a prominent bulge and a relatively unperturbed edge-on disc. 

\section{Conclusions}

We have performed a comprehensive analysis of the spectroscopic data taken with MEGARA of the inner regions of the early-type spiral galaxy NGC~7025. We have measured absorption line indices and have performed full spectral fitting of these data. From these measurements we have derived the star formation history and mass-weighted ages at different galactocentric distances and the corresponding age gradients. Our results allow us to put forward a scenario under which NGC~7025 is a galaxy that evolved secularly until about 3.5-4.5\,Gyr ago, when it experienced a relatively minor merger (mass ratio 1/10) that induced an increase in star formation and perturbed the geometry of its disc and possibly ended up quenching the star formation for the last 2.5\,Gyr. Besides, the presence of this intermediate-age population seems to become more prominent in the outer regions of the bulge, which leads to a  development of a peak in the SFH at those ages and to the mild negative age gradient derived. 

Besides these specific results on NGC~7025, we report on different lessons learned for the ongoing exploitation of the MEGADES survey with GTC. First, our analysis shows that determining the stellar population properties in galaxies by measuring the absorption line indices leads to an incomplete picture, especially if the data do not cover a sufficiently wide spectral range. Nevertheless, the measurement of these absorption line indices do allow verifying the results obtained with more complex tools. Thus, if we aim to unravel the nature of stellar populations more precisely, we must consider full spectral fitting tools. Although this method may be more effective, we must also be aware that the spectral range we are studying can influence the results of our analysis, with a bluer wavelength coverage being more sensitive to the presence of younger populations. Not only the wavelength range under study, but also the signal-to-noise and even the velocity dispersion and resolution of the spectroscopic data used have an impact on the uncertainties in the derivation of the star formation history of composite stellar populations. In the context of the MEGADES survey, we have carried out a number of tests to determine the expected errors (including potential biases) in such SFH derivations as a function of these parameters, namely spectral setup, S/N, $\sigma$ and the SFH itself. In Appendix~\ref{section: appendix} we show the results of these tests for the combination of parameters for the MEGARA observations of NGC~7025 (LR-B, LR-V and LR-B+LR-V setups and a galaxy velocity dispersion of $\sigma$$\sim$250\,km\,s$^{-1}$).

\begin{acknowledgements}
Based on observations made with the Gran Telescopio Canarias (GTC), installed in the Spanish Observatorio del Roque de los Muchachos of the Instituto de Astrof\'isica de Canarias, in the island of La Palma.

IRAF is distributed by the National Optical Astronomy Observatory, which is operated by the Association of Universities for Research in Astronomy (AURA) under a cooperative agreement with the National Science Foundation.

This project used data obtained with the Dark Energy Camera (DECam), which was constructed by the Dark Energy Survey (DES) collaboration. The Legacy Surveys imaging of the DESI footprint is supported by the Director, Office of Science, Office of High Energy Physics of the U.S. Department of Energy under Contract No. DE-AC02-05CH1123, by the National Energy Research Scientific Computing Center, a DOE Office of Science User Facility under the same contract; and by the U.S. National Science Foundation, Division of Astronomical Sciences under Contract No. AST-0950945 to NOAO.

This research has made use of the NASA/IPAC Extragalactic Database, which is funded by the National Aeronautics and Space Administration and operated by the California Institute of Technology.

This research has made use of the SIMBAD database, operated at CDS, Strasbourg, France.

This work has been supported by MINECO-FEDER grants AYA2016-75808-R and RTI2018-096188-B-I00.

JIP acknowledges financial support from the State Agency for Research of 
the Spanish MCIU through the Center of Excellence Severo Ochoa award to 
the Instituto de Astrof\'{\i}sica de Andaluc\'{\i}a (SEV-2017-0709).
\end{acknowledgements}
%-------------------------------------------------------------------
% \newpage
\begin{appendix}
\section{Mock data tests}
\label{section: appendix}

In order to analyse the results to be obtained as part of the MEGADES survey using pPXF we first wanted to check the performance of the software with the different SSPs to be used to fit these spectra. Thus, we created different mock spectra from templates of different ages, 1.00, 3.16, 5.01, 6.31, 7.94, 10.00 and 12.59\,Gyr, and with solar metallicity. To do this, we took the original templates \citep{Vazdekis_2010} and broadened them to different velocity dispersion values, 50, 100 and 250\,km\,s$^{-1}$, and added different (Gaussian) noise levels until we obtained a signal-to-noise ratio of 10, 20, 50, 100 and 150 at the central wavelength of each spectral setup.

These mock spectra were divided into different spectral ranges that covered the same regions of the optical spectrum of each individual MEGARA VPH or some combination of these. Thus, we created mock spectra in the spectral ranges of LR-U, LR-B, LR-V, LR-R, LR-B+LR-V, LR-U+LR-B+LR-V, LR-B+LR-V+LR-R and LR-U+LR-B+LR-V+LR-R. Here we show specifically the results for LR-B, LR-V and LR-B+LR-V, since those are the configurations on which we have focused for NGC~7025. These mock spectra were analysed using the same wild bootstrapping method followed for the observations of NGC~7025, as explained in section~\ref{section: fitting}. 

\begin{figure}[h]
	\centering
    \includegraphics[trim={0 0.5cm 1cm 1.2cm}, clip, width=1\linewidth]{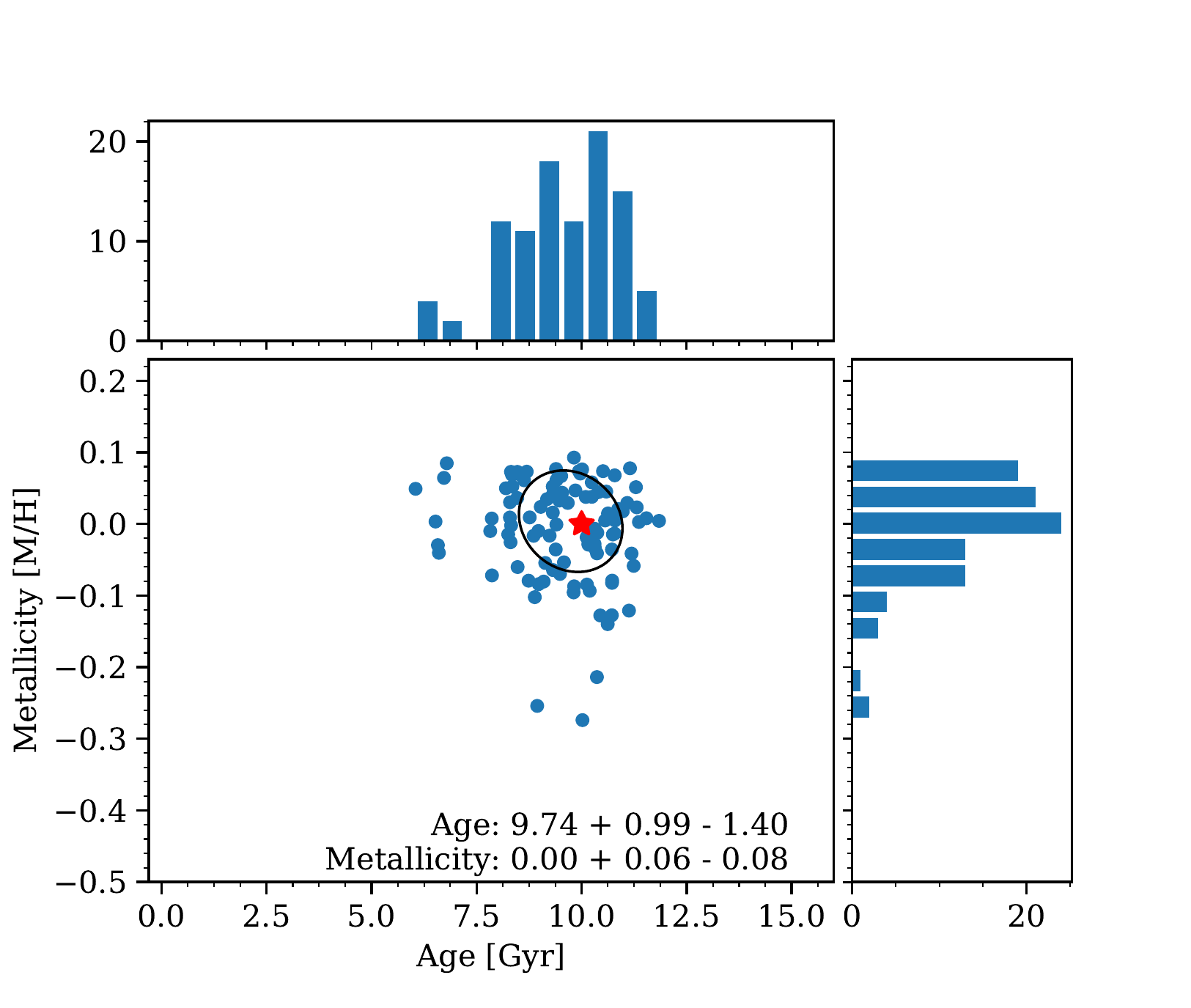}
	\caption{Age and metallicity for MILES model realisations for an age of 10\,Gyr and solar metallicity in the LR-V spectral range with a signal-to-noise ratio of 100 and $\sigma$ of 250\,km\,s$^{-1}$ . The red star shows the input values while the black ellipse encompasses the region with all the solutions within 1$\sigma$.}
	\label{fig:hist1}
\end{figure}

In Figure~\ref{fig:hist1} we report on the results obtained by measuring with pPXF the age and metallicity of the mock spectra realisations made for a SSP model with an age of 10\,Gyr and solar metallicity in the LR-V spectral range with a signal-to-noise ratio of 100 and a line-of-sight velocity dispersion of 250\,km\,s$^{-1}$. Although the scatter of the results is quite large, we recover the input age and metallicity of the original model to 1$\sigma$.

\begin{figure*}[h]
    \center
	\includegraphics[width=0.33\linewidth]{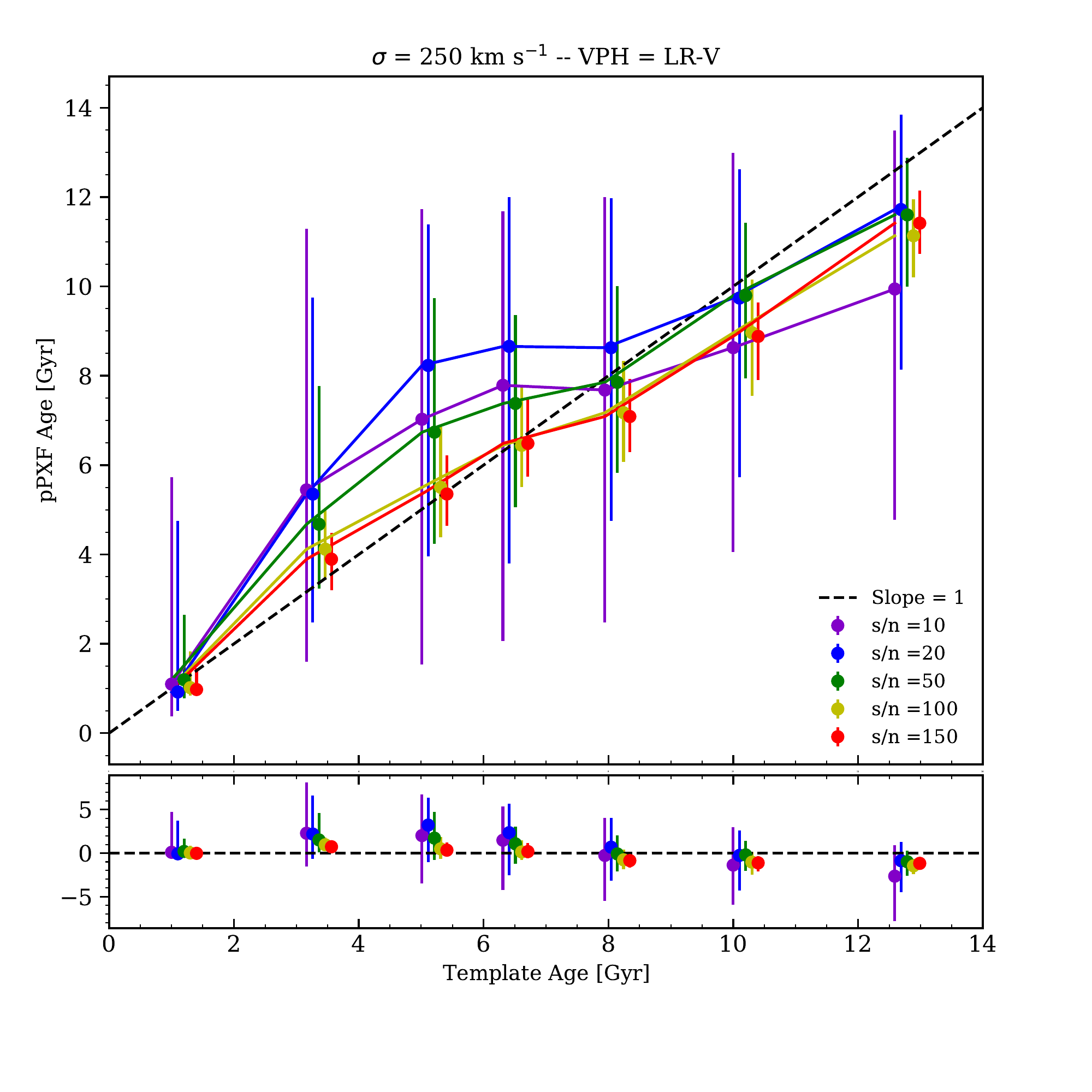}
	\includegraphics[width=0.33\linewidth]{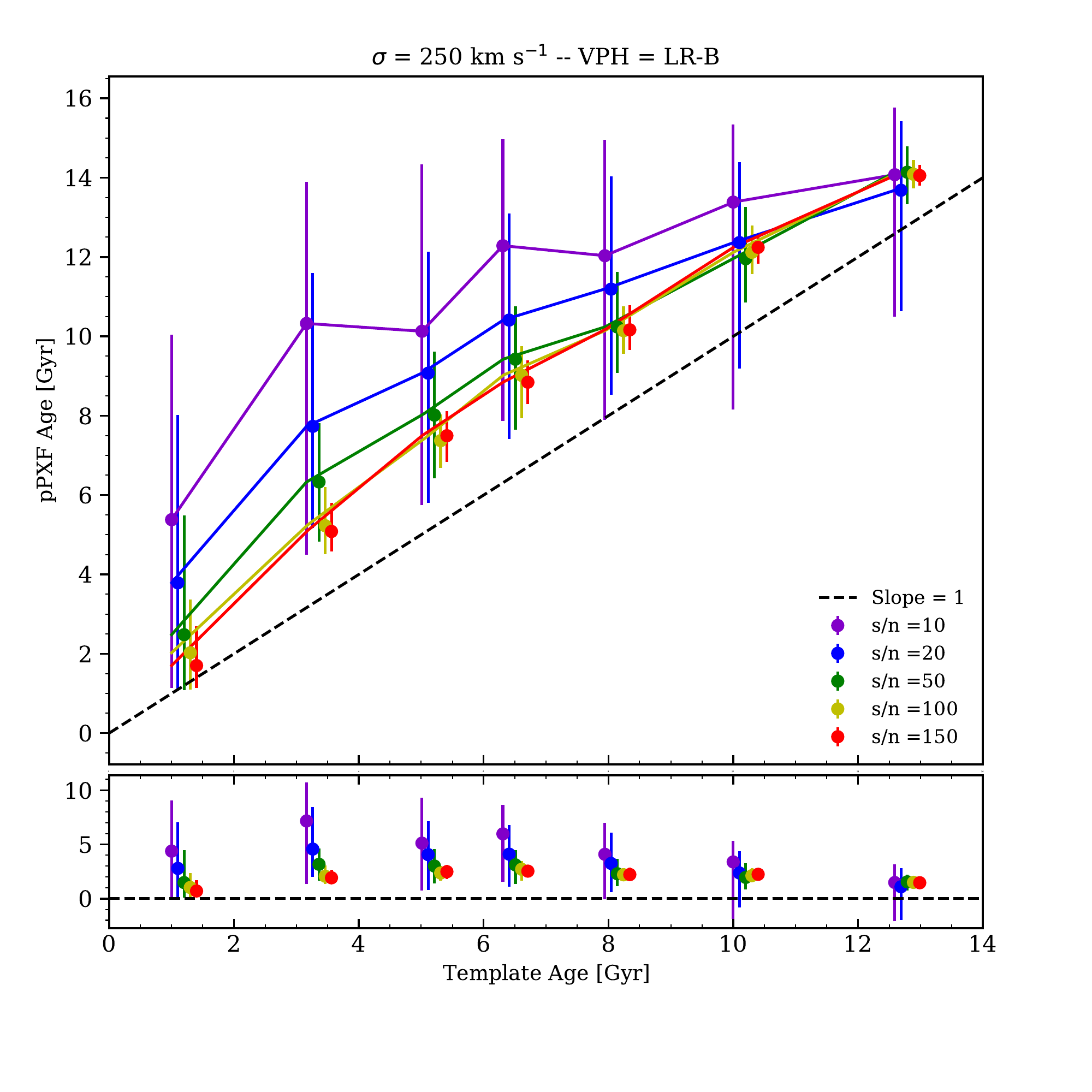}
% 	\caption{Age gradient for LR-V VPH fitted with MILES SSP.}
	\includegraphics[width=0.33\linewidth]{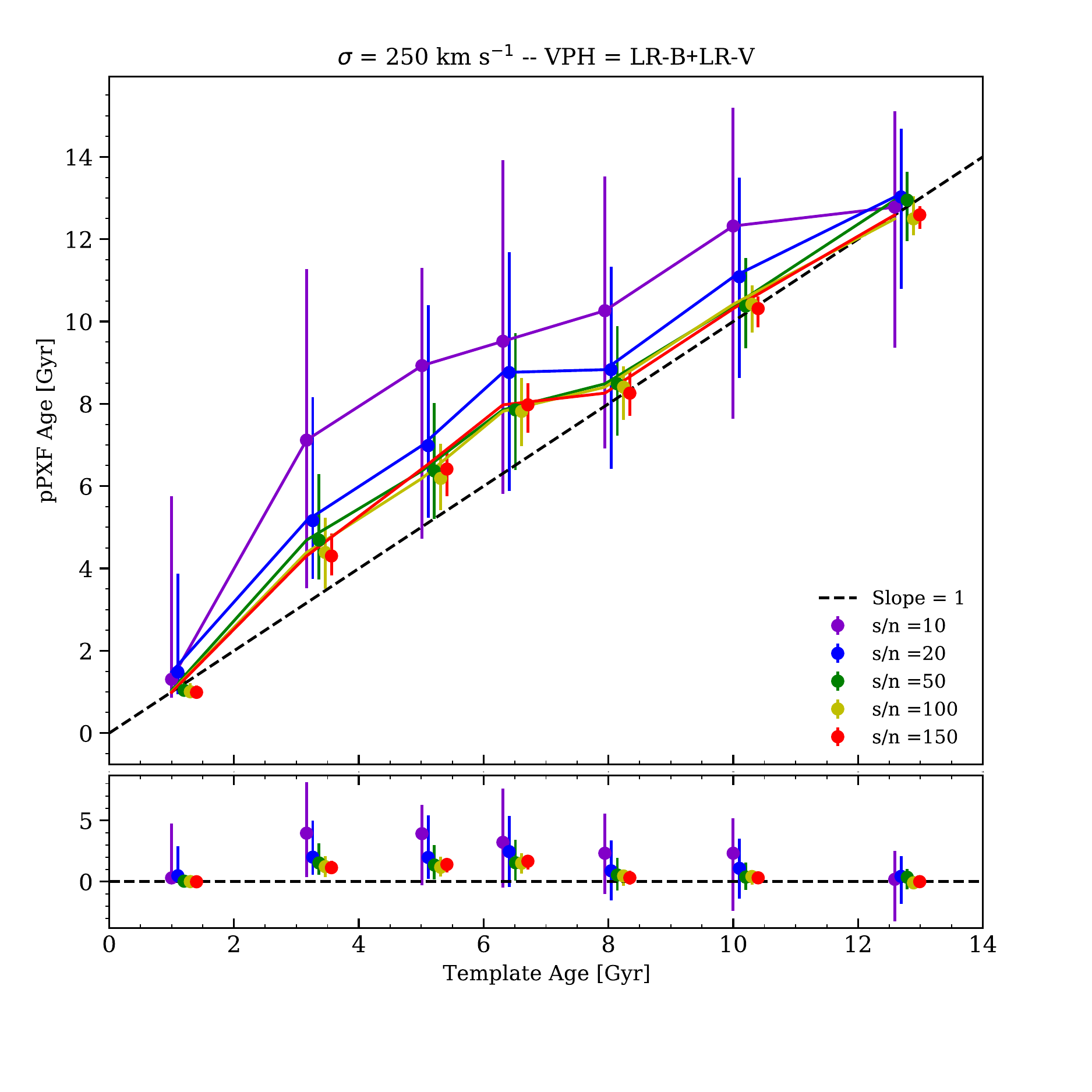}
	\caption{Age of the original SSP models versus the results obtained by analysing with pPXF the spectra of different realisations, all broadened to $\sigma$=250\,km\,s$^{-1}$. We show the results from different spectral ranges, from left to right: LR-V, LR-B and LR-B+LR-V and several signal-to-noise ratios in different colours.}
	\label{fig:diferencia_edad_templates}
\end{figure*}

This is simply one of the examples of the results that could be obtained out of all the possible combinations outlined above which we should now analyse in more detail. Thus, Figure~\ref{fig:diferencia_edad_templates} shows the difference between the age of the original templates and the one predicted by pPXF for different spectral ranges and velocity dispersion values.

We see that, in general, our fitting method (based on pPXF; \citealt{Cappellari_2004}) tends to overestimate the age of the templates, especially in intermediate-age SSPs. As expected, as the signal-to-noise ratio increases, the results improve both in terms of variance and offset. An increase in these differences is also seen as the spectral lines become broader but this effect (not shown here) is not as significant, at least for the range in line widths considered here. These performance tests naturally lead to relative results (such as age gradients) being more robust than those based on absolute age determinations. 

\begin{figure}[h]
	\centering
    \includegraphics[trim={0 0 1cm 1,1cm}, clip,width=1\linewidth]{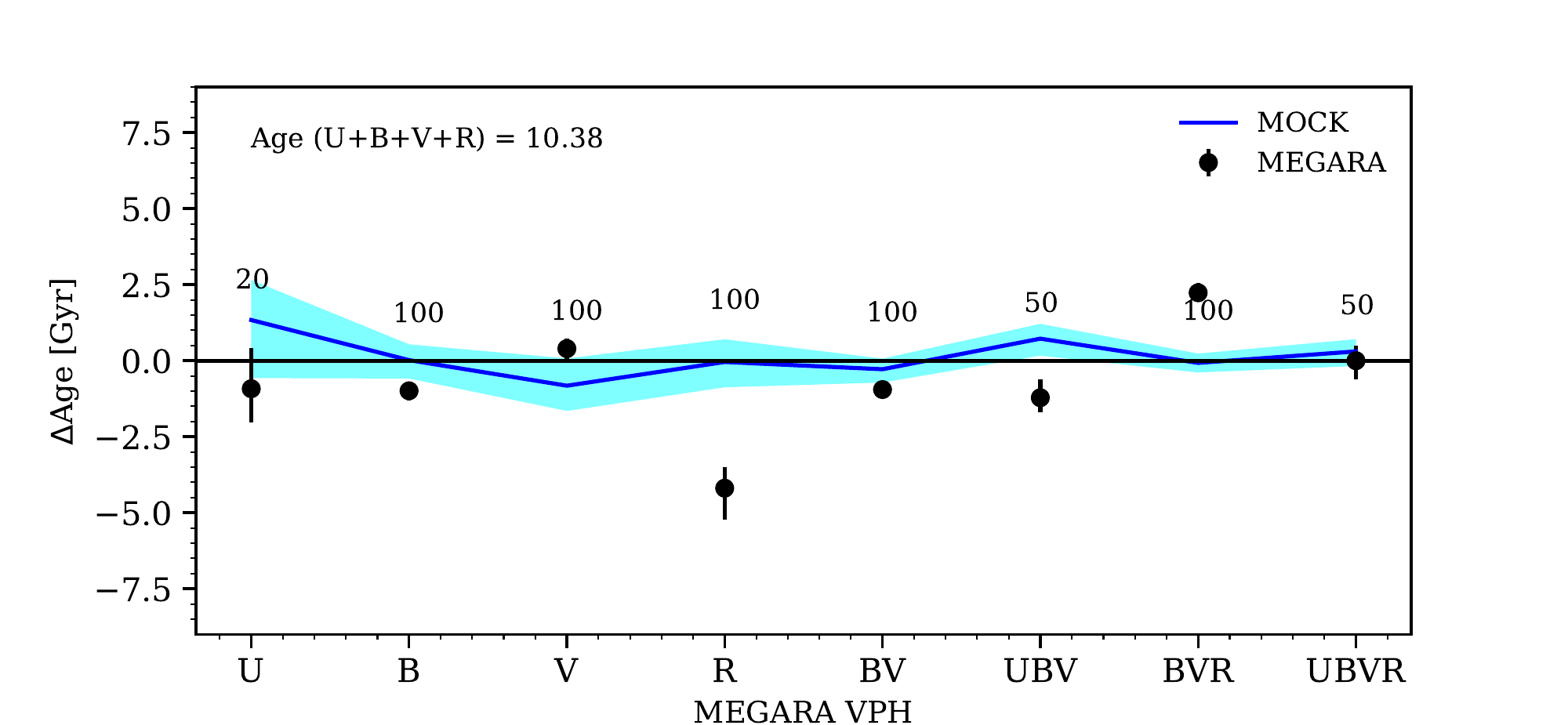}
	\caption{The black points represent the ages obtained using MEGARA data from the innermost elliptical region of NGC~7025 taken with different VPHs combinations, all referenced to the rightmost point, which was obtained using LR-U+LR-B+LR-V+LR-R data. The error bars represent the dispersion of the results at 1$\sigma$. The blue solid line and shaded areas represent the mean and 1-$\sigma$ differences relative to the corresponding SSP mock model age. The numbers labelled in this plot represent the signal-to-noise ratios of the mock models considered, which are also most similar to those reached with our NGC~7025 observations.}
	\label{fig:combine}
\end{figure}

This information is relevant for understanding the limitations of our methodology and, especially, for its application to actual observations of the ongoing MEGADES survey with MEGARA at GTC. In Figure \ref{fig:combine} we show how the age results behave when using MEGARA data for the innermost elliptical aperture measured in NGC~7025 and different VPHs combinations (black scatter points). On the other hand, the blue solid line and corresponding shaded areas represent the average and 1-$\sigma$ results obtained from the analysis of the SSP mock spectra, as described above. All ages are referenced to the age obtained by combining all the VPHs discussed in this paper, that also are the ones consider for its use as part of MEGADES (LR-U, LR-B, LR-V and LR-R). Interestingly, we find that LR-B, LR-V and LR-B+LR-V are the best combinations given the exposure times and signal-to-noise ratios targeted with our MEGADES observations. 

\end{appendix}

\bibliographystyle{aa}
\bibliography{biblio}
\end{document}